\documentclass[iop,twocolappendix]{emulateapj}
\usepackage{color}
\usepackage{threeparttable}
\usepackage{array}
\newcolumntype{R}[1]{>{\raggedleft\let\newline\\\arraybackslash}p{#1}}

\usepackage{graphicx}
\usepackage{float}
\usepackage{amsmath}
\usepackage{xcolor}

\usepackage{hyperref}
\usepackage{enumitem}
\setlist{parsep=0pt,listparindent=\parindent}
    {\pagebreak[4]\global\pdfpageattr\expandafter{\the\pdfpageattr/Rotate 90}}%
    {\pagebreak[4]\global\pdfpageattr\expandafter{\the\pdfpageattr/Rotate 0}}%

\newcommand{\JHU}{Department of Physics and Astronomy, The Johns Hopkins University, Baltimore, MD 21218.}
\newcommand{\STScI}{Space Telescope Science Institute, Baltimore, MD 21218.}
\newcommand{\Kavli}{University of Chicago, Kavli Institute for Cosmological Physics, Chicago, IL, USA.}

\altaffiltext{1}{Department of Astronomy and Astrophysics, University of California, Santa Cruz, CA 92064, USA}
\altaffiltext{2}{Corresponding author: david.jones@ucsc.edu}
\altaffiltext{3}{\Kavli}
\altaffiltext{4}{Hubble, KICP Fellow}
\altaffiltext{5}{\JHU}
\altaffiltext{6}{\STScI}
\altaffiltext{7}{Harvard-Smithsonian Center for Astrophysics, 60 Garden Street, Cambridge, MA 02138, USA}
\altaffiltext{8}{Department of Physics, Harvard University, Cambridge, MA 02138, USA}
\altaffiltext{9}{Astrophysical Institute, Department of Physics and Astronomy, 
251B Clippinger Lab, Ohio University, Athens, OH 45701, USA}
\altaffiltext{10}{Institute for Astronomy, University of Hawaii at Manoa, Honolulu, HI 96822, USA}
\altaffiltext{11}{Department of Physics, Durham University, South Road, Durham DH1 3LE, UK}
\altaffiltext{12}{Astrophysics Research Centre, School of Mathematics and Physics, Queen's University Belfast, Belfast BT7 1NN, UK}
\altaffiltext{13}{Max-Planck-Institut f\"{u}r Astrophysik, Karl-Schwarzschild-Str. 1, DE-85748 Garching-bei-M\"{u}nchen, Germany}
\altaffiltext{14}{Tuorlan Observatorio,  Väisäläntie 20, 21500 Piikkiö, Finland}

\def\numspec{3,930}
\def\numspecgood{3,261}
\def\numsneps{1,169}
\def\numsnetotal{1,364}
\def\numlowz{195}

\def\fp{\textit{Fitprob}}

\def\w{-0.989$\pm$0.057}
\def\wa{-0.513$\pm$0.826}
\def\wbao{-0.984$\pm$0.048}
\def\wbaoho{-1.045$\pm$0.045}
\def\wabao{-0.313$\pm$0.418}
\def\wabaoho{-0.372$\pm$0.452}
\def\massstep{0.102$\pm$0.017}

\begin{document}

\title{Measuring Dark Energy Properties with Photometrically Classified Pan-STARRS Supernovae. II. Cosmological Parameters}
\author{D. O. Jones\altaffilmark{1,2}, D. M. Scolnic\altaffilmark{3,4}, A. G. Riess\altaffilmark{5,6},
  A. Rest\altaffilmark{5,6}, R. P. Kirshner\altaffilmark{7,8}, E.
  Berger\altaffilmark{7}, R. Kessler\altaffilmark{3}, Y.-C. Pan\altaffilmark{2}, R. J. Foley\altaffilmark{2},
  R. Chornock\altaffilmark{9}, C. A. Ortega\altaffilmark{5}, P. J. Challis\altaffilmark{7},
W. S. Burgett\altaffilmark{10},
K. C. Chambers\altaffilmark{10}, P. W. Draper\altaffilmark{11}, 
H. Flewelling\altaffilmark{10},
M. E. Huber\altaffilmark{10}, N. Kaiser\altaffilmark{10}, 
R.-P. Kudritzki\altaffilmark{10}, N. Metcalfe\altaffilmark{11}, 
J. Tonry\altaffilmark{10},
R. J. Wainscoat\altaffilmark{10}, C. Waters\altaffilmark{10},
E.E.E. Gall\altaffilmark{12,13},
R. Kotak\altaffilmark{12,14}, M. McCrum\altaffilmark{12},
S. J. Smartt\altaffilmark{12}, K. W. Smith\altaffilmark{12}}

\begin{abstract}

  We use 1169 Pan-STARRS supernovae (SNe) and
  195 low-$z$ ($z < 0.1$) SNe\,Ia to measure cosmological
  parameters. Though most Pan-STARRS SNe lack spectroscopic
  classifications, in a previous paper (I) we demonstrated that
  photometrically classified SNe can be used to infer unbiased
  cosmological parameters by using a Bayesian methodology that
  marginalizes over core-collapse (CC) SN contamination. Our
  sample contains nearly twice as many SNe as the largest previous
  SN\,Ia compilation. Combining SNe with Cosmic Microwave
  Background (CMB) constraints from Planck, we measure
  the dark energy equation of state parameter $w$ to be
  -0.989$\pm$0.057 (stat$+$sys). If $w$ evolves
  with redshift as $w(a) = w_0 + w_a(1-a)$, we find $w_0 = -0.912 \pm 0.149$
  and $w_a =$ -0.513$\pm$0.826. These results are consistent with
  cosmological parameters from the Joint Lightcurve Analysis
  and the Pantheon sample. We try
  four different photometric
  classification priors for Pan-STARRS SNe and two alternate
  ways of modeling CC\,SN contamination, finding that
  no variant gives a $w$ differing by more
  than 2\% from the baseline measurement.
  The systematic uncertainty
  on $w$ due to marginalizing over CC\,SN contamination,
  $\sigma_w^{\textrm{CC}} = 0.012$, is the third-smallest source
  of systematic uncertainty in this work.  We
  find limited (1.6$\sigma$) evidence for evolution of the
  SN color-luminosity relation with redshift,
  a possible systematic that could constitute a significant
  uncertainty in future high-$z$ analyses.
  Our data provide
  one of the best current constraints on $w$, demonstrating
  that samples with $\sim$5\% CC\,SN contamination can give
  competitive cosmological constraints when the contaminating
  distribution is marginalized over in a Bayesian framework.
\end{abstract}


\section{Introduction}
\label{sec:intro}

The cause of the universe's accelerating
expansion at late times is one of the fundamental
questions in astrophysics today.  Twenty years
ago, distances from Type Ia supernovae (SNe\,Ia)
revealed that the universe was accelerating \citep{Riess98,Perlmutter99}
and the most common interpretation of this cosmic acceleration was that
$\sim$70\% of the energy in the present day universe
must consist of a repulsive ``dark energy''.
In the time since this discovery, large SN datasets
have compiled up to $\sim$750 spectroscopically confirmed SNe\,Ia
and measured the expansion history of
the universe at $z \lesssim 1$ with increasing precision
\citep{Riess04,Kessler09,Hicken09,Conley11,Sullivan11,Suzuki12,Rest14,Betoule14}.
Because SNe\,Ia are observed in the recent cosmic
epochs when dark energy is most dominant, they have
more leverage to measure dark energy than most other
cosmological probes \citep{Weinberg13}.
In conjunction with baryon acoustic oscillation (BAO) and
Cosmic Microwave Background (CMB) constraints
(e.g. \citealp{Eisenstein05,Bennett03,Anderson14,Planck15}),
SNe\,Ia can be used to infer the dark energy equation of state
parameter $w$ (equal to $P/\rho c^2$, the ratio of pressure to density).

The simplest model of dark energy is a cosmological
constant, a vacuum energy that exerts a spatially
and temporally constant negative pressure ($w = -1$).
However, if $w$ is measured to be greater than $-1$
it would be an indication of ``quintessence'' dark energy, a dynamic
scalar field.  A $w$ value of less than $-1$
would imply so-called ``phantom'' dark energy, which requires
extremely exotic physics \citep{Amendola13}.

Nearly all SN\,Ia analyses have measured a dark energy
equation of state consistent with $w = -1$.  The most
precise measurement to date is that of \citet[hereafter B14]{Betoule14},
who combined 740 spectroscopically confirmed SNe\,Ia from
the Sloan Digital Sky Survey (SDSS; \citealp{Alam15}), the
SuperNova Legacy Survey (SNLS; \citealp{Astier06}) high-$z$ SNe
from HST \citep{Riess07} and low-$z$
SNe \citep{Hamuy96,Riess99,Jha06,Hicken09a,Hicken09b,Contreras10,Folatelli10}
to form the Joint Light-curve Analysis (JLA).
JLA SNe\,Ia, when combined
with CMB data from the Planck satellite
and BAO constraints
from \citet{Anderson14} and \citet{Ross15}, yield
$w = -1.006 \pm 0.045$ \citep{Planck15}.

Statistical and systematic uncertainties on the JLA
measurement of $w$ are approximately equal.
Though a great deal of recent progress has been made
to lower systematic uncertainties, including the leading
systematic of photometric calibration error \citep{Scolnic15},
lower uncertainties are also possible 
just by adding more SNe\,Ia.  Although
a significant reduction of the statistical uncertainty now requires
hundreds of additional SNe\,Ia, \textit{thousands} of SNe\,Ia
have already been discovered by Pan-STARRS (PS1; \citealp{Kaiser10}).
Thousands more are currently being discovered by the Dark Energy Survey
(DES; \citealp{Flaugher05}) and tens or hundreds of thousands will be discovered by the
Large Synoptic Survey Telescope (LSST) in the coming decade.

Obtaining spectroscopic classifications for thousands of SNe is
prohibitively expensive.  SN\,Ia spectra cannot be efficiently
obtained with multi-object spectroscopy as they have a sparse density
on the sky: their rate is $\sim$10 yr$^{-1}$ deg$^{-2}$ for
those with $R \lesssim 22$ and spectral classifications must be
obtained within $\sim$2 weeks of maximum light.  At the median PS1 redshift
of $z \sim 0.3$, spectroscopic classifications also necessitate $\sim$1
hour or more of 4m-class telescope time per SN.  In addition, $\sim$30\%
of these SNe\,Ia will fail sample selection requirements \textit{after}
their spectrum has been observed and thus cannot be placed
on the Hubble diagram (\S\ref{sec:data}).  Assuming poor weather
on $\sim$30-50\% of nights, 100 nights of 4m-class telescope time
will result in a cosmologically useful sample of just $\sim$400
SNe\,Ia.  In future surveys, such as LSST, the cost of obtaining
spectroscopy for tens of thousands of SN\,Ia will
far exceed the available resources.

The alternative to spectroscopic classifications is using
classifications based only on photometric SN light curves,
but this method subjects the sample to contamination
by core-collapse (CC) SNe and peculiar SNe\,Ia.
However, if cosmological distances can be measured without bias
in a sample with CC\,SN contamination, photometrically classified SNe\,Ia could
be used to measure $w$ without penalty.
To this end, SN light curve classification algorithms have improved
greatly in the last few years.  The advent of LSST
has provided additional motivation to develop quick, robust classification
methods that rely only on limited photometric data
(e.g \citealp{Saha16}).  Machine learning algorithms in
particular have been found to yield both efficiencies
(few bona fide SNe\,Ia are misclassified) and sample
purities $\gtrsim$96\% in cases where the
classifier can be trained on a representative SN sample
\citep{Sako14,Lochner16}.


The first measurement of $w$ with photometrically classified SNe,
\citet{Campbell13}, used 752 SDSS SNe, most
lacking spectroscopic classifications, to measure cosmological
parameters. They reduced CC\,SN contamination using the PSNID
Bayesian light curve classifier \citep{Sako11}, among other sample cuts,
and estimated that their final sample had 3.9\% CC\,SN contamination.
However, \citet{Campbell13} did not include a systematic uncertainty budget in their
measurements.  Because CC\,SNe are 1-2 mag fainter than SNe\,Ia,
a contamination fraction of just 2\% could shift the mean distance by
0.02-0.04 mag, equivalent to a 5-10\% difference in $w$ over the redshift
range $0 < z < 0.5$.

For this reason, \citet*{Kunz07} proposed the Bayesian Estimation
Applied to Multiple Species (BEAMS) method to simultaneously
determine the SN\,Ia and CC\,SN distributions.
BEAMS models photometrically selected SN samples as a
combination of SNe\,Ia and CC\,SNe, simultaneously fits
for the contributions of each
and marginalizes over nuisance parameters to give
cosmological parameter measurements.  BEAMS is able to
yield cosmological parameter measurements with less bias and nearly
optimal uncertainties \citep*{Kunz07}.  \citet{Hlozek12},
the first measurement of cosmological parameters from photometrically
classified SNe, used the BEAMS method to measure the cosmic
matter density $\Omega_M$ from SDSS SNe lacking spectroscopic classifications,
but again did not include a systematic uncertainty budget in their measurements.
However, the case of systematic uncertainties in BEAMS was explored
theoretically by \citet{Knights13}, who developed a BEAMS formalism
for correlated SN data that gives reliable cosmological parameter
estimation (see also \citealp{Rubin15} for a treatment of
systematic uncertainties that also includes CC\,SN contamination).

We expanded on this work in \citet[hereafter J17]{Jones17}.
J17 undertook a series of Monte Carlo (MC) simulations to test the application
of a BEAMS-like algorithm to a Pan-STARRS photometrically
classified SN sample and made a
first estimate of the systematic uncertainty on $w$ due
to CC\,SN contamination. We found a statistically insignificant
bias of $\Delta_w^{CC} = -0.001\pm0.004$
and a modest systematic uncertainty of 0.014, which we estimated using four
different SN classification methods and three different contamination models.
J17 also includes SN selection effects (i.e. Malmquist bias), which
were not included in the original BEAMS analyses.

In the current work, we apply the J17 methodology to PS1 SNe to
measure cosmological parameters with robust systematic uncertainties.
Previously, only 10\% of PS1 SNe\,Ia $-$ half of the spectroscopically
classified SN\,Ia sample $-$ had been used to measure
cosmological parameters \citep{Rest14,Scolnic14b}.
The present sample is drawn from 350 spectroscopically classified SNe\,Ia
and \numspecgood\ PS1 SNe with spectroscopic host galaxy redshifts.
We anchor our Hubble diagram with a compilation of spectroscopically confirmed
low-$z$ SNe\,Ia from the CfA1-4 and Carnegie Supernova Project samples
\citep{Riess99,Jha06,Hicken09a,Hicken09b,Contreras10,Folatelli10,Stritzinger11}.
We exclude SDSS and SNLS SNe from this sample
in order to give cosmological constraints that are independent
of previous high-$z$ data.
After applying conventional light curve cuts (e.g., B14),
we will show that \numsnetotal\ PS1+low-$z$ SNe remain.
Statistically, we expect $\sim$5\% of these SNe to be CC\,SN contaminants (J17).

A companion paper, \citet[hereafter S17]{Scolnic17},
compiles 1049 spectroscopically classified SNe\,Ia from PS1 and other surveys to
give cosmological constraints.  S17 presents
the PS1 spectroscopic sample, including improvements
to the PS1 photometric pipeline that are used in this work.  This work
also relies heavily on the detailed analysis and simulations of the
low-$z$ sample in S17 and their improvements to the
relative and absolute photometric calibration of all surveys.

The sample of PS1 SNe with host galaxy redshifts was presented in J17,
including a description of our campaign to measure host galaxy redshifts
for $\sim$60\% of all SN candidates.
In \S2 we briefly discuss this sample and present the low-$z$ and PS1
spectroscopically classified SNe that are included in this
analysis.  We also derive bias-corrected distance measurements and
estimate the probability that each SN is Type Ia.  In \S3, we
summarize our cosmological parameter estimation methodology and in \S4, we
discuss contributions to the systematic uncertainty budget.
In \S5, we perform consistency checks on the methodology.  In \S6, we
give measurements of $\Omega_M$ and $w$
from SN\,Ia$+$CMB constraints.  In \S7, we present
combined cosmological constraints after combining SNe with CMB, BAO and local H$_0$
measurements and compare our constraints to B14 and S17.  In \S8,
we examine the test case of measuring $w$ from a SN sample without any
$z > 0.1$ spectroscopic classifications.  Our conclusions are in \S9.

\section{Distances and Photometric Classifications from the Supernova Data}
\label{sec:data}

\begin{deluxetable*}{lcccc}
\tabletypesize{\scriptsize}
\tablewidth{0pt}
\tablecaption{SALT2-Based Data Cuts}
\tablehead{
& \multicolumn{3}{c}{Number of SNe}&Comments\\
&PS1 Host-$z$&PS1 SN-$z$&Low-$z$&}
\startdata
Total candidates&5235&\nodata&\nodata&\nodata\\
Host Sep $R < 5$&4461&\nodata&\nodata&likely host galaxy can be identified\\
Good host redshifts&3147&\nodata&\nodata&\nodata\\
Fit by SALT2&2534&\nodata&\nodata&SALT2 parameter fitting succeeds\\
Not an AGN&2448&174&315&separated from center or no long-term variability\\
$-3.0<x_1<3.0$&1938&168&296&SALT2 light curve shape\\
$-0.3<c<0.3$&1523&160&258&SALT2 light curve color\\
$\sigma_{\textrm{peakMJD}} < 2\times(1+z)$&1490&159&254&uncertainty in time of max. light (rest frame days)\\
$\sigma_{x_1} < 1$&1111&147&253&$x_1$ uncertainty\\
fit prob. $\ge$ 0.001&1053&142&195&$\chi^2$ and N$_{dof}$-based prob. from SALT2 fitter\\
Obs. at $t-t_{pk} > 5$ days&1031&137&195&Observed after maximum at $5 < t-t_{pk} <45$ days\\
E(B-V)$_{MW} < 0.15$&1031&137&195&Milky Way reddening\\
\enddata
\tablecomments{The host-$z$ column includes all PS1 SNe with a spectroscopic
  redshift of their host galaxy.  The SN-$z$ column includes only spectroscopically
  classified PS1 SNe \textit{without} spectroscopic host galaxy
  redshifts.  The reasons for this distinction are due to
  selection biases and are discussed in \S\ref{sec:sim}.}
\label{table:cuts}
\end{deluxetable*}

\subsection{Data}
\label{sec:data}

The PS1 Medium Deep Survey covers 10
7-square degree fields in 5 filters, with typical observing
cadences in a given field of 6 observations per 10 days.
The PS1 SN discovery pipeline is described in detail in \citet{Rest14}.
Likely SNe were flagged based on three signal-to-noise ratio (SNR)
$\ge$ 4 observations in the $griz_{PS1}$ filters and no previous
detection of a SN at that position.
The PS1 survey overview is given in \citet{Chambers16}.

Over its four years of operation, PS1 flagged $\sim$5,200 likely SNe.
Spectroscopic followup was
triggered for $\sim$10\% of SNe, typically those with $r \lesssim 22$ mag,
on a wide variety of spectroscopic instruments (see \citealp{Rest14} and S17).
For 520 of these candidates, spectroscopic observations of the
SN near maximum light allowed their type to be determined and
approximately 350 of these 520 were spectroscopically classified as Type Ia
(S17).

During the last year of PS1, we began a survey to obtain spectroscopic host galaxy
redshifts for the majority of the sample, both those with SN spectra
and those without.  This survey primarily used the
Hectospec multi-fiber instrument on the MMT \citep{Fabricant05,Mink07}.
We also measured redshifts
with the Apache Point Observatory 3.5m
telescope\footnote{\url{http://www.apo.nmsu.edu/arc35m/}} (APO),
the WIYN telescope\footnote{The WIYN Observatory is a joint
facility of the University of Wisconsin-Madison, Indiana
University, the National Optical Astronomy Observatory and the
University of Missouri.}, and for two of the most southern medium deep fields, the
Anglo-Australian Telescope (AAT).
An additional $\sim$600 of our redshifts
come from SDSS \citep{Smee13} or other public
redshift surveys\footnote{Public redshifts are from 2dFGRS \citep{Colless03},
  6dFGS \citep{Jones09}, DEEP2 \citep{Newman13}, VIPERS
  \citep{Scodeggio16}, VVDS \citep{LeFevre05},
  WiggleZ \citep{Blake08} and zCOSMOS \citep{Lilly07}.}.
We chose targets independent of SN type in order
to build a sample without any color or shape selection bias.
Of \numspec\ targets, the host galaxies of
\numspecgood\ SN candidates had strong
enough spectral features and high enough SNR to
yield reliable spectroscopic redshifts.  These data are discussed in detail
in J17.  We estimate that 1.4\% of
these redshifts are incorrect and, as SNe with incorrect
redshifts are indistinguishable from CC\,SNe when placed
on the Hubble diagram, the incorrect redshift fraction will contribute to the
``contamination'' systematic uncertainty for this sample (also discussed in J17).

Though our sample contains a mix of galaxy types (and $\sim$25\%
of hosts are absorption line galaxies),
we are unable to obtain redshifts for SNe in low surface
brightness hosts.  Previous high-$z$ SN searches
\textit{favored} SNe in low surface brightness hosts, which allow
SN spectra with less host galaxy contamination to be obtained.
In the photometrically classified sample, however, including hostless SNe is impossible and
the hosts with spectroscopic redshifts have a median $r$ magnitude of
20.3.  Therefore, the preponderance of
bright, massive host galaxies gives our sample significantly different
SN and host demographics compared to previous high-$z$ data but
makes it more similar to the nature of the current low-$z$ sample,
which primarily consists of SNe\,Ia found by targeting bright galaxies.


After SN discovery and redshift follow-up, the PS1 light
curves were reprocessed with an enhanced version of the discovery
pipeline that included a more realistic (non-Gaussian) PSF model.
The PS1 photometric pipeline has been improved further for
this analysis and the complementary analysis of
S17.  The improvements
include deeper templates, more accurate astrometric alignment,
and better PSF modeling.  The zeropoint
calibration has also been improved by using the Ubercal process
\citep{Schlafly12,Padmanabhan08}.  Ubercal uses repeat observations
of stars in PS1 to solve for the system throughput, atmospheric transparency
and detector flat field in the $griz_{PS1}$ filters.  It has a photometric accuracy
of better than 1\% over the entire PS1 3$\pi$ survey area.
Pipeline improvements are discussed in detail in S17.


We use a compilation of low-$z$ SNe observed over the
last $\sim$20 years to anchor the Hubble diagram.  Nearly all
of these SNe are all included in the JLA analysis, including
the CfA1-3 SN samples \citep{Riess99,Jha06,Hicken09a,Hicken09b}
and Carnegie Supernova Project SNe from the first data release
(CSP; \citealp{Contreras10,Folatelli10})\footnote{See B14 for a detailed
  description of these data and their respective
  photometric systems.}.  We exclude
Calan/Tololo SNe \citep{Hamuy96} as most lie
below the PS1 3$\pi$ survey area and therefore
cannot take advantage of the PS1-based photometric calibration
system we use in this paper (Supercal; \citealp{Scolnic15}).  We also include
the most recent CfA SN compilation (CfA4; \citealp{Hicken12})
and the second CSP data release \citep{Stritzinger11},
which were not included in the JLA analysis but are used in
the \citet{Rest14} and S17 PS1 cosmological analyses.

\subsection{SALT2 Model}
\label{sec:salt2}

To derive distances from PS1 and low-$z$ SNe, 
we use the SALT2 light curve model \citep[hereafter G10]{Guy10}
to measure the light curve parameters of SNe\,Ia.
We apply the most recent version of SALT2 (SALT2.4), which was
re-trained by B14 to include additional high-$z$
SNe and improve the photometric calibration.

We then use the measured SALT2 light curve parameters to
restrict our sample to SNe with shapes and colors consistent
with normal SNe\,Ia ($-0.3<c<0.3$, $-3 < x_1 < 3$) and well-measured
shapes ($\sigma_{x_1} < 1$) and times of maximum light.
Although the SALT2 shape and color cuts are slightly asymmetric with respect to
the mean of the SN\,Ia populations \citep{Scolnic16}, they are chosen
primarily because they are the range within which the SALT2 model is valid.
As measuring cosmological parameters
from SNe without spectroscopic
classifications adds the potential for
new biases to this work, we also strive for consistency with
previous cosmological analyses whenever possible.  For
this reason, our cuts are nearly identical to
those of B14 with two exceptions.  The first is that we add a cut on
the $\chi^2$ and degrees of freedom of the SALT2 light curve
fit (SALT2 fit probability $>$0.001) that was applied by \citet{Rest14}.  This cut
serves to remove CC\,SNe as well as SNe\,Ia with poor light curve fits.
The second is that we require light curves to have at least one observation
$>$5 days and $<$45 days after maximum, a cut that removes a total of 22 SNe.
Without this cut, it is possible that some light curve fits would
have a multi-peaked probability distribution function
for several SALT2 light curve parameters (an issue raised by \citealp{Dai16}).
The cuts on $x_1$ uncertainty and
time of maximum light uncertainty also serve to remove the biases that could
arise from multi-peaked PDFs.  We have not made a similar cut on
the color uncertainty; although this uncertainty is often
high, it should not bias the SN distances (and any bias would be
removed by our bias correction procedure; \S\ref{sec:bias}).

After fitting, we also remove a maximum of two light curve epochs that lie $>$3$\sigma$
from the best-fit SALT2 model.  1.3\% of light curve epochs between
$-15 < t_{max} < 45$ days are 3$\sigma$ outliers.
We then re-run SALT2 with these data removed.
The purpose of this procedure is to
remove photometric data affected by un-flagged image or subtraction
defects without removing so many data points that CC\,SN light curves
begin to resemble those of SNe\,Ia.  Light curve outlier removal increases the
number of SNe passing the SALT2 fit probability cut by $\sim$10\%
(giving a slightly larger sample size than the one presented in J17)
but does not noticeably increase the CC\,SN contamination.

The SALT2 cuts (Table \ref{table:cuts}) reduce the PS1 spectroscopically confirmed SNe\,Ia
sample by $\sim$30\%.  They reduce the number of PS1 SNe\,Ia
without spectroscopic classifications by 60\%, as these SNe
have lower average signal-to-noise ratios (SNRs\footnote{These SNe more
frequently fail the shape uncertainty cut.  In PS1,
  SNe with $x_1$ uncertainty $<$ 1 have a mean SNR at maximum light
  of 15.6.  SNe with $x_1$ uncertainty $>$ 1 have a mean SNR at
  maximum light of 8.3.}) and a lower fraction with SN\,Ia-like shapes and colors.
Once shape, color and $\sigma_{x_1}$ cuts have been applied,
the time of maximum uncertainty cut and the fit probability cut
remove similar fractions of SNe for both photometric and spectroscopic samples.
The Milky Way extinction cut removes no PS1 SNe, as the Medium Deep fields were
chosen to be in regions of the sky with low Milky Way E(B-V).
The number of SNe remaining after each sample cut is shown in Table \ref{table:cuts}.

Table \ref{table:cuts} also includes selection criteria that apply only to photometrically classified SNe.
These include a requirement that the host galaxy can be identified reliably (using the normalized separation
between the SN and a galaxy center, $R$; \citealp{Sullivan06}).  We also remove potential AGN by
by discarding SN candidates with both evidence for long-term variability and positions within
0.5\arcsec\ of their host centers.

Once light curve parameters have been measured by using the SNANA fitting program to implement the
SALT2 model, we use the Tripp estimator \citep{Tripp98} to infer the SN
distance modulus from these light curve parameters:

\begin{equation}
\mu = m_B - \mathcal{M}\ + \alpha \times x_1 - \beta \times c +\Delta_M + \Delta_B.
\label{eqn:salt2}
\end{equation}

\noindent $x_1$ is the light curve stretch parameter, $c$ is
the light curve color parameter, and $m_B$ is the log of
the light curve amplitude (approximately the peak SN magnitude
in $B$).  The distance to a given
SN also depends on the global nuisance parameters $\alpha$,
$\beta$, and $\mathcal{M}$.  $\mathcal{M}$ $-$ a combination of the
absolute SN magnitude and the Hubble constant $-$ $\alpha$, and
$\beta$ are typically marginalized over when fitting to the
cosmological parameters (e.g. B14, \citealp{Conley11}).
$\Delta_M$ is a correction based on the mass of the SN host galaxy,
discussed in \S\ref{sec:hostmass}, and $\Delta_B$ is the
distance bias correction, caused by SN selection effects.
We use simulations to determine an initial $\Delta_B$ and apply
it to the data (\S\ref{sec:bias}) before
measuring $\alpha$, $\beta$, and $\Delta_M$.  After
$\alpha$ and $\beta$ have been measured, we re-determine
$\Delta_B$ using the measured $\alpha$ and
$\beta$ as input in the simulations.  The simulated/measured $\alpha$
and $\beta$ are given in \S\ref{sec:firstconstraints}.


After light curve fitting with the SALT2 model, even SNe\,Ia with low photometric
uncertainties have a $\gtrsim$10\% scatter
in shape- and color-corrected magnitude.
This is traditionally referred to as
the intrinsic dispersion, $\sigma_{int}$ \citep{Guy07}.
$\sigma_{int}$ is defined as
the global uncertainty that must be added in quadrature to the distance
errors $\sigma_{\mu}$ of each SN such that the reduced $\chi^2$ of the Hubble
residuals equals 1.  This is not added to the uncertainty but kept
as a free parameter in the cosmological parameter estimation.
SN\,Ia uncertainties also include redshift uncertainty and
lensing uncertainty ($\sigma_{lens} = 0.055z$; \citealp{Jonsson10}).



\subsubsection{Host Galaxy Masses}
\label{sec:hostmass}

\begin{figure*}
\includegraphics[width=7in]{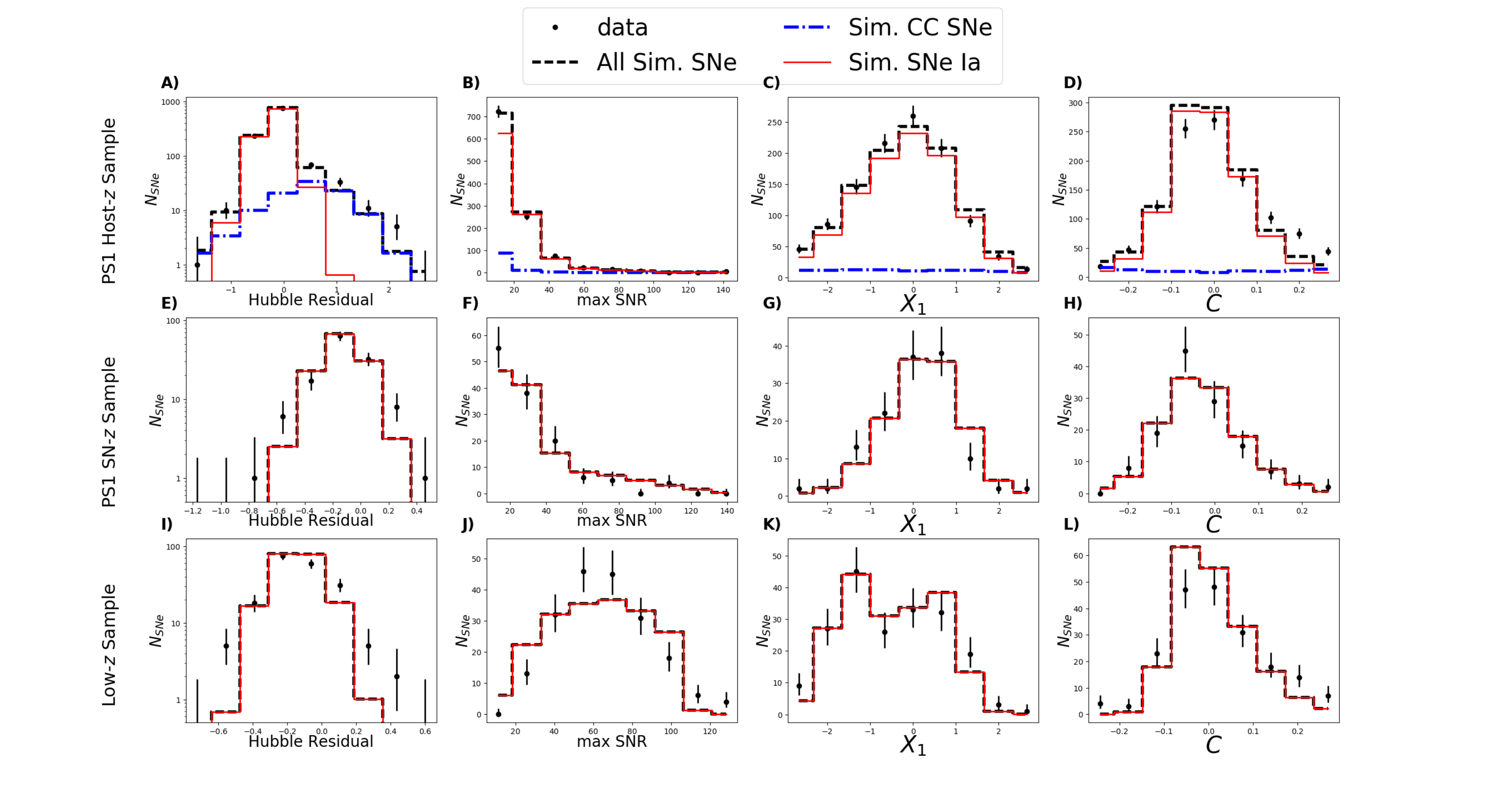}
\caption{Simulations of the PS1 host galaxy redshift sample (host-$z$ sample), the PS1
  SN redshift sample (SN-$z$ sample), and the low-$z$ SN sample
  compared to the real SNe used to measure cosmological parameters
  in this work.  The PS1 host-$z$ sample consists of $\sim$9\% CC\,SN
  contamination, the details of which are discussed in J17 (CC\,SN
  contamination is not relevant for distance bias correction).  }  
\label{fig:sim}  
\end{figure*}

\begin{figure}
\includegraphics[width=3.5in]{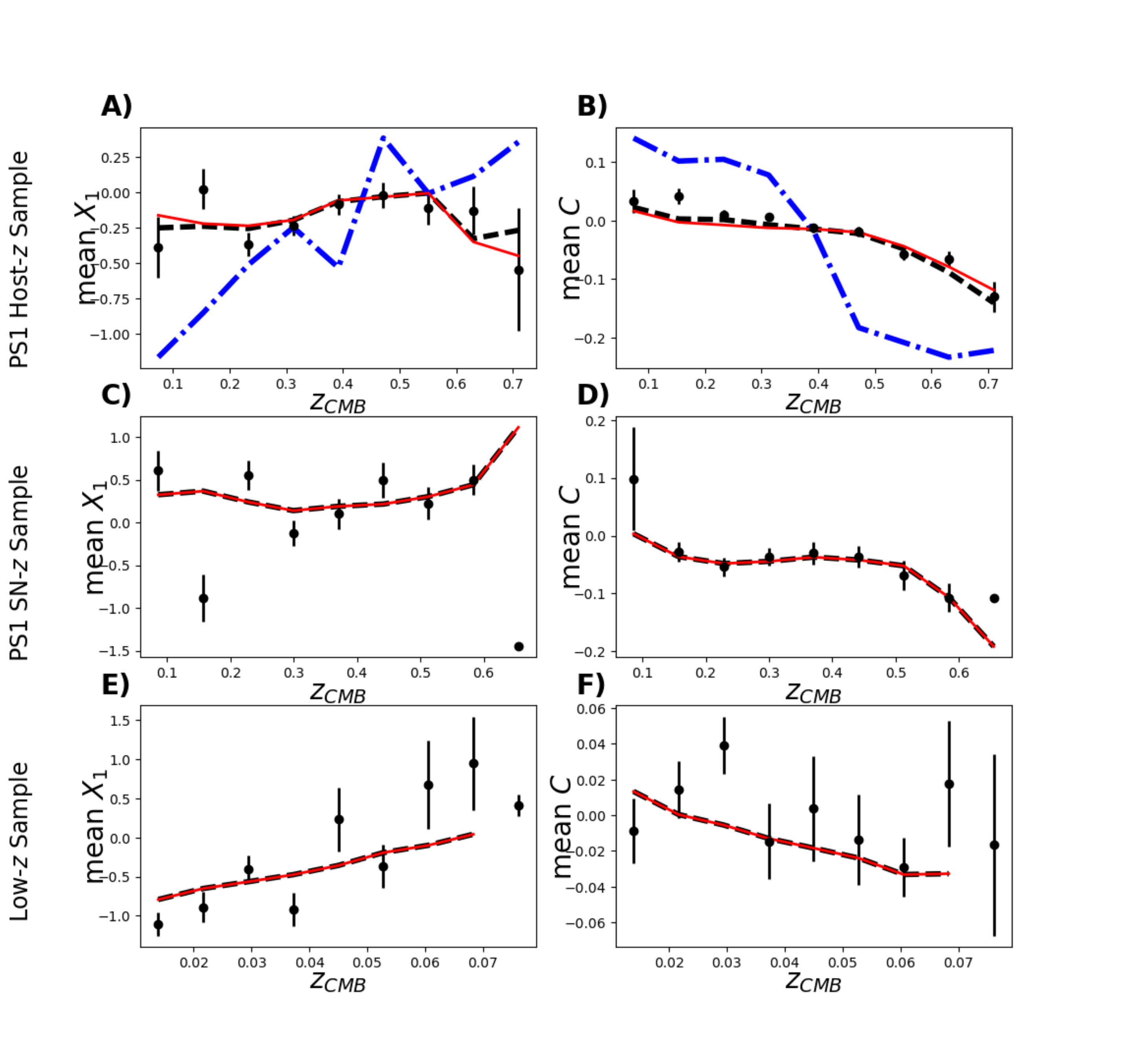}
\caption{Similar to Figure \ref{fig:sim} (see Fig. \ref{fig:sim} legend),
  but showing the dependence of $x_1$ and $c$ on redshift for each survey.}
\label{fig:simztrend}  
\end{figure}

It has been shown that after shape and
color correction, SNe\,Ia are $\sim0.05$-0.1 mag brighter in high mass
host galaxies
(log($M_{\ast}/M_{\odot}) > 10$) than in lower-mass
host galaxies at the same redshifts
($\Delta_M$; \citealp{Kelly10,Lampeitl10,Sullivan10}).
$\Delta_M$ has recently been measured at $>$3$\sigma$ significance
in photometrically classified SN samples even though such
samples (including PS1) have strong selection biases toward high-mass host
galaxies \citep{Campbell16,Wolf16,Uddin17}.  Although the underlying physics
behind the mass step are unclear, a simple step function appears to fit
the SN data well (B14).

Computing $\Delta_M$ robustly requires measuring the
host galaxy masses of every SN
in a self-consistent way.  We therefore measured host
masses using the SED-fitting method of \citet{Pan14}
with PS1 and low-$z$ host galaxy photometry.
For the low-$z$ sample, we use $ugrizBVRIJHK$ photometry from
2MASS \citep{Skrutskie06} and SDSS.
For PS1,
we use SExtractor \citep{Bertin96} to measure the photometry from PS1
templates.  The PS1 templates
are comprised of $\sim$3 years of co-added PS1 data, omitting only the
year in which the SN\,Ia occurred.

The likely host
of each SN is assumed to be the galaxy with the lowest $R$ parameter
relative to the SN position, as
discussed in J17.  The $R$ parameter defines a separation
between the SN and a galaxy center and is normalized by the
size of the galaxy in the direction of the SN\footnote{We predict
that for $\sim$1\% of SNe, this method will incorrectly determine
the host galaxy, but in J17 we determined that this fraction of
mismatches does not bias the cosmology.}.  If the nearest host has
$R > 5$ (i.e, the SN spectrum gives the only redshift), we assume
the true host was undetected following \citet{Sullivan06}.

We use the low-$z$ and PS1 host galaxy photometry
to estimate $M_{\ast}$ with the \texttt{Z-PEG}
SED-fitting code \citep{LeBorgne02}, which in turn is based
on spectral synthesis models from \texttt{PEGASE.2} \citep{Fioc97}.
Galaxy SED templates correspond to spectral types SB, Im, Sd, Sc, Sbc,
Sa, S0 and E.  We simultaneously marginalize over E(B-V), which is allowed
to vary from 0 to 0.2 mag.  Uncertainties are determined from
the range of model parameters that are able to fit the data with
similar $\chi^2$, and are typically $\sim$0.1-0.3 dex.

Undetected galaxies of spectroscopically classified SNe\,Ia
are placed in the log($M_{\ast}/M_{\odot}) < 10$ bin.
At $z \gtrsim 0.5$, we cannot be sure that SN hosts
have log($M_{\ast}/M_{\odot}) < 10$,
and we therefore add a systematic uncertainty of 0.07 mag
in quadrature to those distance uncertainties (similar to B14).



\subsection{Supernova Selection Bias}
\label{sec:distbias}

\subsubsection{Simulating Pan-STARRS and Low-$z$ Supernovae}
\label{sec:sim}

A magnitude-limited sample of SNe will have a distance
bias, caused by SN selection effects, that can be determined from rigorous simulations of the survey
(see e.g., B14, \citealp{Scolnic14b}, \citealp{Conley11}).
We use the SNANA software \citep{Kessler10}
to simulate SNe\,Ia based on the SALT2 model,
with detection efficiencies, zeropoints, PSF sizes, sky noise, and
other observables from the real PS1 and low-$z$ surveys.
We generate the simulations using the values of $\alpha$ and $\beta$
measured from our data as input ($\alpha = 0.161$ and $\beta = 3.060$; \S\ref{sec:checks}).

We use three survey simulations in this analysis:  simulations of
the set of PS1 SNe with redshifts from their host galaxies (the host-$z$
sample), the set of PS1 SNe\,Ia without host redshifts and with \textit{only}
redshifts from SN spectroscopy (the SN-$z$ sample; these SNe
have also been spectroscopically classified),
and the compilation of low-$z$ SNe\,Ia.
It is important that we use distinct simulations for SNe with and without
host redshifts; because SN spectroscopy is only attempted for
bright SNe ($r_{pk} \lesssim 22$), a lower magnitude limit than
the PS1 survey detection limit comes into play for the SN-$z$ sample.
The SN-$z$ sample includes only the portion of our data without host
galaxy redshifts and thus is comprised almost entirely of
$r_{pk} < 22$, spectroscopically classified
SNe in faint hosts ($r_{\textrm{host}} \gtrsim 22$).  On the other hand, the
host-$z$ sample is nearly an ideal, magnitude-limited SN sample,
but it consists only of SNe in brighter ($r \lesssim22$-23) hosts.
Even after shape and color correction, SN Ia luminosity
is a function of the biased host galaxy
properties in these samples, and we must correct for these biases 
using the $\Delta_M$ parameter (variants given
in the systematic error analysis, \S4).  All PS1 simulations
include photometric noise from the host galaxy, as discussed in J17.
Simulations of the PS1 host-$z$
sample are presented in J17 (including CC\,SN contamination, which
we discuss in detail in J17), while the SN-$z$ and low-$z$ samples are
presented in S17.  The sizes of each of the three SN subsamples are
given in Table \ref{table:cuts}.

The host-$z$ sample is also host galaxy magnitude-limited.
Because SN shape and color correlate with host galaxy
brightness (e.g. \citealp{Childress13}),
the SN shape and color distribution in the host-$z$ sample
has a $z$ dependence that is difficult to
model.  Similarly, the SN-$z$ sample consists of
spectroscopically classified SNe for which host galaxy
redshifts could not be measured, and therefore will also
have a biased, $z$-dependent host galaxy distribution.
Because of this, we add one additional component
to the host-$z$ and SN-$z$ simulations: we allow the means of
the simulated SALT2 parameters
$x_1$ and $c$ to evolve slightly with redshift to
better match the data.
We discuss the details and impact of this method in Appendix \ref{section:x1cevol},
and find that it changes the distance bias by up to $\sim$20 mmag
in the highest redshift bins but by less than 5 mmag on average.

The low-$z$ surveys are exceptionally difficult to model due
to the heterogeneous nature of the surveys, their multiple photometric
systems and analysis pipelines, their semi-arbitrary spectroscopic selection
functions, and their targeting of NGC galaxies.  Furthermore, the
cadence and depth of the search is often unknown.  Because of
this, we simulate both a ``magnitude-limited'' variant and a
``volume-limited'' variant of the low-$z$ survey.  We treat
the magnitude-limited variant as the baseline simulation for bias corrections.
The volume-limited variant
matches the observed data with a ``host galaxy targeting''
selection function $-$ fraction of hosts observed as a function of redshift
$-$ instead of a ``spectroscopic follow-up'' selection function
(the fraction of SNe followed as a function of magnitude).
Similarly to the PS1 simulations, we use
redshift-dependent $x_1$ and $c$ distributions due
to the redshift-dependent host galaxy properties ($x_1$/$c$ and
host properties are correlated; \citealp{Childress13}).
These simulations are discussed in more detail in S17.

For each survey, the simulations are compared to the data
in Figures \ref{fig:sim} and \ref{fig:simztrend}.  The distributions of $x_1$, $c$
and their redshift dependences are consistent with the data,
as is the distribution of SN SNRs at maximum light.  Discrepancies on the
red tail of the $c$ distribution could be due to small inaccuracies
in the CC\,SN simulations (J17).
The biggest discrepancies between simulations and data are found in the
low-$z$ simulations due to the difficulty of modeling those searches
and follow-up programs as discussed above.

\subsubsection{Using Simulations to Correct for Selection Bias}
\label{sec:bias}


\begin{figure}
  \includegraphics[width=3.25in]{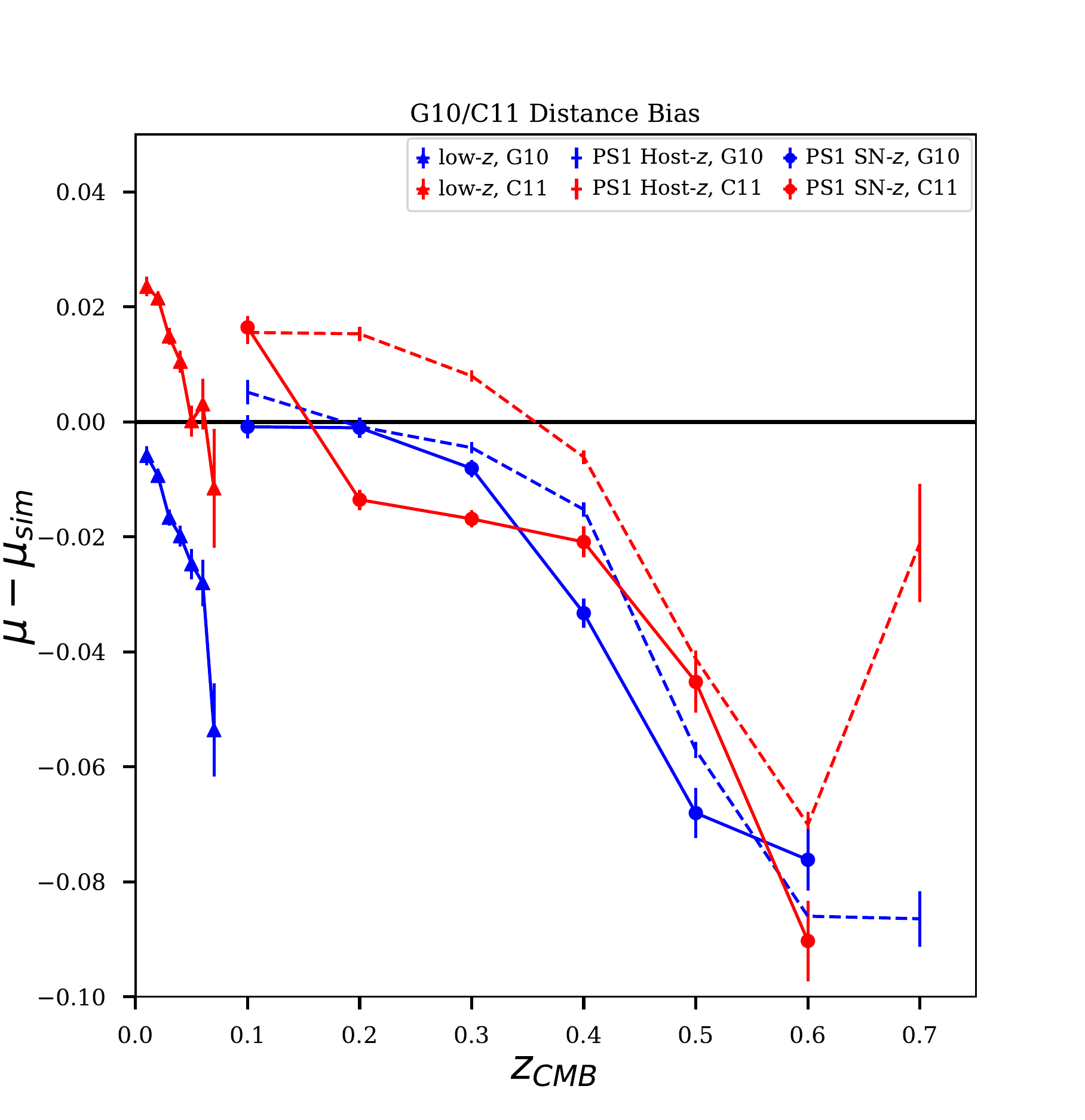}
\caption{Difference in SN\,Ia distance bias for
  the G10 and C11 scatter models.  Low-$z$ SNe have a bias of up to 0.035
  mag while PS1 SNe have a bias of up to $\sim$0.1 mag at the highest
  survey redshifts.  At the highest survey redshifts, where few SNe\,Ia
  can be discovered, the C11 model predicts a drastically
  difference bias from the G10 model due to selection and measurement
  biases in the SALT2 $c$ parameter.}
\label{fig:bias}
\end{figure}

Due to their intrinsic dispersion, SNe\,Ia
discovered in magnitude-limited surveys appear increasingly
luminous at greater distance $-$ even after
shape and color correction.  Even the low-$z$ SN\,Ia
surveys used here may be biased toward preferentially
selecting brighter SNe\,Ia for spectroscopic follow-up
(see B14, their Figure 5).  The bias in distance is given by the
SNANA simulations discussed above and is defined by
\citep{Mosher14}:

\begin{equation}
\Delta_B(z) = \langle \mu_{fit} - \mu_{sim} \rangle_z.
\label{eqn:distbias}
\end{equation}

\noindent For low-$z$ surveys, the bias
can be up to $\sim$0.035 mag ($z > 0.05$), while PS1 has
distance biases of nearly 0.1 mag at $z > 0.5$.  

Uncertainty in the intrinsic dispersion model is the dominant uncertainty
in the bias corrections.
The uncertainty is encapsulated by two primary scatter models that
are both consistent with the data.  First, the G10 SALT2 model
assumes that 70\% of the
$\sim$0.1 mag intrinsic dispersion in derived SN\,Ia distances is
uncorrelated with the shape or color of the SN (achromatic dispersion).  An alternative
model is that of \citet[hereafter C11]{Chotard11}. C11
find an equally good fit to SN data by assuming
75\% of SN dispersion can be attributed to chromatic variation.

The host-$z$ and SN-$z$ biases are very similar, which
is surprising given that SNe in the SN-$z$ sample are much brighter
on average than those in the host-$z$ sample.  The reason is that
the lower average SNR of the host-$z$ sample exacerbates
a bias caused by the $x_1$ uncertainty cut.
At a given SNR, SNe with narrower (measured) light curve shapes
are given lower $x_1$ uncertainties by SALT2.  This introduces
a non-intuitive bias in the case where many $x_1$ uncertainties
are near the cutoff point (for inclusion in our sample) of $\sigma_{x_1} = 1$.
As discussed in J17 (Figure 8), a $\sigma_{x_1} < 1$ sample
cut biases the recovered values of $x_1$ by up to $\alpha(x_1 - x_{1,sim}) = -0.1$
at high-$z$.  The size of this bias is similar to the size of the $m_B$ bias
of spectroscopically confirmed SNe\,Ia ($\sim$0.05 mag
at $z \sim 0.5$).

The SALT2 nuisance parameter $\beta$ is 25\% higher in
the C11 model than the G10 model \citep{Scolnic16},
and these two models can give very different predictions for the
distance bias as a function of redshift
(Figure \ref{fig:bias}).
Due to the chromatic nature of
the C11 dispersion, the C11 bias
is a strong function of the ($z$-dependent) SN
$c$ distribution in a given survey.
This is especially apparent when examining the difference
between the G10/C11 biases for the different samples.  
Low-$z$ and photometrically classified SNe have median $c$ between -0.01
and 0.01, giving an average $\beta^{C11}c - \beta^{G10}c = 0.015$ mag for
low-$z$ and 0.003 mag for PS1 photometrically classified SNe.
In contrast, PS1 spectroscopically confirmed SNe\,Ia have
a median $c$ of -0.04, giving an average difference of
$\beta^{C11}c - \beta^{G10}c = $ -0.028 mag in distance.
Unfortunately,
there are not enough spectroscopically classified SNe\,Ia
to distinguish between the G10/C11 scatter models in our data,
and the differences between these two model predictions
will contribute to our systematic error budget.

\subsection{Photometric Classification}
\label{sec:photclass}

\begin{figure}
  \includegraphics[width=3.25in]{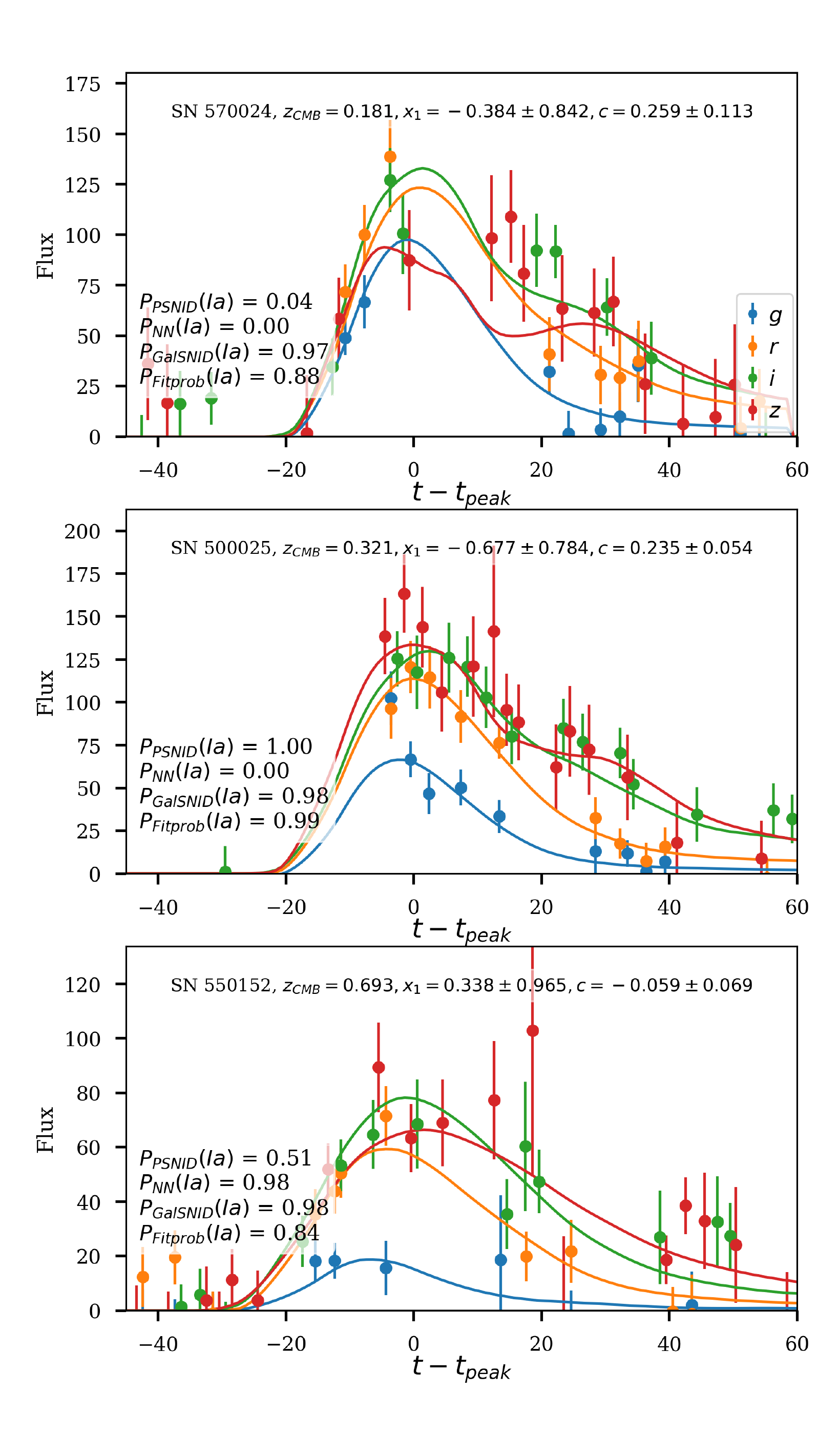}
  \caption{Three PS1 light curves with ambiguous classifications
    included in our sample.  The curves show their best-fit SALT2
    light curve fits.}
  \label{fig:lcex}
\end{figure}

In the previous sections, we made SALT2-based cuts and
distance bias corrections to our data without requiring any knowledge as to which
of the SNe in the photometrically classified sample were SNe\,Ia.
We now use PSNID \citep{Sako14} to classify each SN in this sample
as Type Ia, Ib/c or II based on its light curve.
PSNID matches observed SN light curves to simulated SN\,Ia
and CC\,SN light curves.  The comparison of data to templates
gives a $\chi^2$ and prior-based probability that
a given SN is Type Ia.  We use the version of PSNID that has been
implemented in SNANA\footnote{Version 10.52g.}.  For SNe\,Ia we use the SALT2 model
as the PSNID SN\,Ia template and for CC\,SNe, PSNID
marginalizes over 51 CC\,SN templates
when classifying SNe.  We include a grid of host galaxy reddening values
for each template (because templates have not been corrected for host galaxy reddening,
we allow just $0 < A_V < 1$ of additional reddening).

Although PSNID classifications will be used for the
baseline version of our cosmological analysis (\S\ref{sec:methods}),
we also use three alternate classification methods.  These include
two light curve-based methods, Nearest Neighbor
(NN; \citealp{Sako14}; \citealp{Kessler17}) and \fp.
The NN classifier uses the proximity of SN light curve
parameters to the SALT2 $x_1$, $c$, and redshift
of simulated CC and Ia SNe
to determine the likely SN type.  \fp\ is
the fit probability from the SALT2 light curve fit multiplied
by a redshift-dependent SN type prior.  This prior is
based on simulations, which give the expected
fractions of CC\,SNe and SNe\,Ia at each redshift
(J17, Appendix B).  One additional method, GalSNID
(\citealp{Foley13}; J17), takes advantage of the paucity of CC\,SNe in low star
formation environments to estimate the SN type probability
from only host galaxy properties.  \fp\ and GalSNID
are less accurate classifiers (J17) but are also less subject to the uncertainties in CC\,SN
simulations.  In J17, we suggest that uncertainties
in the shape of CC\,SN luminosity functions and the dearth of CC\,SN templates
for several subtypes necessitate the use of methods that are less reliant
on simulations.

\begin{figure}
  \includegraphics[width=3.25in]{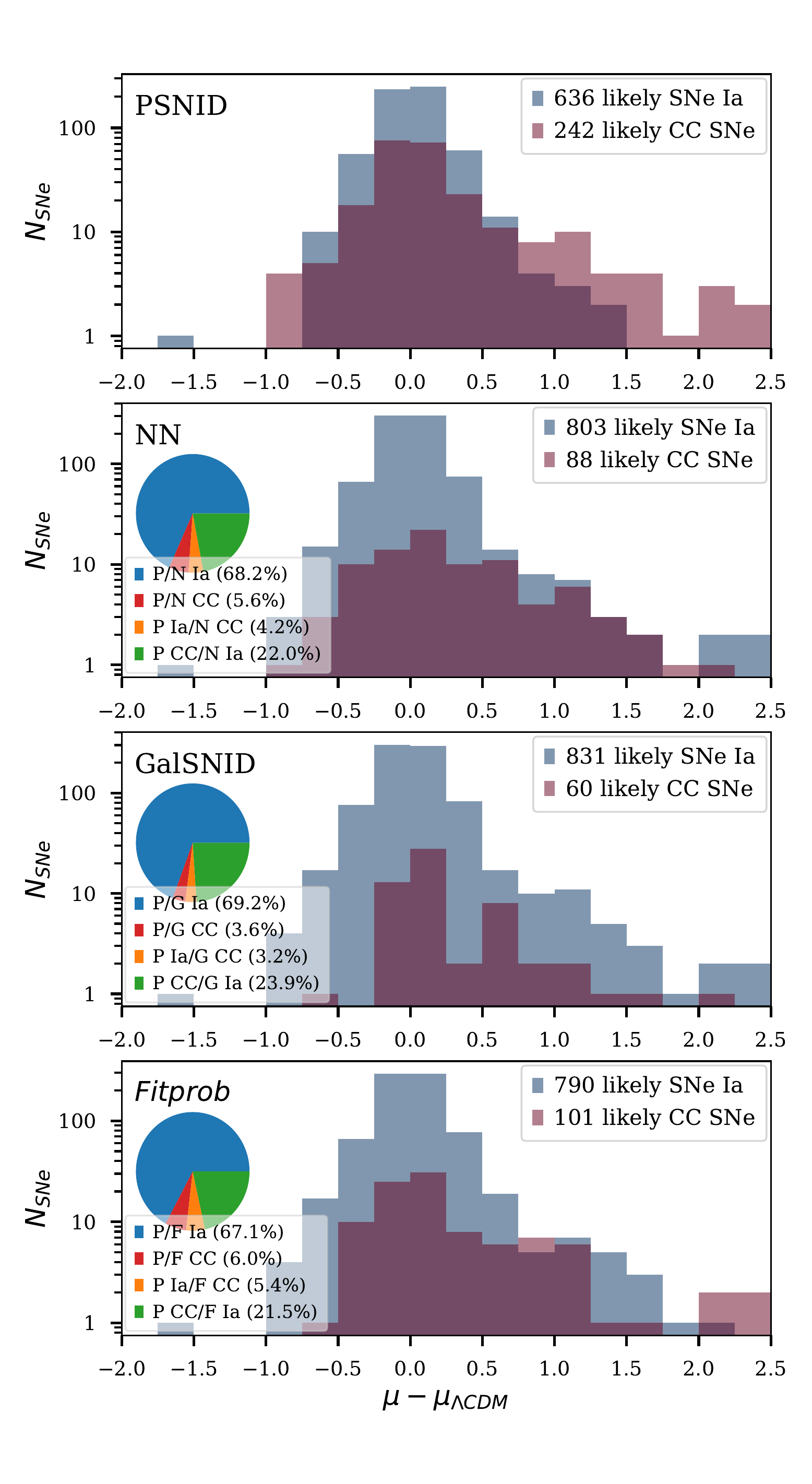}
  \caption{For SNe without spectroscopic classifications,
    log-scaled histograms of Hubble residuals for
    likely PS1 SNe\,Ia (P(Ia) $>$ 0.5; blue) and likely PS1
    CC\,SNe (P(Ia) $<$ 0.5; red)
    from each classifier considered in this work.  $Fitprob$
    classifies the most real SNe\,Ia as CC\,SNe,
    while GalSNID likely classifies
    the most real CC\,SNe as SNe\,Ia.  In spite of large classification
    differences, the SN\,Ia distances 
    given by different classifiers will be shown to be consistent with
    each other and with the spectroscopically
    confirmed PS1 sample (\S\ref{sec:checks}).  The pie charts show the level
    of agreement/disagreement between each classifier and
    PSNID, where P, N, G, and F indicate PSNID, NN,
    GalSNID, and \fp\ classifications.  In these pie charts,
    we label SNe with P(Ia) $>$ 0.5 as Ia and SNe with P(Ia) $<$ 0.5 as CC.}
  \label{fig:class}
\end{figure}

\begin{deluxetable*}{lrrR{0.22in}R{0.22in}rrrrp{0.4in}r}
\tablecaption{PS1 Coordinates and Light Curve Parameters}
\tablehead{SN&$\alpha$&$\delta$&$z_{CMB}^{SN}$&$z_{CMB}^{Host}$&$t_{peak}$&$x_1$&$c$
&$m_B$&P$_{PS}$(Ia)\tablenotemark{a}&log(M$_{Host}$/M$_{\odot}$)}

\startdata
010196&12:16:49.602&46:14:06.33&\nodata&0.369&55246.4(0.2)&-1.457(0.876)&0.228(0.075)&22.970(0.076)&\nodata&11.192(0.155)\\
010203&08:40:02.784&43:26:32.85&0.088&0.087&55230.6(0.1)&1.234(0.163)&-0.061(0.028)&18.135(0.046)&\nodata&10.384(0.009)\\
010204&08:41:36.065&43:24:02.18&0.477&0.477&55241.1(0.4)&-1.727(0.839)&-0.032(0.048)&22.841(0.015)&0.9957&10.519(0.315)\\
010218&09:54:32.47&01:56:37.53&\nodata&0.577&55248.7(0.5)&0.509(0.768)&-0.098(0.056)&23.212(-0.002)&0.8756&9.737(0.151)\\
010222&12:16:56.796&47:17:21.68&\nodata&0.408&55241.3(0.4)&0.139(0.523)&0.028(0.036)&22.386(0.024)&0.9998&10.164(0.301)\\
010230&12:21:10.815&47:48:13.43&\nodata&0.303&55246.3(0.4)&-0.659(0.362)&-0.042(0.035)&21.768(0.043)&1.0000&10.395(0.157)\\
010430&08:45:12.962&43:52:36.68&\nodata&0.327&55258.2(0.5)&-1.447(0.428)&-0.029(0.068)&21.886(0.090)&0.9622&12.070(0.194)\\
020026&12:15:12.803&46:02:40.40&\nodata&0.321&55276.6(0.4)&0.314(0.325)&-0.127(0.031)&21.485(0.036)&1.0000&10.824(0.041)\\
020033&12:17:03.99&46:04:22.76&\nodata&0.530&55268.5(1.5)&-0.888(0.962)&-0.065(0.074)&22.858(0.041)&\nodata&11.737(0.013)\\
020034&12:18:12.126&46:05:10.35&\nodata&0.199&55272.1(0.3)&-1.165(0.431)&0.085(0.034)&20.828(0.042)&\nodata&11.037(0.049)\\
020047&09:56:56.438&01:36:49.80&\nodata&0.266&55260.2(0.8)&-0.952(0.884)&0.111(0.151)&22.832(0.224)&0.9850&10.528(0.150)\\
020075&14:18:54.904&53:59:57.08&0.156&0.157&55286.8(0.4)&2.350(0.465)&0.207(0.053)&21.395(0.085)&\nodata&8.795(0.217)\\
020104&09:58:36.773&02:17:37.70&\nodata&0.306&55277.1(0.6)&0.701(0.533)&-0.080(0.026)&21.507(0.040)&0.9314&10.905(0.257)\\
020123&12:22:25.600&48:02:28.66&\nodata&0.498&55265.3(1.4)&-0.668(0.967)&-0.018(0.087)&22.856(0.062)&0.9915&10.532(0.382)\\
020148&10:38:21.978&57:23:24.30&0.102&0.102&55259.7(0.2)&-1.103(0.241)&-0.023(0.036)&18.980(0.059)&\nodata&11.306(0.065)\\
020194&09:58:34.811&00:49:52.10&\nodata&0.246&55277.9(0.5)&-1.012(0.734)&0.079(0.058)&21.420(0.069)&\nodata&10.965(0.133)\\
020198&10:46:52.536&57:07:45.51&0.361&0.361&55286.2(0.7)&-0.805(0.891)&-0.083(0.081)&22.284(0.119)&0.5443&10.034(0.033)\\
020200&14:12:11.636&53:27:47.48&0.116&0.116&55291.1(0.2)&0.005(0.166)&0.110(0.037)&20.513(0.062)&\nodata&10.400(0.186)\\
030005&12:25:23.414&47:29:11.16&\nodata&0.420&55284.7(0.6)&-1.345(0.617)&-0.002(0.076)&22.315(0.083)&0.9810&11.270(0.018)\\
030007&14:09:23.651&53:37:06.96&\nodata&0.260&55293.8(0.4)&-0.669(0.337)&0.067(0.050)&21.516(0.076)&1.0000&10.030(0.279)\\
030068&12:14:39.906&48:05:21.86&\nodata&0.296&55294.7(0.3)&-0.310(0.789)&-0.012(0.076)&22.474(0.096)&0.9989&11.457(0.102)\\
030216&14:14:56.573&54:12:41.36&\nodata&0.198&55304.1(0.9)&-2.157(0.532)&0.010(0.041)&21.768(0.069)&\nodata&10.799(0.094)\\
030245&12:26:08.645&46:30:52.82&\nodata&0.581&55290.2(0.8)&-0.633(0.686)&-0.188(0.076)&22.748(0.017)&1.0000&10.450(0.451)\\
030252&12:17:28.972&48:05:38.05&\nodata&0.326&55319.1(0.6)&0.746(0.552)&0.094(0.055)&22.240(0.091)&1.0000&8.784(0.193)\\
030263&12:20:47.701&48:10:01.13&\nodata&0.299&55312.0(0.2)&0.717(0.275)&-0.049(0.030)&21.168(0.047)&1.0000&10.116(0.310)\\
040121&10:39:04.003&58:35:25.74&\nodata&0.322&55309.5(0.4)&0.561(0.368)&-0.030(0.038)&21.527(0.049)&1.0000&9.409(0.281)\\
040139&14:17:09.899&53:05:11.39&\nodata&0.267&55324.3(0.9)&-0.211(0.306)&0.146(0.036)&21.599(0.057)&1.0000&10.224(0.281)\\
040147&14:15:40.447&54:13:43.85&\nodata&0.244&55317.4(0.4)&-0.927(0.293)&0.100(0.043)&21.316(0.073)&\nodata&10.459(0.148)\\
040151&12:21:49.674&46:27:04.69&\nodata&0.256&55326.3(0.2)&-1.576(0.326)&0.014(0.032)&21.473(0.043)&1.0000&11.207(0.015)\\
040163&12:22:04.649&47:00:36.58&\nodata&0.416&55321.3(0.2)&-0.248(0.470)&-0.070(0.051)&22.302(0.046)&1.0000&9.822(0.632)\\
040168&12:27:10.791&47:11:23.08&\nodata&0.206&55325.0(0.2)&-0.342(0.541)&0.106(0.047)&20.953(0.053)&\nodata&10.898(0.084)\\
040169&10:49:29.313&58:45:59.05&\nodata&0.421&55332.6(0.3)&0.208(0.937)&-0.009(0.064)&22.716(0.059)&0.6477&10.609(0.241)\\
040170&12:14:21.336&47:50:35.25&\nodata&0.190&55330.4(0.3)&-0.935(0.187)&-0.009(0.028)&20.484(0.039)&\nodata&11.449(0.046)\\
040176&12:20:58.358&45:56:04.95&\nodata&0.348&55327.3(1.7)&1.351(0.778)&0.025(0.042)&21.788(0.047)&1.0000&10.992(0.053)\\
040313&12:16:25.19&48:21:56.92&\nodata&0.266&55335.8(0.6)&-0.394(0.727)&0.235(0.056)&22.598(0.083)&0.9996&9.791(0.105)\\
040316&14:11:23.481&52:26:04.60&\nodata&0.443&55324.0(0.2)&0.286(0.782)&0.216(0.075)&22.913(0.054)&0.6527&10.464(0.140)\\
040318&14:17:19.799&53:06:45.28&\nodata&0.300&55334.4(0.2)&0.109(0.617)&-0.013(0.050)&22.065(0.062)&0.6791&10.588(0.272)\\
040343&10:39:09.733&58:40:39.35&\nodata&0.343&55334.0(0.7)&-0.019(0.645)&0.107(0.055)&22.250(0.055)&0.9789&10.898(0.022)\\
040377&10:40:51.886&58:52:53.50&\nodata&0.352&55302.4(1.1)&1.248(0.629)&0.125(0.067)&22.158(0.071)&0.9959&9.750(0.467)\\
040434&12:21:24.625&45:53:41.62&\nodata&0.654&55315.6(0.9)&-1.624(0.838)&-0.059(0.089)&23.100(0.048)&\nodata&10.861(0.018)\\
040473&10:58:22.122&58:28:59.27&\nodata&0.161&55302.6(1.4)&0.056(0.571)&-0.007(0.130)&19.982(0.146)&\nodata&10.964(0.029)\\
040477&16:20:34.012&54:48:24.17&\nodata&0.346&55332.0(0.6)&0.696(0.418)&-0.023(0.032)&21.532(0.031)&1.0000&11.003(0.037)\\
040511&16:07:40.02&55:07:29.94&\nodata&0.314&55343.1(0.3)&-0.165(0.355)&0.110(0.038)&21.828(0.052)&1.0000&11.382(0.007)\\
040512&16:08:01.033&54:13:24.24&\nodata&0.315&55337.8(0.3)&-1.713(0.334)&-0.088(0.036)&21.797(0.045)&1.0000&10.726(0.029)\\
040530&16:10:49.811&54:49:06.59&\nodata&0.265&55345.8(0.3)&-0.034(0.249)&-0.084(0.031)&21.106(0.045)&1.0000&9.589(0.234)\\
\enddata
\tablecomments{Table 2 is published in its entirety in the electronic 
edition of the {\it Astrophysical Journal}.  A portion is shown here 
for guidance regarding its form and content.}
\tablenotetext{a}{P(Ia) probabilities used in the likelihood model.  These are set to P(Ia) $=$ 1 for
spectroscopically classified SNe\,Ia, and set to the probabilities given by PSNID for photometrically
classified SNe.  SNe without a P(Ia) were unable to be classified by PSNID.}
\label{table:lcparams}
\end{deluxetable*}

Figure \ref{fig:lcex} shows classification
probabilities for three PS1
SNe with ambiguous types.  For SN 570024 (top panel),
two of three light curve-based classification methods agree that this SN
is most likely a CC\,SN due to its poor SALT2 model fit in the $z$-band.
GalSNID, however, finds that this is most likely a bona fide Ia
due to the lack of strong star formation indicators in its host galaxy
spectrum.  For SN 500025 (middle panel), PSNID and $Fitprob$ agree that the
SN is of Type Ia due to the low $\chi^2$ of its light curve fit.
However, the NN classifier finds it most likely to be a CC\,SN
due to its red SALT2 color.  For SN 550152 (bottom panel),
the shapes/colors are consistent with a SN\,Ia but the light curve
fit $\chi^2$ is too high to definitively prefer a SN\,Ia.
This diversity in classification methodologies and outcomes
will help our systematic uncertainty budget to
account for the possibility of cosmological bias due to mistyped SNe.

Figure \ref{fig:class} illustrates the classification
probabilities.
We show the PS1 Hubble residual histograms for
likely SNe\,Ia and likely CC\,SNe without spectroscopic classifications
as determined by each of the
four classification methods considered in this work.
As a diagnostic, if we assume all SNe with Hubble residual $>$1 are CC\,SNe,
we find that PSNID classifies 80\% of these CC\,SNe correctly while NN
classifies 60\% correctly.  \fp\ and GalSNID classify 70\% and 20\% correctly, respectively.
We note that PSNID is unable to classify all SNe, rejecting
13 SNe as too noisy or uncertain for classification.  We
revisit the effect of different classifiers
on our results in \S\ref{sec:checks}.


\begin{deluxetable*}{lrrrrrr}
\tabletypesize{\scriptsize}
\tablewidth{0pt}
\tablecaption{PS1 Host Galaxies}
\tablehead{SN&Host $\alpha$&Host $\delta$&$z_{CMB}^{Host}$&Normalized Sep.\tablenotemark{a}&T\&D $R$\tablenotemark{b}&$z_{\textrm{source}}$}

\startdata
010196&12:16:49.577&46:14:06.27&0.369&0.624&7.340&MMT/Hecto\\
010203&08:40:02.725&43:26:33.14&0.087&1.044&23.880&MMT/Hecto\\
010204&08:41:36.025&43:24:02.56&0.477&1.741&8.080&MMT/Hecto\\
010218&09:54:32.455&01:56:38.21&0.577&2.468&4.280&MMT/Hecto\\
010222&12:16:56.817&47:17:22.39&0.408&1.777&7.280&MMT/Hecto\\
010230&12:21:10.792&47:48:13.10&0.303&1.413&9.420&MMT/Hecto\\
010430&08:45:12.976&43:52:38.13&0.327&1.440&\nodata&SDSS\\
020026&12:15:12.784&46:02:41.30&0.321&1.287&5.040&MMT/Hecto\\
020033&12:17:04.116&46:04:20.33&0.530&2.815&8.830&MMT/Hecto\\
020034&12:18:12.000&46:05:10.87&0.199&1.525&7.380&MMT/Hecto\\
020047&09:56:56.503&01:36:49.99&0.266&1.103&4.200&MMT/Hecto\\
020075&14:18:54.883&53:59:57.18&0.157&1.780&13.710&MMT/Hecto\\
020104&09:58:36.813&02:17:37.53&0.306&0.119&14.020&MMT/Hecto\\
020123&12:22:25.692&48:02:29.95&0.498&2.874&4.190&MMT/Hecto\\
020148&10:38:21.820&57:23:23.82&0.102&0.683&\nodata&SDSS\\
020194&09:58:34.812&00:49:51.78&0.246&0.393&8.830&MMT/Hecto\\
020198&10:46:52.540&57:07:45.40&0.361&0.448&12.460&MMT/Hecto\\
020200&14:12:11.216&53:27:50.95&0.116&3.171&18.190&MMT/Hecto\\
030005&12:25:23.424&47:29:10.85&0.420&0.523&7.480&MMT/Hecto\\
030007&14:09:23.635&53:37:07.06&0.260&1.275&11.920&MMT/Hecto\\
030068&12:14:39.72&48:05:22.10&0.296&2.365&8.320&MMT/Hecto\\
030216&14:14:56.627&54:12:43.21&0.198&1.582&8.360&MMT/Hecto\\
030245&12:26:08.666&46:30:52.84&0.581&0.785&4.610&WIYN/Hydra\\
030252&12:17:28.923&48:05:38.05&0.326&2.510&6.510&WIYN/Hydra\\
030263&12:20:47.774&48:10:00.74&0.299&0.370&4.970&MMT/Hecto\\
040121&10:39:03.921&58:35:27.14&0.322&1.853&4.550&MMT/Hecto\\
040139&14:17:09.885&53:05:11.20&0.267&0.864&23.550&MMT/Hecto\\
040147&14:15:40.454&54:13:43.82&0.244&0.194&8.630&MMT/Hecto\\
040151&12:21:49.768&46:27:06.13&0.256&2.506&16.440&MMT/Hecto\\
040163&12:22:04.663&47:00:37.89&0.416&2.993&5.520&MMT/Hecto\\
040168&12:27:10.792&47:11:22.92&0.206&0.632&16.550&MMT/Hecto\\
040169&10:49:29.349&58:45:59.01&0.421&0.788&8.350&MMT/Hecto\\
040170&12:14:21.256&47:50:39.52&0.190&3.676&21.880&MMT/Hecto\\
040176&12:20:58.375&45:56:04.84&0.348&0.455&11.590&MMT/Hecto\\
040313&12:16:25.161&48:21:56.79&0.266&1.481&7.600&MMT/Hecto\\
040316&14:11:23.433&52:26:04.15&0.443&1.471&11.610&MMT/Hecto\\
040318&14:17:19.807&53:06:45.00&0.300&0.646&12.360&MMT/Hecto\\
040343&10:39:09.748&58:40:39.15&0.343&0.628&5.380&MMT/Hecto\\
040377&10:40:51.832&58:52:53.22&0.352&0.643&15.250&MMT/Hecto\\
040434&12:21:24.595&45:53:41.62&0.654&1.559&4.260&MMT/Hecto\\
040473&10:58:22.401&58:29:00.95&0.161&2.872&17.760&MMT/Hecto\\
040477&16:20:33.996&54:48:24.00&0.346&0.888&11.110&MMT/Hecto\\
040511&16:07:39.907&55:07:28.07&0.314&0.684&14.020&MMT/Hecto\\
040512&16:08:01.035&54:13:25.14&0.315&1.734&6.230&MMT/Hecto\\
040530&16:10:49.922&54:49:06.72&0.265&1.514&19.990&MMT/Hecto\\
\enddata
\tablenotetext{a}{Separation of the SN from the center of its host galaxy,
  normalized by the size and orientation of the host (the $R$ parameter; \citealp{Sullivan06}).
  The isophotal radius of a galaxy corresponds to $R \simeq 3$.}
\tablenotetext{b}{The \citet{Tonry79} cross-correlation parameter,
  computed by comparing the host galaxy
  spectrum to a template spectrum to determine the host redshift.
  Redshifts with $R > 4$ are treated as reliable in this work, though 1.4\% of
  all redshifts are expected to be spurious as discussed in J17.}
\tablecomments{Table 3 is published in its entirety in the electronic 
edition of the {\it Astrophysical Journal}.  A portion is shown here 
for guidance regarding its form and content.}
\label{table:hostparams}
\end{deluxetable*}

\section{Cosmological Parameter Estimation Methodology}
\label{sec:methods}

In the previous section we measured the SALT2 light
curve parameters, host galaxy masses,
SN type probabilities, and bias corrections
that will be used to generate distances from PS1 and low-$z$
SNe\,Ia.  For each SN in our final sample,
these parameters are given in Table 2.
Host galaxy coordinates and redshift information is given
in Table 3. 
Light curves
and host galaxy spectra are available at
\dataset[10.17909/T95Q4X]{https://doi.org/10.17909/T95Q4X}.
From this point forward, we use all PS1 and
low-$z$ data combined $-$ data with and without
spectroscopic classifications $-$ to obtain the best possible measurements
of cosmological parameters.
We will use the PSNID classifications to generate our
baseline, statistics-only cosmological parameter measurements, and will incorporate
the other classification methods into our systematic uncertainty budget.
To reduce CC\,SN contamination, we apply one additional cut on
a classifier-by-classifier basis before estimating cosmological
parameters: we remove SNe with P(Ia) $<$ 0.5.
Therefore, 1109 likely SNe\,Ia will be used in our baseline
cosmological analysis, and between 1263 and 1304 SNe will be
used for the alternate classification methods.

For some readers, the most interesting question
might be whether future cosmological analyses, such as those of DES or LSST,
can robustly measure $w$ \textit{without} a spectroscopically
classified SN sample as part of the data.
We explore this question in \S\ref{sec:nospec}.

With these data, we measure cosmological
parameters from \numsneps\ PS1 SNe and
\numlowz\ low-$z$ SNe\,Ia in two steps: (1) marginalizing
over CC\,SNe and reducing the data to a set of
distance measurements at 25 redshifts
(log-spaced between 0.01 $< z <$ 0.7)
and (2) using those distances, redshifts, uncertainties and covariances
to infer cosmological parameters
with the cosmological Monte Carlo software (CosmoMC; \citealp{Lewis02}).
CosmoMC allows us to easily include the latest CMB, BAO, and/or
H$_0$ priors in our cosmological constraints.  This
two-step procedure is similar to that of B14 (see their Appendix E).

\subsection{The Likelihood Model}
\label{sec:likemodel}

\begin{deluxetable*}{lccc}
\tabletypesize{\scriptsize}
\tablewidth{0pt}
\tablecaption{Free Parameters in the Likelihood Model}
\tablehead{&N$_{\textrm{params}}$&prior&Comments}

\startdata
$f(\vec{z}_{b,Ia})$&25&\nodata&$z$-dependent model of SN\,Ia corrected magnitudes\\
$g(\vec{z}_{b,CC})$&5&2$\pm$3 + $(\mu_{\Lambda CDM}(\vec{z}_{b,CC}) - \mathcal{M})$&$z$-dependent model of CC\,SN corrected magnitudes\\
$\Sigma_{Ia}$&1&0.1$\pm$0.1&SN\,Ia dispersion\\
$\Sigma_{CC}(z_{b,cc})$&5&2$\pm$2&CC\,SN dispersion\\
$\Delta_M$&1&0.07$\pm$0.07&host mass step\\
$\alpha$&1&0.155$\pm$0.05&SALT2 nuisance parameter $\alpha$\\
$\beta$&1&2.947$\pm$0.50&SALT2 nuisance parameter $\beta$\\
$A$&1&1.0$\pm$0.2&re-normalization parameter for P(Ia)\\
$S$&1&0.0$\pm$0.2&shift parameter for P(Ia)\\
\enddata

\tablecomments{List of free parameters and their priors in the BEAMS likelihood
  model.  $z_{b,Ia}$ denotes redshift control points for the SN\,Ia model and
  $z_{b,CC}$ denotes redshift control points for the CC\,SN model.  The
  central values of the $\alpha$ and $\beta$ priors are the best-fit values
  using PS1 spectroscopically confirmed SNe\,Ia alone.}
\label{table:freeparam}
\end{deluxetable*}

The SN likelihood model used here is discussed
and tested comprehensively in J17 and is based on the
Bayesian Estimation Applied to Multiple Species (BEAMS)
algorithm presented in \citet*{Kunz07}\footnote{Our code
  is available online at
  \url{https://github.com/djones1040/BEAMS}}.
 We summarize the model below.

To measure distances from SNe\,Ia, we sample a
posterior distribution $P(\theta|D)$ that is
proportional to a set of priors $P(\theta)$ and
the product (over $N$ SNe) of the likelihoods
of the model given the data for each individual SN.
$D$ is the data, while $\theta$ is the set of free parameters
in the model.  The specific free parameters comprising $\theta$ are discussed
in the paragraphs below and summarized in Table \ref{table:freeparam}.

We use a three-Gaussian
form of the SN likelihood, $\mathcal{L}$.  SNe\,Ia are represented by two
Gaussians: one for SNe\,Ia in low-mass hosts, $\mathcal{L}_i^{Ia,M<10}$, and one
for SNe\,Ia in high-mass hosts, $\mathcal{L}_i^{Ia,M>10}$.
CC\,SNe are represented by the third Gaussian, $\mathcal{L}_i^{CC}$
(alternative CC\,SN models are given in \S\ref{sec:ccsn}):

{\scriptsize
  \begin{equation}
    \begin{split}
      P(\theta|D) &\propto P(\theta) \times \prod_{i=1}^N(\mathcal{L}_i^{Ia,M<10} + \mathcal{L}_i^{Ia,M>10} + \mathcal{L}_i^{CC}),\\
      \mathcal{L}_i^{Ia,M<10} &= \frac{\mathrm{P_i(M < 10)}\mathrm{P_i(Ia)}}{\sqrt{2\pi(\sigma_{i,Ia}^2+\Sigma_{Ia}^2)}}\exp\Big[-\frac{(m^{corr}_{i,Ia}+\Delta_M - f(z_i))^2}{2(\sigma_{i,Ia}^2+\Sigma_{Ia}^2)}\Big],\\
  \mathcal{L}_i^{Ia,M>10} &= \frac{\mathrm{P_i(M > 10)}\mathrm{P_i(Ia)}}{\sqrt{2\pi(\sigma_{i,Ia}^2+\Sigma_{Ia}^2)}}\exp\Big[-\frac{(m^{corr}_{i,Ia}-f(z_i))^2}{2(\sigma_{i,Ia}^2+\Sigma_{Ia}^2)}\Big],\\
  \mathcal{L}_i^{CC} &= \frac{\mathrm{P_i(CC)}}{\sqrt{2\pi(\sigma_{i,CC}^2+\Sigma_{CC}(z_i)^2)}}\exp\Big[-\frac{(m^{corr}_{i,CC}-g(z_i))^2}{2(\sigma_{i,CC}^2+\Sigma_{CC}(z_i)^2)}\Big].
\end{split}
  \label{eqn:beamslike}
\end{equation}
}

\noindent $m^{corr}_{i,Ia}$ and $m^{corr}_{i,CC}$ (in the exponential terms)
are shape- and color-corrected magnitudes for the $i$th SN
that we compute from the SALT2 parameters $m_B$, $x_1$, $c$, and $\Delta_B$ using the Tripp estimator.
They are functions of nuisance parameters $\alpha$ and $\beta$
(Eq. \ref{eqn:salt2}; $m^{corr}_{i,Ia} = \mu_i + \mathcal{M}$).
Because we only wish to measure SALT2 nuisance parameters from
SNe\,Ia, we allow separate values of $\alpha$ and $\beta$ in
the Ia and CC components of the likelihood.
$m^{corr}_{i,Ia}$ values are computed using free parameters
$\alpha_{Ia}$ and $\beta_{Ia}$.  $m^{corr}_{i,CC}$ values
use $\alpha_{CC}$ and $\beta_{CC}$, which are fixed to
the values for SNe\,Ia given
by B14 (allowing these to be free parameters does
not improve the cosmological results).
$\sigma_{i,Ia}$ and $\sigma_{i,CC}$ are the uncertainties on the
corrected magnitudes of the $i$th SN using ($\alpha_{Ia}$, $\beta_{Ia}$) or 
($\alpha_{CC}$, $\beta_{CC}$), respectively.

$\Delta_M$, the mass step, is a free parameter that adjusts the $m^{corr}_{i,Ia}$
of SNe\,Ia in low-mass hosts to match those in high-mass hosts. In
the $\mathcal{L}_i^{Ia,M<10}$ and $\mathcal{L}_i^{Ia,M>10}$ terms in Eq. \ref{eqn:beamslike},
P$_i(M > 10)$ and P$_i(M < 10) = 1 - $P$_i(M > 10)$ are the probabilities
from host masses and host mass measurement uncertainties that a
given SN has a host galaxy with mass $>$10 dex or $<$10 dex, respectively.
We treat the uncertainties as Gaussian, an approximation that predominantly affects only the minority
($\sim$25\%) of SNe that have host masses within 1$\sigma$ of
log($M_{\ast}/M_{\odot}$) $=$ 10.  In previous cosmological
analyses (e.g. B14), the uncertainties on log($M_{\ast}/M_{\odot}$)
were neglected.

If the SN host galaxy has been misidentified, this could contribute
to the systematic uncertainties on cosmological parameters.  But
for the photometrically classified sample, misidentified host galaxies would have incorrect redshifts and are therefore
treated as part of the contaminating distribution ($\mathcal{L}_i^{CC}$).  They then contribute to
the ``contamination'' systematic, as discussed in J17.  For spectroscopically classified SNe without
host galaxy redshifts, we expect only $\sim$2 SNe\,Ia to have misidentified host galaxies
(based on the 1.2$\pm$0.5\% fraction of mismatched host galaxies computed in J17).

$f(z_i)$ is the variable of interest for cosmological parameter estimation.  It is the
continuous, $z$-dependent model for the
SN\,Ia corrected magnitudes $-$ the mean of the SN\,Ia Gaussian $-$ and is allowed to vary across
the redshift range of the survey ($0.01 < z < 0.7$).  We
evaluate the model at any $z$ across this redshift range by choosing a fixed set of 25 log-spaced
redshift ``control points'' ($\vec{z_b}$; $\Delta\mathrm{log_{10}}(z) = 0.077$) at which the corrected SN\,Ia magnitudes
$f(\vec{z}_{b}) = \mu(\vec{z}_{b}) + \mathcal{M}$ are free parameters.
For any redshift $z_i$, we interpolate between the redshift control
points below ($z_b$) and above ($z_{b+1}$):

\begin{equation}
\begin{split}
\mu(z_i) = (1 - \xi)\mu_b + \xi\mu_{b+1}\\
\xi = \mathrm{log}(z_i/z_b)/\mathrm{log}(z_{b+1}/z_b),
\end{split}
\end{equation}

\noindent where $\mu_b$ is the distance modulus at redshift $z_b$.
Interpolating with a simple linear model instead of $\Lambda CDM$
produces differences of $<$1 mmag at all redshifts.  The
SN\,Ia dispersion $\Sigma_{Ia}$ plays the same role as the
intrinsic dispersion and is kept fixed at all redshifts.

The $z$-dependent mean and standard deviation of the CC\,SN
Gaussian model ($g(z_i)$ and $\Sigma_{CC}(z_i)$) are
interpolated between 5 log-spaced redshift control points.
Unlike SNe\,Ia, the dispersion of the heterogeneous
CC\,SN population changes with
redshift due to to strong detection biases at high $z$.

Each Gaussian is multiplied by the prior probability
(P$_i($Ia) and P$_i($CC$) = 1 - $P$_i($Ia)) that a given SN
is or is not of type Ia.  We use the PSNID classifier to
estimate these probabilities.  Alternative classification methods
are included as part of our systematic error budget (\S\ref{sec:ccsn}).

For SNe with photometric classifications,
our method allows the type priors to be shifted and re-normalized to
account for incorrect classifications (see J17).
For spectroscopically classified SNe\,Ia, we set the prior
probabilities, P$_i$(Ia), equal to one and do not allow them to be adjusted.
We include broad Gaussian priors (Table \ref{table:freeparam}) on all free parameters with the exception
of $f(\vec{z_b})$, the SN\,Ia corrected magnitudes.  We apply no priors (i.e. flat priors) to
$f(\vec{z_b})$ to avoid any possibility of
cosmological bias.

We estimate the free parameters by sampling
the log of the posterior with
a Markov Chain Monte Carlo (MCMC) algorithm.
As in J17, we use the Parallel-Tempered Ensemble Sampler
from \texttt{emcee} as our MCMC method \citep{Foreman13}.

\subsection{Constraining Cosmological Parameters}
\label{sec:cosmoest}

\begin{figure*}
  \includegraphics[width=7in]{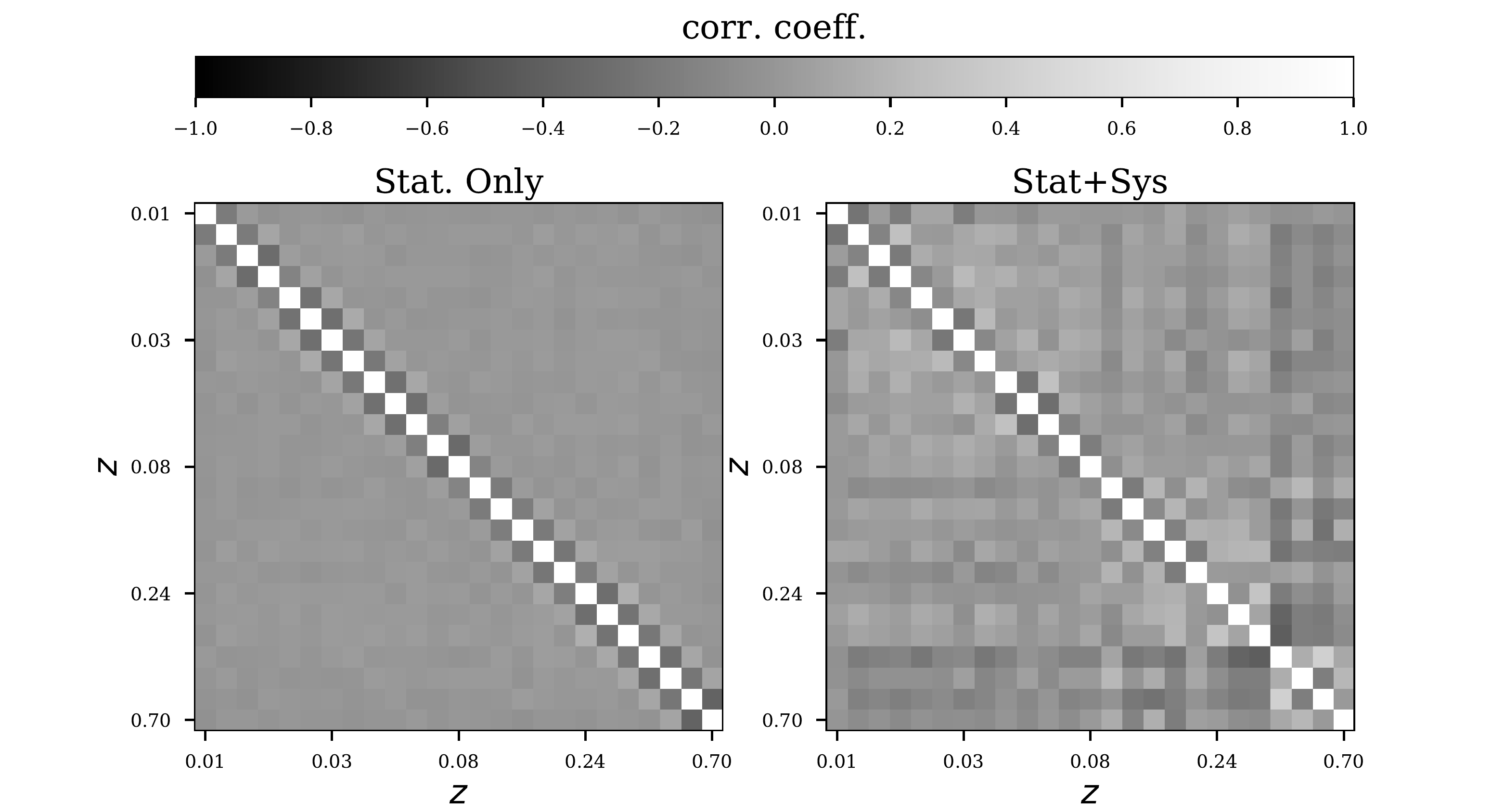}
  \caption{Statistics-only and stat+sys correlation matrices from
    the PS1$+$low-$z$ SN sample.  The statistics-only correlation
    matrix shows the strong anti-correlation between neighboring bins.
    The stat+sys correlation matrix shows larger-scale correlations
    due to systematic uncertainties and
    large uncertainties in the bins with minimal data ($z \sim 0.1-0.2$).
    The correlation matrix is equal to $C_{ij}/\sqrt{C_{ii}C_{jj}}$
    for covariance matrix $C$.}
  \label{fig:corrmat}
\end{figure*}

\begin{figure*}
\includegraphics[width=7in]{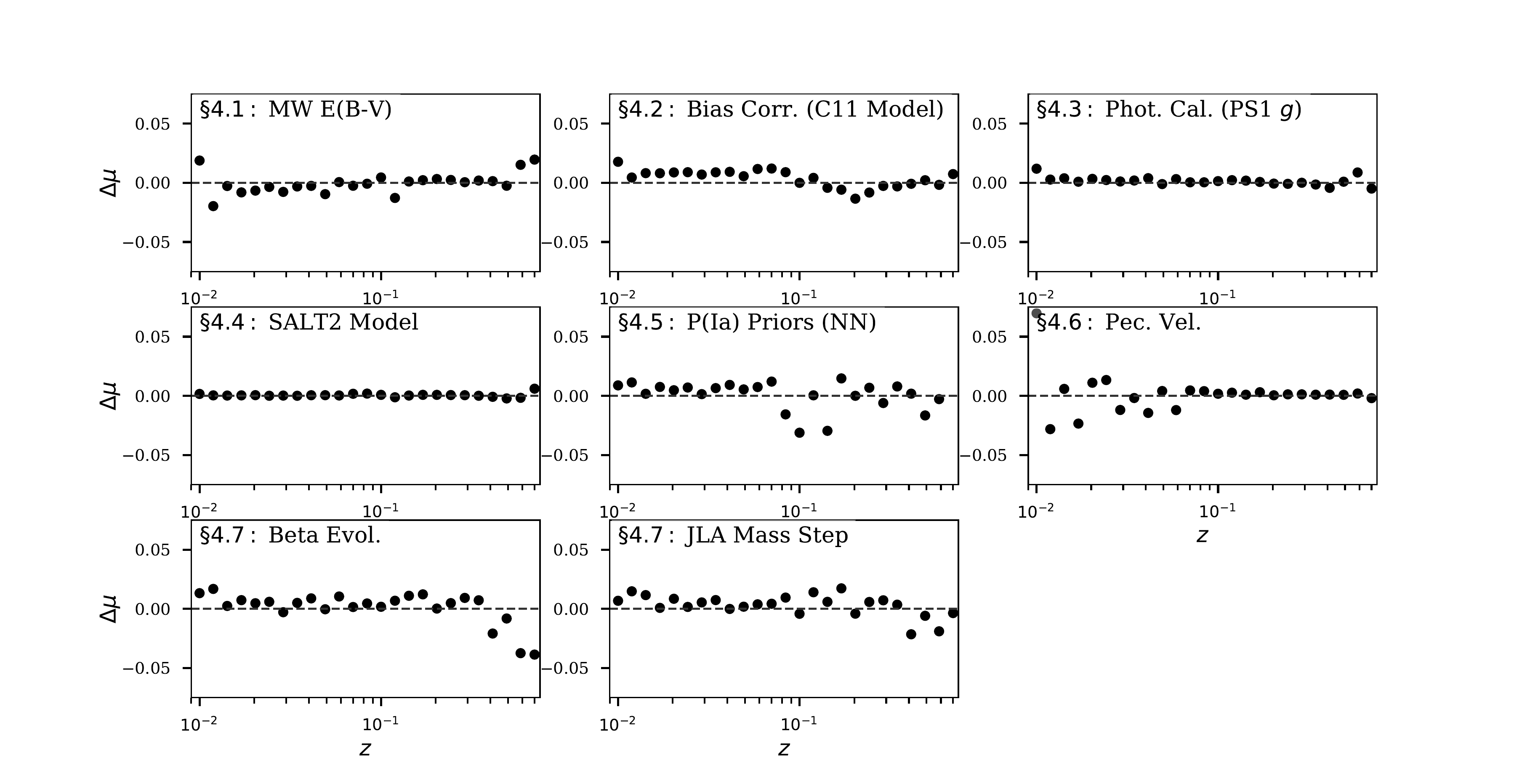}
\caption{The average change in distance modulus $\Delta \mu$ from
  an example of each type of systematic uncertainty in this analysis.  Deviations
  at $z \simeq 0.01$ and $\simeq 0.1$ are primarily due to low SN statistics in these
  bins and have little effect on the cosmological constraints.}
\label{fig:sys}
\end{figure*}

From the methods presented above, we infer the corrected magnitudes
of SNe\,Ia at 25 redshift control points, $f(\vec{z}_{b})$,
using the baseline SN light curve parameters,
bias corrections, and J17 methodology.  We also measure the set of
$f(\vec{z}_{b})$ for each systematic
uncertainty (\S\ref{sec:syserr}).  From these values, a
systematic error covariance matrix $C_{sys}$ is
created \citep{Scolnic14b,Conley11}:

\begin{equation}
C_{sys}^{jk} = \sum_{n=1}^{N}\frac{\partial f(z_j)}{\partial S_n}\frac{\partial f(z_k)}{\partial S_n}\sigma(S_n^2).
\label{eqn:syscov}
\end{equation}
  
\noindent The sum is over all $N$ systematics, and
$\frac{\partial f(z_j)}{\partial S_n}$ is the change
in corrected magnitude after applying a single systematic
$S_n$ to the individual light curves.
$\sigma(S_n)$ is the size of each systematic uncertainty.  The systematic
covariance matrix is then combined with the
statistical covariance matrix:

\begin{equation}
C_{tot} = D_{stat} + C_{sys}.
\end{equation}

\noindent  Note that the
statistics-only covariance matrix, $D_{stat}$,
includes both diagonal and off-diagonal components
because the magnitudes $f(\vec{z_b})$ are anti-correlated
with the neighboring magnitudes
$f(\vec{z_{b+1}})$ and $f(\vec{z_{b-1}})$:

\begin{equation}
D_{stat}^{ij} = \sum_{k}^{N_{MCMC}}\frac{(f_k(z_{b,i}) - \overline{f(z_{b,i})})(f_k(z_{b,j}) - \overline{f(z_{b,j})})}{N_{MCMC}}.
\end{equation}
  
\noindent $N_{MCMC}$ is the length of the MCMC chain that samples free parameters
$f(\vec{z_b})$.  $f_k(z_{b,i})$ is the value of $f$ at the $i$th control point
from the $k$th MCMC sample.  $\overline{f(z_{b,i})}$ is the mean of $f$ at the
$i$th control point from the full MCMC chain.
Figure \ref{fig:corrmat} shows the reduced correlation matrices from statistical
uncertainties alone (left) and statistical and systematic uncertainties
combined (right).  The statistics-only correlation matrix shows significant anti-correlations
between neighboring control points, while the systematic uncertainties add larger-scale
correlations between the control points (see Figure \ref{fig:sys} in \S\ref{sec:syserr}).

We then use the cosmological Monte Carlo software (CosmoMC;
\citealp{Lewis02}) to measure cosmological parameters by minimizing
the following $\chi^2$:

\begin{multline}
\chi^2 = (\mu'(\vec{z_b}) - \mu_{\Lambda CDM}(\vec{z_b}; \Omega_M, w, ...))^{\dagger} C_{tot}^{-1} \\ (\mu'(\vec{z_b}) - \mu_{\Lambda CDM}(\vec{z_b}; \Omega_M, w, ...)),
  \label{eqn:chi2}
\end{multline}
  
\noindent where $\mu'(\vec{z_b}) = f(\vec{z_b}) - \mathcal{M}$ (we marginalize over $\mathcal{M}$ using CosmoMC).
The vector of model distances, $\mu_{\Lambda CDM} = 5$log$(d_L) - 5$,
is a function of the cosmology:

\begin{equation}
  \begin{split}
    d_L(z,w,\Omega_M,\Omega_{\Lambda},\Omega_K) = (1 + z)\frac{c}{H_0}\int_0^z\frac{dz}{E(z)},\\
    E(z) = [\Omega_M(1 + z)^3 + \Omega_k(1 + z)^2 + \Omega_{\Lambda}(1 + z)^{3(1+w)}]^{1/2}.
  \end{split}
\end{equation}

\noindent $\Omega_M$ is the cosmic matter density, $\Omega_{\Lambda}$
is the dark energy density, and $\Omega_k$ is the curvature of space.
$w$ is the redshift-independent dark energy equation of state parameter
($z$-dependence will be added in \S\ref{sec:firstconstraints}).

\section{Systematic Uncertainties}
\label{sec:syserr}

The SNe in this sample are affected by systematic uncertainties
that can broadly be attributed to 8 sources of error: Milky Way extinction,
distance bias correction, photometric calibration, SALT2 model calibration,
sample contamination (primarily by CC\,SNe),
low-$z$ peculiar velocity corrections,
the redshift dependence of SN nuisance parameters,
and the dependence of SN\,Ia luminosities on their host galaxies.
Figure \ref{fig:sys} illustrates the redshift dependence
of each type of systematic uncertainty.  We discuss each of
these uncertainties in detail below.


\subsection{Milky Way Extinction}

Milky Way extinctions for each SN are given by \citet{Schlafly11},
who use the colors of stars with spectra in SDSS to derive a 14\% correction to
the reddening maps of \citet*{Schlegel98}.  We assume a conservative,
fully correlated 5\% uncertainty on the E(B-V) measurements of \citet{Schlafly11}, which could be caused by
selection biases in the SDSS stars chosen for spectroscopic follow-up or the use of
stars that lie in front of some fraction of the Galactic dust \citep{Schlafly11}.

\subsection{Distance Bias Correction}
\label{sec:biassys}

Two effects lead to systematic uncertainties in distance
bias corrections.  The dominant effect is the difference
between the G10/C11 distance bias predictions.
As discussed in \S\ref{sec:bias}, the difference between the G10 and C11 dispersion models
is up to $\Delta\mu(z) \sim 0.03$ mag.
As there is no \textit{a priori}
reason to choose one dispersion model over the other,
we choose to adopt the average of the two
bias predictions for our baseline distance bias
correction.  The systematic error then becomes half
the difference between the G10/C11 bias.

A secondary effect is that uncertainty in the
survey detection limit or 
spectroscopic follow-up selection function
can cause the simulated
distance bias to be inaccurate.  We adjust the detection
efficiency (for the PS1 host-$z$ sample)
and the spectroscopic selection efficiency (for the
PS1 SN-$z$ sample) such that the SNR at maximum
light for simulated SNe matches the data with a 
$\sim$20\% higher reduced $\chi^2$ 
(a 1$\sigma$ difference).
These efficiencies are well-constrained by the data;
the detection efficiency adjustment for the host-$z$ sample,
for example, corresponds to lowering the magnitude limit of the survey
by $\sim$4 mmag.

The low-$z$ distance bias is measured
from low-$z$ simulations that lack reliable
detection and spectroscopic selection efficiencies.
For these simulations, we use the ``volume-limited'' simulations
discussed in \S\ref{sec:sim} as the selection bias systematic.  The volume limited variant
has $<$0.01 mag distance bias using the G10 scatter model (small biases due
to the correlation of Hubble residuals with $x_1$ and $c$ still
arise; \citealp{Scolnic16}),
and a bias of $\sim$0.02 mag using the C11 model because
$\beta_{fit}-\beta_{sim} = 0.7$.
The systematic uncertainty due to the detection limit and
spectroscopic follow-up selection function is subdominant to the
G10/C11 systematic uncertainty.


\subsection{Photometric Calibration Uncertainties}
\label{sec:calib}

In this work, the systematic uncertainties in the photometric
calibration are the same as the S17 analysis.  They are due to
uncertainty in the survey filter functions, uncertainty
in the calibration of HST CALSPEC standard stars,
and uncertainty in the calibration of the PS1/low-$z$ photometric
systems relative to HST.

Uncertainties in the survey filter functions are modeled
as uncertainties in the zeropoints and
effective wavelengths of each filter.  PS1 has a
effective central wavelength uncertainty of 7\AA\ per filter
\citep{Scolnic15}.  The low-$z$
filter uncertainties are typically $\sim$6$-$7\AA\ but
are survey- and filter-dependent.  They can be as high as
25$-$37\AA\ (exact values are given in \citealp{Scolnic15};
see their Table 1).

The relative calibration uncertainties are given by
the Supercal method.  Supercal uses the excellent (sub-1\%)
relative calibration of PS1 across 3$\pi$ steradians to compare
the photometry of tertiary standard stars in previous SN surveys
to the photometry of these same stars on the PS1 system.  Typical corrections are
on the order of 1\%, but can be up to 2.5\% for $B$ band low-$z$
data.  Uncertainties in the Supercal procedure are typically
3-4 mmag per filter but can be up to 10 mmag for low-$z$
surveys such as CfA1.

Finally, there is uncertainty in the AB magnitude system itself
as measured using HST CALSPEC standard stars.  We follow B14
by assuming a global 0.5\% slope uncertainty for the
flux as a function of wavelength, which was
determined by comparing white dwarf models to the HST data
\citep{Bolin14,Betoule13}.  In total, we include 62 individual systematic
uncertainties to describe the uncertainty in the
photometric calibration.  Most are due to the relative calibration: there is
one systematic for the filter zeropoint
and the filter $\lambda_{eff}$ $\times$ number of surveys
$\times$ number of filters per survey.

\subsection{SALT2 Model Calibration Uncertainties}
\label{sec:calib}

The training of the SALT2 model is subject to
the same sources of photometric
calibration uncertainty discussed above.  B14 created
variants of the SALT2.4 light curve
model by applying zeropoint and filter
function shifts to the training data and subsequently re-training SALT2.
These account for 10 individual systematics, which
are averaged to give the SALT2 model systematic error.
These uncertainties are discussed in \S5.4 of B14.

Re-training SALT2 using the improved calibration from
Supercal will lower the SALT2 systematic uncertainty
in future analyses.  However, we do not re-train the SALT2
light curve model for this analysis, as the SALT2 training
data are not public.


\subsection{Core-Collapse Supernova Contamination}
\label{sec:ccsn}

Systematic error due to marginalizing over the contamination
in our sample is a new source of uncertainty caused by our
use of photometrically classified SNe.
J17 predict that the PS1 host-$z$ sample contains $\sim$9\%
CC\,SNe.
Our method of measuring
distances from SNe\,Ia while marginalizing over CC\,SNe
is subject to biases in two areas: inaccurate
prior probabilities that a given SN is of type Ia and
differences between the CC\,SN model and the true distribution
of CC\,SNe.  The systematic error estimation from CC\,SN
contamination was presented in detail in J17 and relies on
varying these components.


We use the four methods of estimating prior
probability discussed previously (\S\ref{sec:photclass}) and 
three parametric models for the CC\,SN distribution.
The baseline likelihood model for CC\,SNe,
$\mathcal{L}_i^{CC}$ (Eq. \ref{eqn:beamslike}),
is a Gaussian with a mean and
standard deviation $-$ $g(z_i)$ and $\Sigma_{CC}(z_i)$ for the $i$th SN
$-$ that are both functions of redshift.  The two alternate CC\,SN
parametric models are a two-Gaussian model and a skewed Gaussian model.
We demonstrated in J17 that these models typically
agree well with single Gaussian
results; all three CC\,SN distributions tend to be much
broader than the SN\,Ia distribution, therefore
encompassing most outliers regardless of whether the functional
form is an exact representation of the CC\,SN data.

Because several of these variants are highly covariant
with one another, we group the different contamination variants
into two systematics: one using the results from
SN classifiers trained on simulated CC\,SN data and
a second using ``un-trained'' classifiers.  The trained
classifiers include NN and PSNID.  The ``trained''
systematic is the average change in SN distances when
either the NN classifier is used, or when the PSNID classifier
is used with alternate CC\,SN models.  \fp\ and GalSNID are not
trained on simulations, and
so we include the average of the \fp\ and GalSNID distances
as a second systematic.  The untrained classifiers are
not optimal methods, but are included here as an alternative
to classifiers that depend on simulations with
limited CC\,SN templates and known biases.  If each variant were instead
treated as an individual systematic, our final uncertainty
would only increase by just 2\% and the final value of $w$
would be higher by just 0.003.

Finally, we found in J17 that BEAMS could yield
results with less bias if $\alpha$ and $\beta$ are
fixed to their known values from spectroscopically
classified samples.  For a single-Gaussian CC\,SN model with PSNID,
we include this variant in our systematic uncertainty
budget by forcing $\alpha$ and $\beta$ to be equal to the values
measured from spectroscopically confirmed PS1+low-$z$ SNe.
The shape and color distributions
in the full PS1 sample are different than those
in the PS1 spectroscopically classified sample,
which could mean that $\alpha$ and
$\beta$ are in fact not the same in the full sample as in the spectroscopically
classified sample \citep{Scolnic16}.
However, because it is not possible to distinguish between true differences in $\alpha$/$\beta$
and differences caused by the known $\alpha$/$\beta$ biases when marginalizing
over CC\,SNe (J17), this variant is a necessary addition
to the error budget.

\subsection{Peculiar Velocity Correction}
\label{sec:pecv}

The magnitude of SN peculiar velocities, due to bulk flows and nearby
superclusters, becomes $\gtrsim$5\% of the Hubble flow
at $z \lesssim 0.03$.
We correct for peculiar velocities using the nearby galaxy
density field measured by the 2M++ catalog
from 2MASS \citep{Lavaux11}.  The uncorrelated uncertainty
associated with each correction is $\pm$250 km s$^{-1}$ (S17).
The peculiar velocity model is parameterized by the
equation $\beta_I = \Omega_M^{0.55}/b_I$, where $b_I$
describes the light-to-matter bias.
($\beta_I$ is unrelated to the SALT2 nuisance parameter).
\citet{Carrick15} measure $\beta_I = 0.43 \pm 0.021$.
We adopt a conservative 5$\sigma$ ($\pm$0.1) systematic on $\beta_I$ for
our peculiar velocity systematic uncertainty.

\subsection{SN\,Ia Demographic Shifts}
\label{sec:evo}

\begin{figure*}
\includegraphics[width=7in]{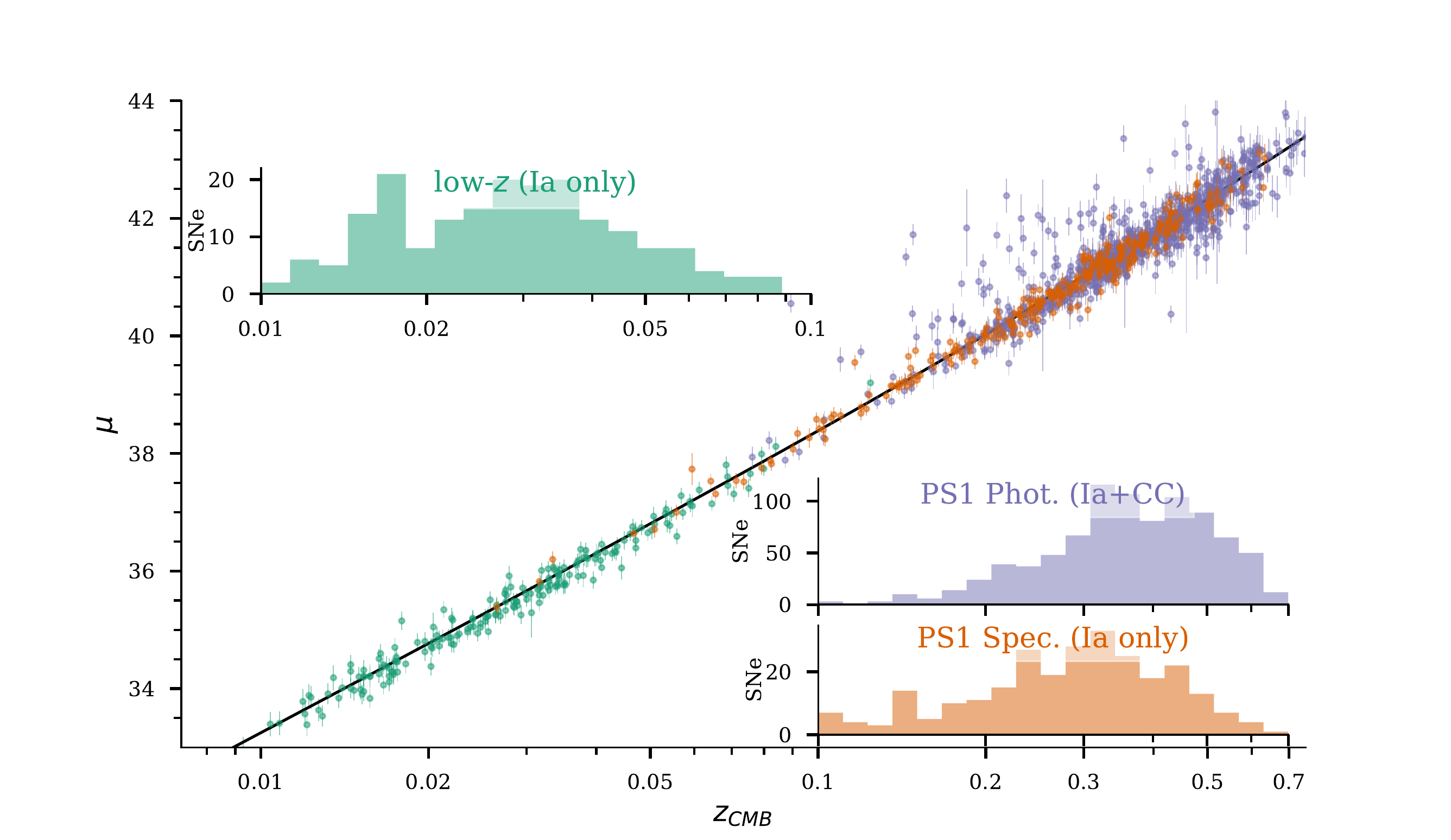}
\caption{The PS1$+$low-$z$ Hubble diagram with low-$z$ SNe\,Ia,
  spectroscopically classified SNe\,Ia and photometrically
  classified SNe.  The data that appear much fainter than
  $\Lambda$CDM (black line) are likely CC\,SN contaminants.  We use
  \numsnetotal\ SNe to measure cosmological
  parameters.}
\label{fig:hubble}
\end{figure*}

\begin{figure}
\includegraphics[width=3.25in]{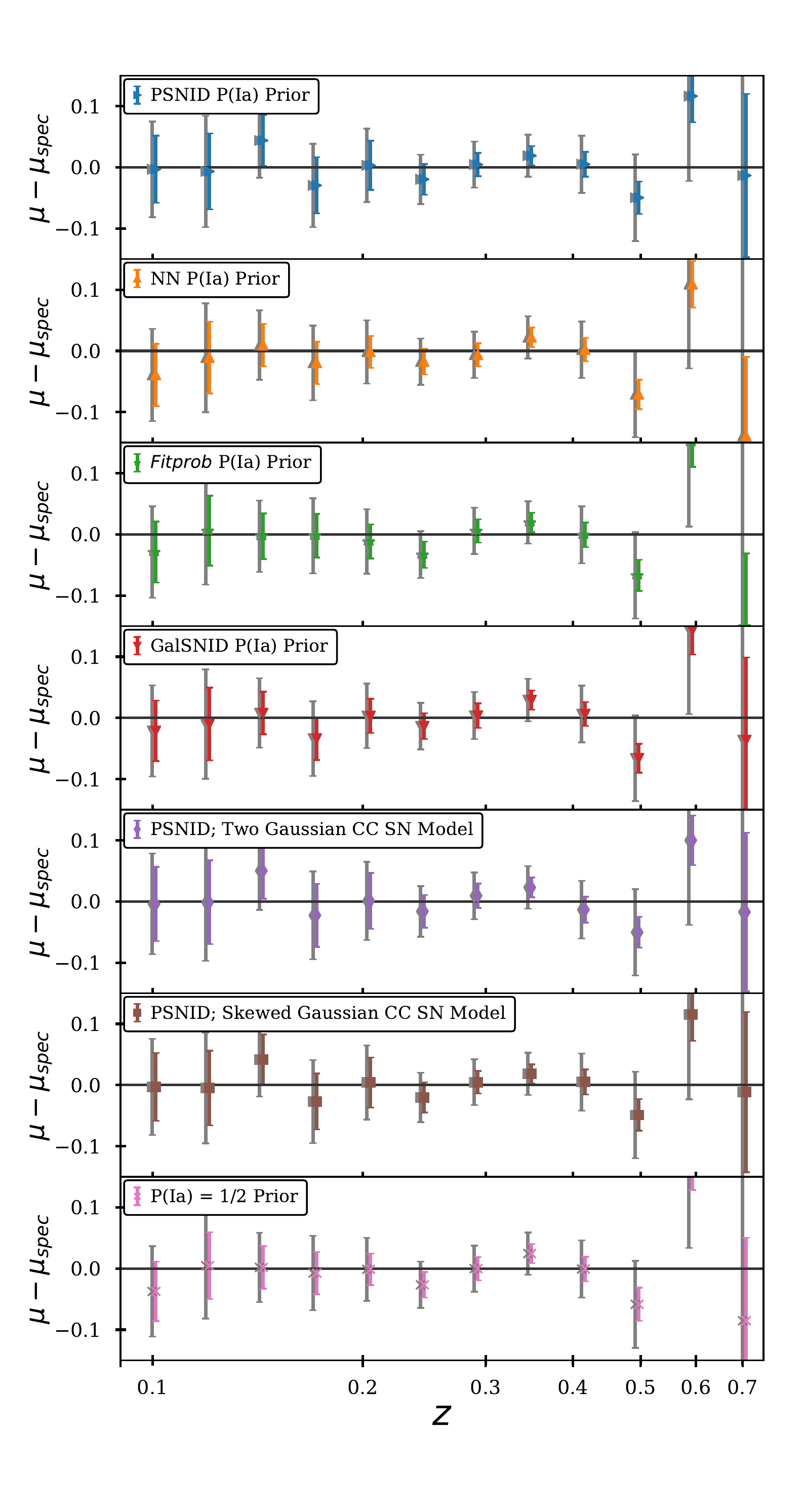}
\caption{The difference in binned distance from the full photometric
  sample compared to binned distances from the subset of
  $\sim$280 spectroscopically
  classified PS1 SNe\,Ia.  Grey (large) error bars are the uncertainties on
  spectroscopic and photometric distances added in quadrature,
  while the smaller errors are from
  the photometric sample alone (small redshift offsets
  are added to the photometric points for visual clarity).  Binned distances are
  consistent between methods,
  with a small bump at $z \sim 0.35$ that could be due to high
  CC\,SN contamination at this redshift but is also consistent
  with statistical fluctuation.  For comparison to the predicted
  biases from simulations, see Figure 11 of J17.}
\label{fig:beamsdist}
\end{figure}

\begin{figure*}
\includegraphics[width=7in]{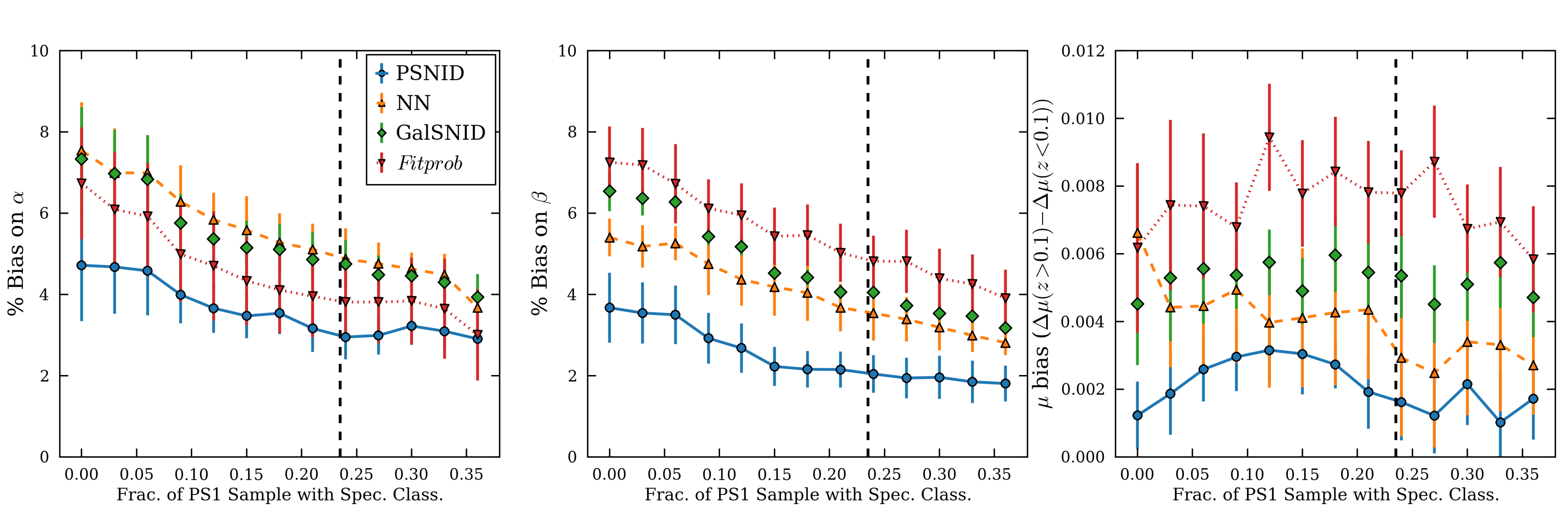}
\caption{From simulations, the bias on $\alpha$, $\beta$, and distance
  due to marginalizing over CC\,SNe as a function of the fraction of
  spectroscopically classified SNe\,Ia in the data.  In this work, $\sim$24\%
  of the PS1 sample is spectroscopically classified (vertical lines),
  giving a predicted reduction in $\alpha$/$\beta$ bias of
  $\sim$30-40\%.  The typical reduction in bias for a single distance
  bin is also $\sim$30-40\%, although the average distance bias at $z > 0.1$
  relative to $z < 0.1$ is largely unchanged (within the errors)
  with additional spectroscopic classifications.}
\label{fig:nuisancebiasplot}
\end{figure*}

Though SN\,Ia have been shown to be excellent standardizable
candles at low-$z$, it has been suggested that the relationship
between their luminosities, colors, or host galaxy properties
may change with redshift.  We address these possibilities
by adding three systematic tests.  For these tests, we add
additional parameters to our model for
estimating cosmological parameters (\S\ref{sec:methods}).  The
first is to allow a linear evolution of the mass step as
a function of redshift.  Mass step evolution was proposed by
\citet{Childress14}, and could be observed if the mass step
is caused by physical differences in SNe\,Ia with different progenitor
ages.  The second is to allow a linear
evolution in the SALT2 color-standardization
parameter, $\beta$, as
a function of redshift.  This was suggested as a possible
concern by \citet{Conley11}.  The third is evolution in SALT2 $\alpha$.
$\Delta_M$, $\alpha$ and $\beta$ in Equations
\ref{eqn:salt2} and \ref{eqn:beamslike} then become:

\begin{equation}
  \begin{split}
  \Delta_M &= \Delta_{M,0} + \Delta_{M,1} \times z,\\
  \alpha &= \alpha_0 + \alpha_1 \times z.\\
  \beta &= \beta_0 + \beta_1 \times z.
  \end{split}
  \label{eqn:evol}
\end{equation}

\noindent $\Delta_{M,0}$, $\Delta_{M,1}$, $\beta_0$, $\beta_1$, $\alpha_0$ and $\alpha_1$
are free parameters.  They are measured simultaneously with SN\,Ia
distances in \S\ref{sec:checks}.  Because we find no hint
of mass step evolution or $\alpha$ evolution, we
include only $\beta$ evolution as a systematic uncertainty in our final
measurement (see \S\ref{sec:checks}).

We also include a $\Delta_M$ variant that shifts the divide between
``low-mass'' and ``high-mass'' hosts by 0.15 dex relative to the
standard divide at log($M_{\ast}/M_{\odot}) = 10$, following the uncertainty
on the location of the step measured by S17.  Finally,
because possible bias in $\Delta_M$ due to marginalizing
over CC\,SN contamination was not estimated in J17,
we add one variant where $\Delta_M$ is fixed to the
value measured by B14 (0.07$\pm$0.023 mag).

We note that because our sample preferentially contains bright host
galaxies, our results are sensitive to uncertainty in the
relation between host galaxy properties and SN luminosity.  However,
because most low-$z$ SNe originate from SN searches that specifically
targeted bright galaxies, the PS1 photometric data are in some ways
more similar to the existing low-$z$ data than previous high-$z$ datasets.
In this way, our results might be less biased by the uncertainty in
the relationships between SNe and their host galaxies than previous analyses.


An additional potential systematic is the relation between SN\,Ia corrected
magnitudes and their local host galaxy environments.
Several papers have recently asserted that SN\,Ia corrected
magnitudes are correlated with their local star formation environments
on a scale of $\sim$1-3 kpc (the LSF step;
\citealp{Rigault13,Rigault15}).  Due to the
$\sim$1\arcsec\ PSF of PS1 and the lack of ultraviolet or
$u$-band observations for much of our sample, it is impossible to
measure robust local star formation rates over the PS1 redshift range.
However, \citet*{Jones15b} re-examined the evidence for
the LSF step, finding that the re-training
of SALT2 in B14/G10 reduced or eliminated many of the biases
in the SALT2 model.  \citet*{Jones15b} found no evidence for a
LSF step in the B14 low-$z$ sample.
\citet{Roman17} also recently measured a strong dependence
of SN\,Ia luminosities on local $U-V$ color but find that this
effect is expected to change $w$ by just 0.006 relative to the
standard $\Delta_M$ correction.  Though our data are not optimal
for investigating local properties, we plan to use PS1 data
to more robustly determine the relationship between SNe\,Ia and their
global or semi-local host galaxy properties in future work.

\section{First Results and Consistency Checks}
\label{sec:checks}

The PS1+low-$z$ Hubble diagram is shown in Figure
\ref{fig:hubble} and light curve parameters for our full sample
are given in Table \ref{table:lcparams}.
There are $\sim$3.5 times as many photometrically classified
SNe as there are spectroscopically classified SNe.
The binned SN\,Ia distance
uncertainties from the full sample
are an average of 40\% lower than from
spectroscopically classified SNe\,Ia alone
(statistical uncertainties only).
At $0.2 < z < 0.5$,
where $\sim$75\% of the PS1 data lie, uncertainties are $\sim$45\% lower.
This is in spite of the fact that the photometrically
classified SNe have lower average SNRs; the median SNR at peak is 22 for all PS1 SNe, compared
to 38 for spectroscopically classified
SNe.  We also don't expect that marginalizing over CC\,SNe has inflated
the binned distance uncertainties.
In J17 we used simulated data to find that our method of marginalizing over CC\,SNe
increases the statistical uncertainty on binned SN\,Ia
distances by just 3\%.

Our likelihood model (Eq. \ref{eqn:beamslike}) is simultaneously used to
measure $\alpha$, $\beta$, and the dispersion $\Sigma_{Ia}$,
which are given in Table \ref{table:nuisance}.  These measurements use the
baseline classification method, PSNID, and the one-Gaussian CC\,SN model,
while the alternate methods contribute to
the systematic errors in the middle column.  We measure
$\alpha = 0.165\pm0.019$ (stat+sys), which is consistent with the
value measured by S17 from low-z, PS1, SDSS, and SNLS spectroscopically
confirmed SNe ($\alpha=0.156\pm0.006$).  \citet{Zhang17} also find
$\alpha = 0.165\pm0.010$ for low-$z$ SNe.  However,
we note that this value is higher than measured by B14 by $\sim$1$\sigma$
($\sim2.5\sigma$ from statistical uncertainties alone)
and the reason for this difference is unclear.

The uncertainty on the SN\,Ia dispersion, $\Sigma_{Ia} = 0.082\pm0.067$, is extremely high.
This is a consequence of removing P(Ia) $<$ 0.5 SNe from the sample before cosmological
parameter estimation and allowing
SN type probabilities to be shifted and re-normalized by the likelihood model.  If P(Ia) $<$ 0.5 SNe
are included, we find that $\Sigma_{Ia}$ is better constrained, with a value of 0.106$\pm$0.032
(stat. errors only), consistent with $\Sigma_{Ia} = 0.118$ from spectroscopically
classified SNe\,Ia alone.  We note that in spite of the large uncertainty on $\Sigma_{Ia}$,
the distance uncertainties are slightly smaller when P(Ia) $<$ 0.5 SNe are removed.
Removing P(Ia) $<$ 0.5 SNe changes the statistics-only measurement of $w$
by just 0.3\%.

As a test, if the sample is analyzed without
BEAMS, i.e. treating all SNe as SNe\,Ia, $\Sigma_{Ia}$ increases by 71\% to 0.187.
In general, the systematic error on all nuisance parameters
is higher than it would be in an analysis of spectroscopically
classified SNe\,Ia, due to the predicted biases on those
parameters when marginalizing over CC\,SN contamination
(J17).  Fortunately, J17 found that
biases of 3-6\% on nuisance parameters do not give
similar fractional biases on binned distances
or $w$ (0.5\%$\pm$0.4\% bias on $w$ for the baseline method).

In J17, we predicted that our method of marginalizing over CC\,SNe
would bias $\alpha$ and $\beta$ by $+$3\%.  This gives a prediction that the
$\alpha$ and $\beta$ measured here will be higher than
the $\alpha$ and $\beta$ measured from spectroscopically
confirmed SN\,Ia alone.  Table \ref{table:nuisance}
shows that this may indeed be the case; $\alpha$ is 6\% higher and $\beta$
is 3\% higher than the values from spectroscopically
confirmed PS1$+$low-$z$ SNe\,Ia
(though at $<$1$\sigma$ significance if we neglect the partial correlations
between these two samples).  However, we also expect higher
measured values of $\beta$ due to the redder colors
of the full PS1 sample \citep{Scolnic16}.

We also measure the mass step at 6$\sigma$ significance and at nearly
8$\sigma$ from statistical errors alone (we report systematic uncertainties
that neglect the host mass variants).
Our measurement of $\Delta_M =$ \massstep\ is consistent with the B14 measurement of 0.07$\pm$0.023.
It is also consistent with the $\Delta_M$ that we measure from the low-$z$ sample
alone, $\Delta_M = 0.110\pm0.038$.
Interestingly, the host mass step $\Delta_M$ is higher
in the full PS1$+$low-$z$ sample than in the sample of spectroscopically
classified SNe\,Ia alone (1.1$\sigma$ significance from statistical
uncertainties alone, though these measurements
are not independent).  It's unclear if this difference could be due to
statistical fluctuation, a bias from the method, or the presence of
broader light curve shapes and redder colors in the full
sample.  $x_1$ and $c$ correlate with
both host mass and Hubble residual \citep{Scolnic16} and could increase the
size of the step (S17).  We will use simulations to investigate whether our
method of marginalizing over CC\,SN contamination could bias determinations
of the host mass step in future work.

\subsection{Impact of Different Classification Methods}
\label{sec:classeffect}

\begin{deluxetable}{lcclccc}
\tabletypesize{\scriptsize}
\tablewidth{0pt}
\tablecaption{Nuisance Parameters}
\tablehead{&\multicolumn{3}{c}{All SNe}&&\multicolumn{2}{c}{Spec. Class. SNe}\\
  &&$\sigma_{stat}$&$\sigma_{stat+sys}$&&&$\sigma_{stat}$}

\startdata
$\alpha$&0.165&0.006&0.019&&0.155&0.009\\
$\beta$&3.028&0.067&0.152&&2.944&0.092\\
$\Sigma_{Ia}$&0.082&0.067&0.101&&0.118&0.008\\
$\Delta_M$&0.102&0.013&0.017\tablenotemark{a}&&0.064&0.020\\
\enddata
\tablenotetext{a}{The systematic uncertainty excludes
  the analysis variants that change the location, size and
  $z$-dependence of $\Delta_M$.}
\tablecomments{Nuisance parameters from PS1+low-$z$ SNe.  The
  systematic uncertainty on $\beta$ is likely overestimated
  due to the biases from the GalSNID and \fp\ classification methods
  discussed in J17.  The exceptionally large
  uncertainty on $\Sigma_{Ia}$ is due to our decision to
  exclude SNe with P(Ia) $<$ 0.5 and to allow P(Ia)
  to be shifted and re-normalized, but we have verified that this choice
  has a minimal effect on the final cosmological parameters.}
\label{table:nuisance}
\end{deluxetable}

Regardless of which classifier is used, uncertainties on
binned distances from the full PS1 sample are much smaller than
the uncertainties on binned distances from
spectroscopically classified SNe\,Ia alone (by $\gtrsim$40\%).
The binned SN\,Ia distance measurements from each classifier are also
remarkably consistent (Figure \ref{fig:beamsdist}).
Nearly all distances are within 1$\sigma$ of distances
derived from the PS1 spectroscopically classified SN\,Ia sample.
Additionally, binned distances from $0.2 \lesssim z \lesssim 0.5$,
where 75\% of our data lie,
show few discrepancies between the different methods.
Even the test case
of using an uninformative
prior of P(Ia) $=$ 1/2 for all photometrically classified SNe
(bottom panel) yields distances within 1$\sigma$ of the
spectroscopic sample in all bins but one.
We note that close agreement is predicted by J17; even
in a sample without spectroscopically confirmed SNe\,Ia,
J17 predict biases of $<$10 mmag due to
the method.  We will revisit this prediction in
\S\ref{sec:nospec} to test whether our methodology
remains robust and consistent in the case of an ``ideal'' photometrically
classified SN sample; i.e., a sample without spectroscopic
classifications.

The nuisance parameters
$\alpha$ and $\beta$, as measured using different classification
priors, are more consistent than expected
from J17.  When using different classification priors,
$\alpha$ and $\beta$ vary by
30-50\% less than the simulation-based predictions
in J17 (in this work we observe differences of
$\Delta\beta \sim 0.07$ and $\Delta\alpha \sim 0.004$
between the four different classification methods).
In Figure \ref{fig:nuisancebiasplot}, we provide a possible explanation
for why our results are more consistent than expected.
J17 simulations included no subset of
spectroscopically classified PS1 SNe, while our data consist of $\sim$24\%
spectroscopically classified PS1 SNe.  Because of this, we
used simulations of the PS1 host-$z$ and SN-$z$ samples (\S\ref{sec:sim})
to predict the effect of adding spectroscopically classified subsets of
SNe to the data.  We find that the predicted biases
on $\alpha$ and $\beta$ due to marginalizing over
CC\,SNe decrease by 30-40\% when the
PS1 data consist of 24\% spectroscopically classified SNe.

Similarly, the biases on individual distance bins
decrease by $\sim$30-40\% when 24\%
of PS1 SNe are spectroscopically classified.
For PSNID priors, Figure \ref{fig:nuisancebiasplot} shows that
the predicted (weighted) average bias in distance modulus
at $z > 0.1$ relative to $z < 0.1$ is just 2 mmag.

\subsection{Evolution of Nuisance Parameters}
\label{sec:nuisance}

\begin{figure}
\includegraphics[width=3.25in]{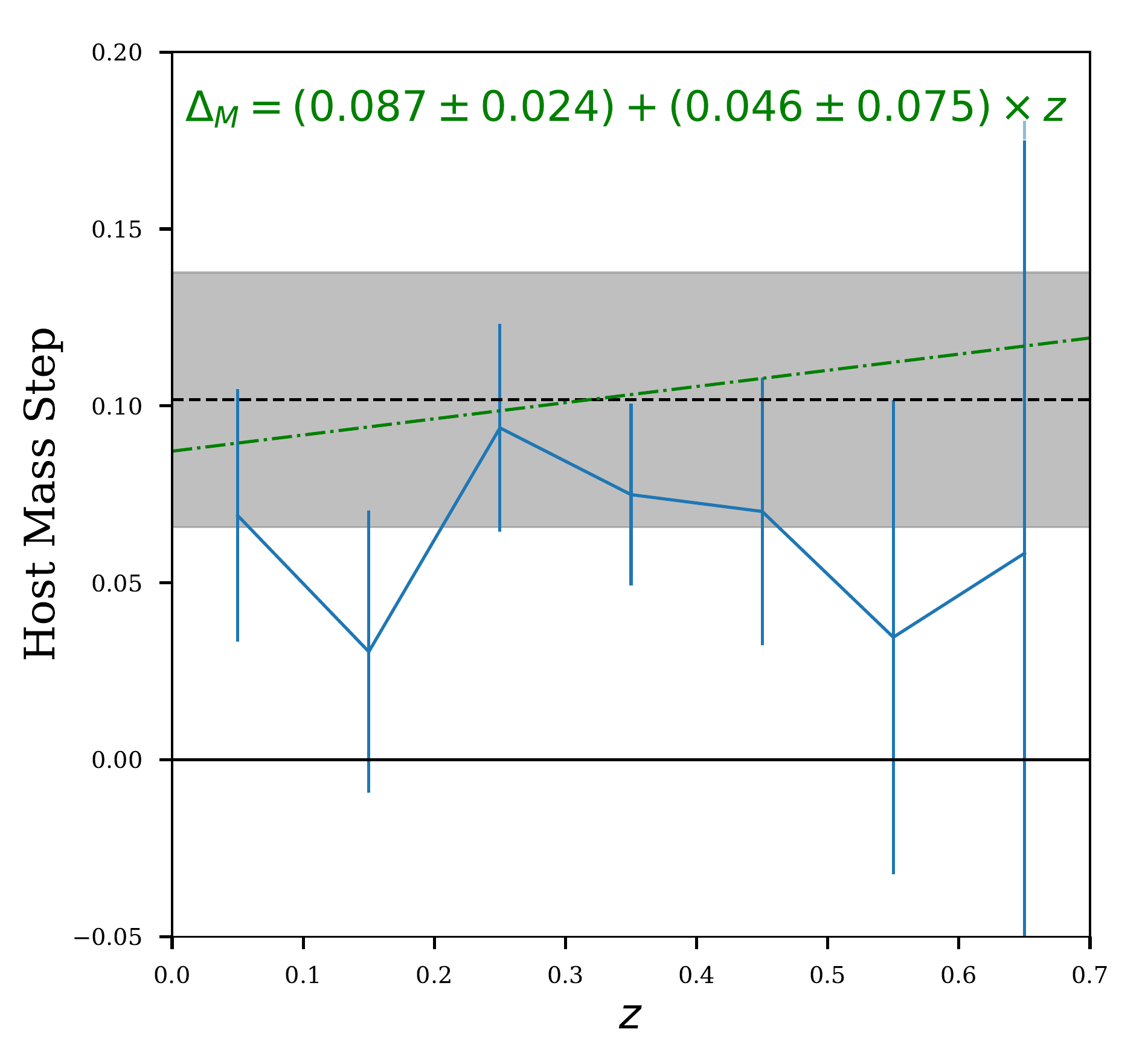}
\caption{Evolution of the host galaxy mass step with redshift after
  2.5-$\sigma$ clipping to remove most CC\,SNe.  Binned
  points are shown with the best fit global mass step (black) and linear
  trend (green) from marginalizing over CC\,SNe.}
\label{fig:mass}
\end{figure}

Using Equation \ref{eqn:evol} to add linear mass step ($\Delta_M$) evolution
to BEAMS, we find no evolution in $\Delta_M$ as a function of
redshift (we use the baseline classifier, PSNID).
However, our uncertainties are large, $\sim$0.08 mag,
due to lack of low-mass hosts at high redshift\footnote{S17,
  however, finds evidence of mass step evolution.
  The discrepancy could be due to the larger SNLS redshift range
  and additional SNe\,Ia in low-mass hosts at $z > 0.5$.}.
In Figure \ref{fig:mass}, we estimate the redshift dependence
of the mass step with a 2.5$\sigma$ clip of Hubble residuals
($-0.45 \lesssim$ HR $\lesssim 0.45$) to
remove most CC\,SNe and then plot the maximum likelihood mass step in
redshift bins of 0.1.  This is an incomplete removal of CC\,SN
contamination, but doubles as a simple sanity check on BEAMS.
We find no statistically significant evidence for mass step
evolution.

\begin{figure}
\includegraphics[width=3.25in]{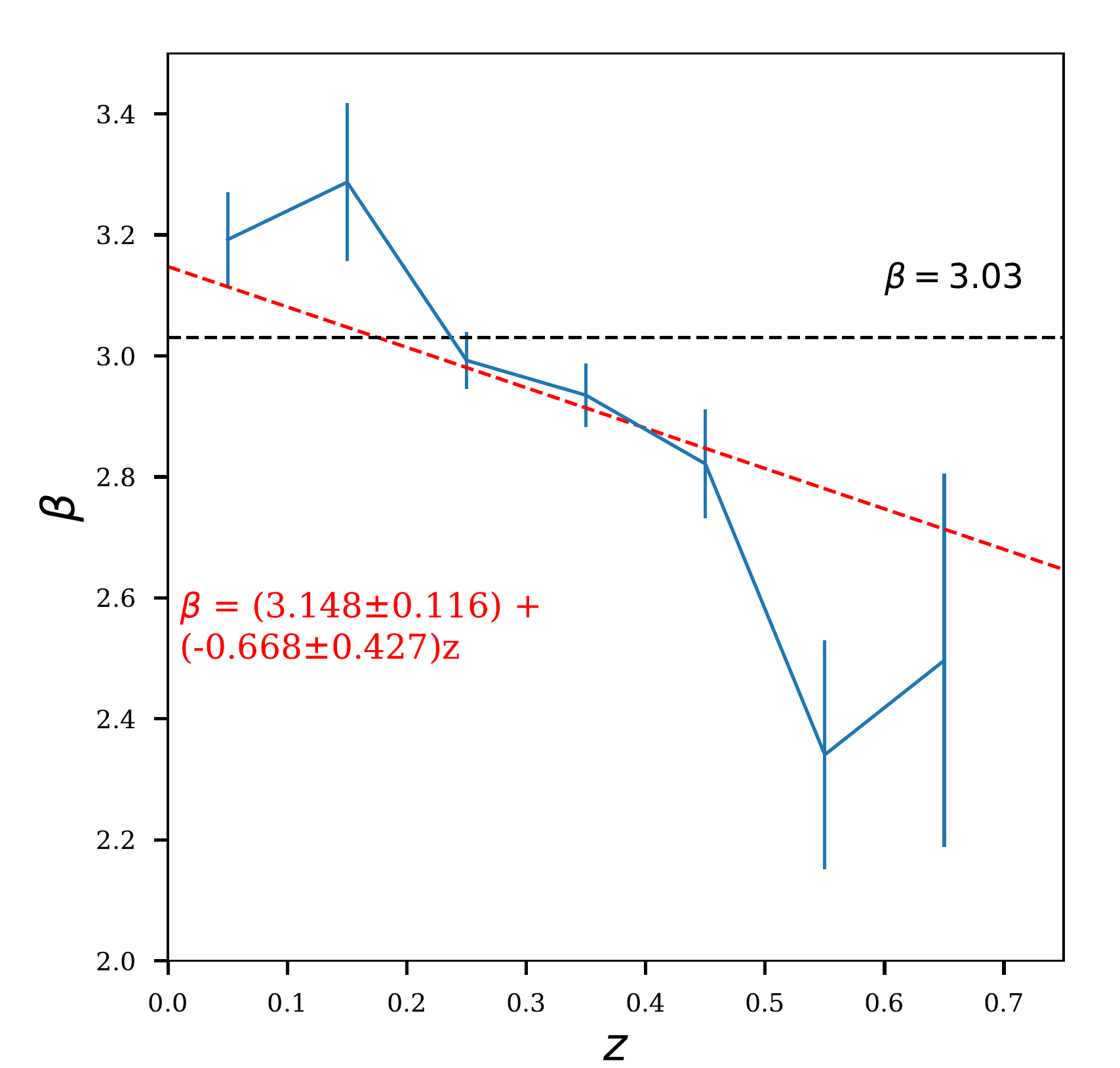}
\caption{Binned evolution of the SALT2 nuisance parameter $\beta$
  with redshift after 2.5-$\sigma$
  clipping to remove most CC\,SNe.  The best
  fit $\beta$ (black) and linear trend (red) are computed by
  marginalizing over CC\,SNe with
  the full likelihood model.}
\label{fig:beta}
\end{figure}

We do see 1.6$\sigma$ evidence for evolution of the $\beta$ parameter,
however (Figure \ref{fig:beta}).
Fortunately, this does not constitute a large
contribution to our
systematic error budget as it predominantly affects the highest
survey redshifts where few SNe are found (Figure \ref{fig:sys}).
Evidence for $\beta$ evolution
was seen in SNLS data \citep{Conley11}, though its significance
is attributed to selection effects in B14.  S17 find just 1$\sigma$
evidence for $\beta$ evolution ($\beta = (3.139\pm0.099)+ z \times (-0.348\pm0.289)$),
a measurement that includes SNe at redshifts up to $\sim$2 \citep{Riess17}.
Though there are not enough SNe\,Ia at
$z > 1.5$ to constrain a changing value of $\beta$, larger
high-$z$ datasets may be able to confirm or discount $\beta$ evolution.
We caution that blue ($c < 0$) SNe\,Ia have lower observed $\beta$
(SNe primarily appear blue due to noise and selection
biases; \citealp{Scolnic16}),
and our high-$z$ data are dominated by blue SNe (Figure \ref{fig:sim}).
However, our methodology does not recover any significant evolution
of $\beta$ when tested on simulated SN samples
with a constant input $\beta$.  In 10 simulated SN samples, 5 using the G10 model and
5 using the C11 model,
we found just a single sample showing $>$1$\sigma$ evidence of negative $\beta$ evolution
with redshift (the simulation had a $\beta$ slope with significance of 1.2$\sigma$).
If $\beta$ does change with $z$,
it could suggest an evolution in dust properties or the
evolution of SN progenitors with redshift and could contribute
significantly to the systematic error budget at $z > 0.5$.

We also checked for $\alpha$ evolution using the same parametric form
as Equation \ref{eqn:evol}, and find $\alpha(z) = 0.157\pm0.01 + z*(0.018 +/- 0.040)$.
Because we find that $\alpha$ evolution is not statistically significant, 
we have not included it in our systematic uncertainty budget.

\section{Cosmological Constraints from Supernova and CMB Data}
\label{sec:firstconstraints}

\begin{deluxetable}{lccc}
\tabletypesize{\scriptsize}
\tablewidth{0pt}
\tablecaption{Summary of Systematic Uncertainties on $w$}
\tablehead{Error&$\Delta w$\tablenotemark{a}&$\Delta\sigma_w$\tablenotemark{b}&Rel. to $\sigma_w^{stat}$}
\startdata
All Sys.&0.033&0.043&1.137\\
Phot. Cal.&0.007&0.021&0.558\\
Bias Corr.&0.012&0.019&0.518\\
Mass Step&0.006&0.017&0.449\\
Beta Evol.&0.012&0.016&0.428\\
MW E(B-V)&0.009&0.015&0.390\\
CC\,SN Contam&-0.001&0.012&0.332\\
SALT2 Model&0.001&0.008&0.207\\
Pec. Vel.&0.002&0.007&0.182\\
\enddata
\tablenotetext{a}{Difference in measured $w$
  relative to the final value of $w$ with all systematics included.}
\tablenotetext{b}{The additional uncertainty
  added in quadrature from each source of systematic
error.  The statistical uncertainty on $w$ is 0.0375.}
\label{table:syserr}
\end{deluxetable}


We first constrain $\Omega_M$ using the SN\,Ia
data alone and assuming
a flat $\Lambda$CDM cosmology.  We find
$\Omega_M = 0.319 \pm 0.040$, consistent with
B14 (0.295$\pm$0.034).
These results are independent of, but in good agreement with, the
Planck constraints on $\Omega_M$ ($\Omega_M = 0.308\pm0.012$).

We combine these data with CMB constraints from the
Planck full-mission data \citep{Planck15}.  In
contrast to the \citet{Planck14} constraints used in
B14, the full-mission Planck data does not require WMAP
polarization measurements.  Planck provides the full
likelihoods for the CMB data, which can then be
combined with SNe\,Ia using CosmoMC.
Planck data greatly improve our constraints on $w$
using the CMB temperature power spectrum, which gives a
precise constraint on the cosmic matter density
at $z \sim 1090$.  Constraints from a matter-dominated
cosmic epoch are largely independent of an evolving or
non-cosmological constant dark energy, which affects
cosmic evolution only at the late times probed by SNe\,Ia and BAO
measurements.

With Planck priors, we measure $w = $ \w\ (stat+sys).  Systematic
uncertainties on this measurement are 14\% higher than
statistical uncertainties (Table \ref{table:syserr}).
Though we have 85\% more SNe than B14 and 31\% more SNe than S17,
our uncertainty is approximately the same as B14 and 39\% higher than S17.
There are three primary reasons for this.  First, we have
fewer independent surveys to reduce the photometric
calibration systematic.  Second, we have estimated a more
conservative systematic uncertainty on the
selection bias correction than B14.  Lastly, PS1 photometrically classified
SNe have much lower SNR (for PS1, SNR at maximum is
an average of 17 for photometrically classified
SNe and 39 for spectroscopically classified SNe\,Ia),
and PS1 SNe, unlike SNLS SNe, cannot be found at $z \sim 0.7 - 1$.


We also use these data to constrain the two-parameter redshift evolution of
$w$ using the most common parameterization:

\begin{equation}
w = w_0 + w_a z/(1+z).
  \label{eqn:wa}
\end{equation}

\noindent Eq. \ref{eqn:wa} is a first order Taylor
series expansion of $w$ as a function of
scale factor $a$ \citep{Linder03}.  We find
$w_0 = -0.912\pm0.149$ and $w_a = $\wa.  These constraints are
slightly better than those of B14, which is due to our use of the
most recent chains from Planck.  We also find much tighter constraints
on $w_a$ after combining with BAO (\S\ref{sec:allconstraints}).

\subsection{Systematic Uncertainties on $w$}
\label{sec:sysw}

Contributions to the systematic uncertainties on
$w$ are summarized in Table \ref{table:syserr}.
The photometric calibration
systematic, the largest source of systematic
uncertainty in most previous analyses (e.g. R14, B14),
remains the largest systematic uncertainty in this work
($\sigma_w^{cal} = 0.021$) but is now almost the same magnitude
as the selection bias.
The calibration has been significantly improved by the
Supercal procedure and continued improvements will
come from a new network of white dwarf
standards \citep{Narayan16}.

The second largest systematic uncertainty is due to the selection
bias ($\sigma_w^{bias} = 0.020$).  $\sigma_w^{bias}$
is dominated by the difference between the G10
and C11 scatter models and the uncertain spectroscopic selection function
of the low-$z$ surveys.  It may be that re-training SALT2
assuming the C11 scatter model, e.g. \citet{Mosher14},
will reduce this systematic in the future.

\begin{deluxetable}{lrr}
\tabletypesize{\scriptsize}
\tablewidth{0pt}
\tablecaption{$w$ with Different Photometric Classification Priors and CC\,SN Models}
\tablehead{Method&$\Delta w$&$\Delta \sigma_w$}
\startdata
PSNID&\nodata&\nodata\\
\tableline\\
PSNID, Skewed Gaussian CC Model&-0.004&0.000\\
PSNID, 2-Gaussian CC Model&0.018&0.011\\
NN&0.016&0.000\\
GalSNID&0.008&0.000\\
Fitprob&-0.007&0.000\\
Spec. $\alpha$/$\beta$&-0.008&0.000\\
\enddata
\tablecomments{$w$ from each CC\,SN model and photometric classification
  prior, relative to the baseline case of using PSNID classification priors and
  a single, $z$-dependent Gaussian to model the CC\,SNe.  The final line is
  the change in $w$ when $\alpha$ and $\beta$ are fixed to the values
  measured from spectroscopically confirmed SNe\,Ia.}
\label{table:photclass}
\end{deluxetable}

\begin{figure*}
\includegraphics[width=7in]{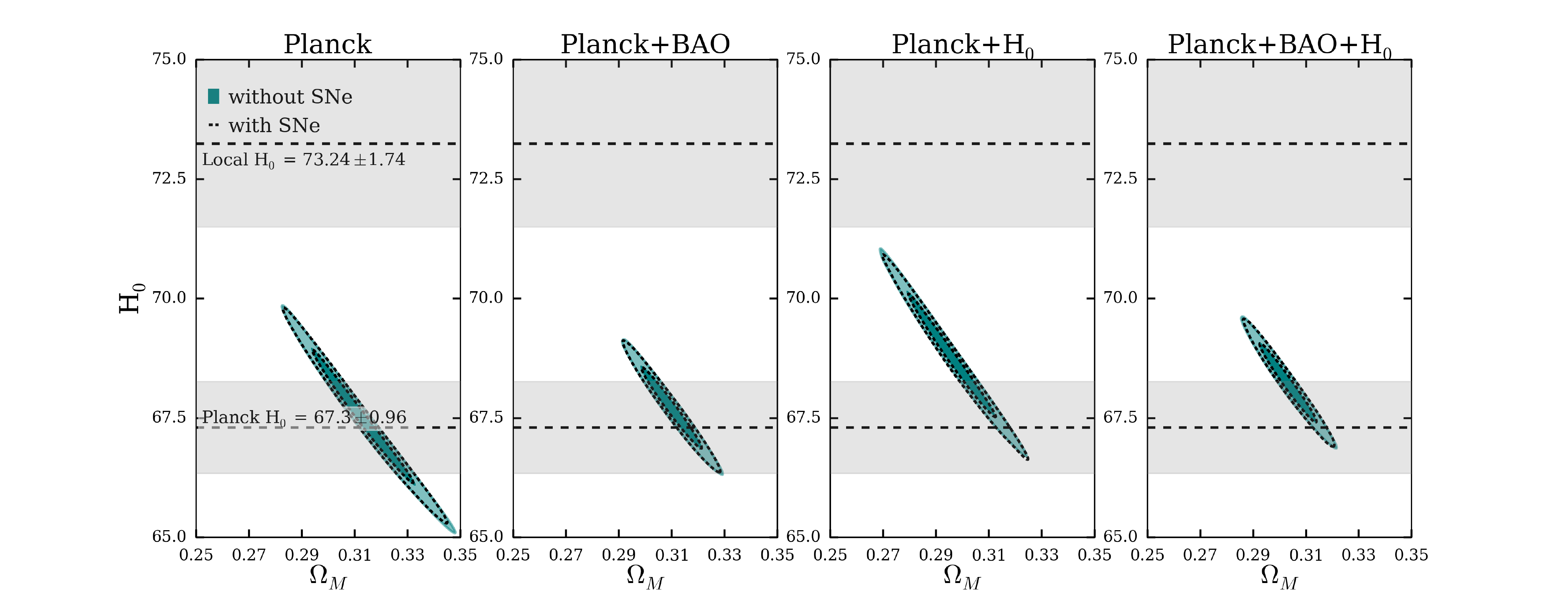}
\caption{Discrepant constraints on H$_0$ from CMB, BAO,
  and local measurements assuming $\Lambda$CDM.  SNe\,Ia
  disfavor a scenario in which exotic dark energy can
  resolve these conflicts.}
  \label{fig:h0}
\end{figure*}

\begin{figure*}
\includegraphics[width=7in]{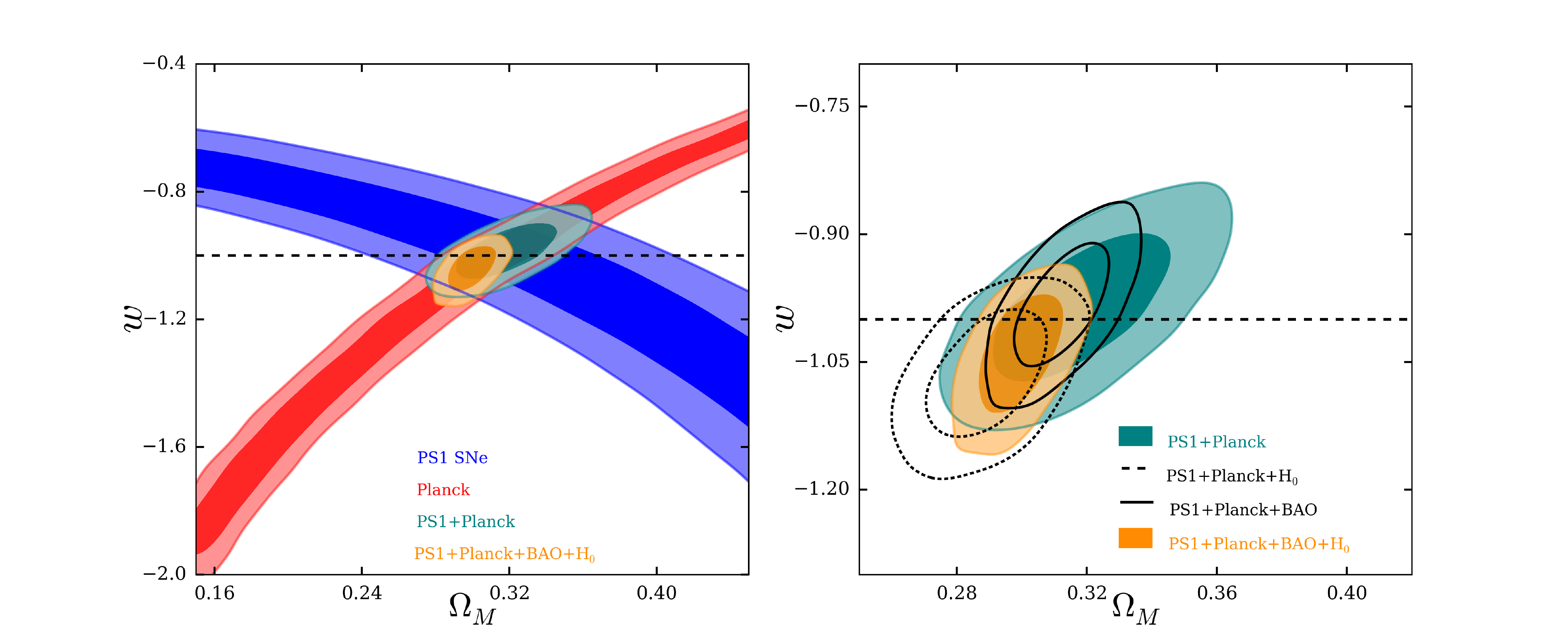}
\caption{Constraints on $w$ and $\Omega_M$ from
  PS1+low-$z$ SNe in conjunction with other probes.}
  \label{fig:wcdm}
\end{figure*}

\begin{figure}
\includegraphics[width=3.25in]{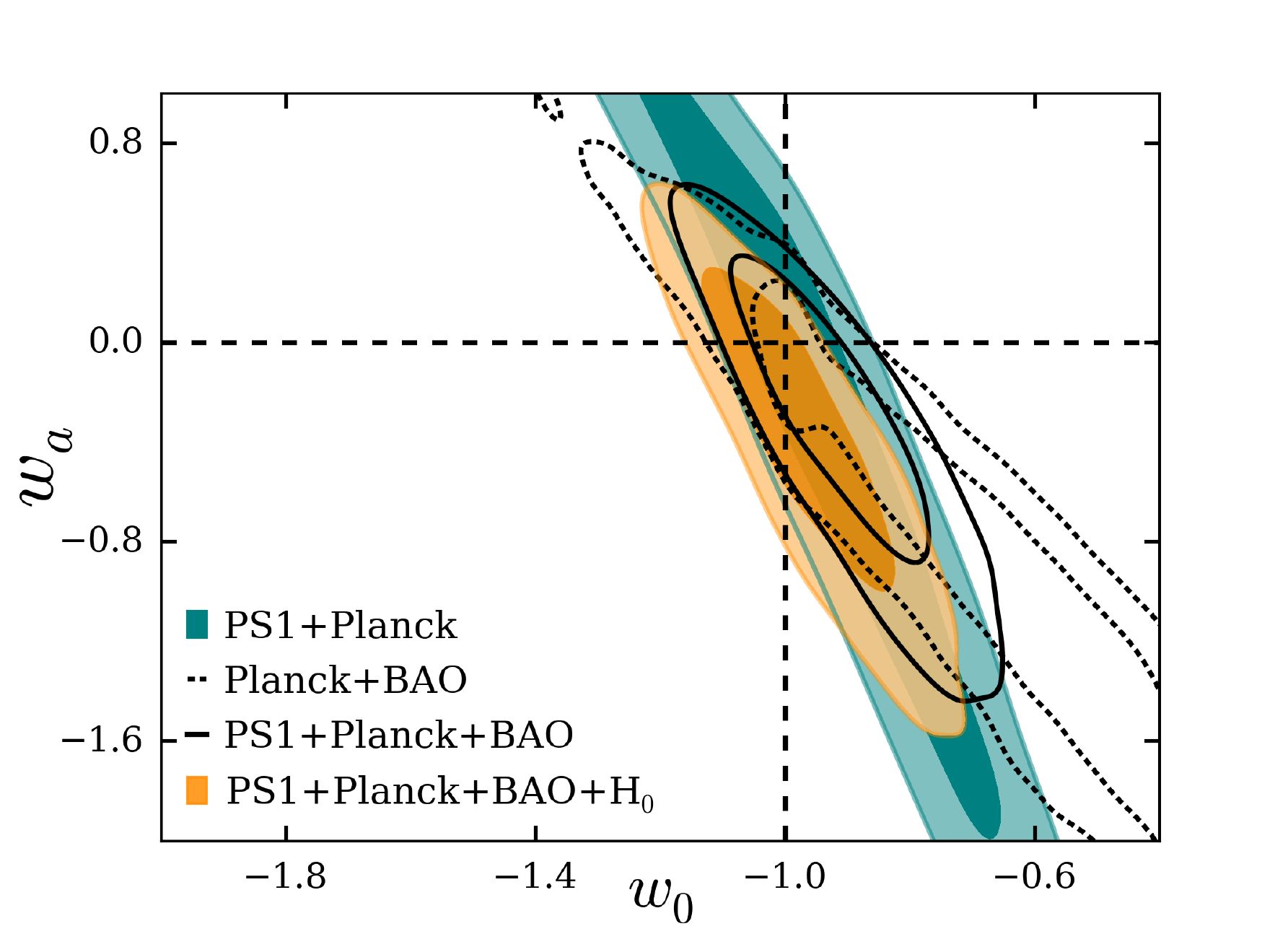}
\caption{Constraints on $w_0$ and $w_a$ from
PS1+low-$z$ SNe, Planck, BAO, and H$_0$.}
  \label{fig:wzcdm}
\end{figure}

The systematic due to marginalizing over
CC\,SNe, $\sigma_w^{CC} = 0.012$, is the third-smallest
systematic, nearly equal in size to the Milky Way
extinction systematic and
smaller than the systematics pertaining to the host mass
step and $\beta$ evolution.
Table \ref{table:photclass}
shows the value of $w$ measured from each
classification prior and CC\,SN parameterization
discussed in \S4.  All measurements of $w$ are within 2\%
of the baseline method.  We note that it is likely
the NN and PSNID classifiers are more accurate
than the other two classifiers used in this work.  However, both
NN and PSNID are directly dependent on CC\,SN templates
and simulations for training, neither of which are likely
representative of the true CC\,SN population (see J17 for more discussion).  We
include the alternative \fp\ and GalSNID classifiers as
they are less subject to the uncertainty in CC\,SN
simulations, but note that excluding them would significantly reduce
the systematic uncertainty due to CC\,SN contamination.

The dispersion of measured $w$ from different BEAMS variants
is nearly $\sim$25\% lower than predicted by J17, in spite of
the fact that, unlike J17, we did not fix $\alpha$ and $\beta$ to the
values from the spectroscopic sample (except for the final
variant listed in Table \ref{table:photclass}).  This may be due to
sample-to-sample variations, but is more likely explained by
tighter constraints on $\Omega_M$ from the full Planck chains
compared to the J17 approximation ($\Omega_M$ prior of 0.30$\pm$0.02)
and the fact that a sizeable portion ($\sim$24\%) of our high-$z$ data are
spectroscopically classified SNe\,Ia.
With simulations, we found that a subset of SNe with known types can
greatly help the BEAMS method to constrain distances
and SN\,Ia nuisance parameters (\S\ref{sec:checks}).  If the amount of CC\,SN
contamination was overestimated in J17, that could
also help to explain the lower contamination systematic.
The magnitude of the CC\,SN contamination systematic
can be further reduced by improved validation of
classifiers and a better understanding of the diversity of
CC\,SNe, their luminosity functions, and the inclusion of
additional CC\,SN templates in classifier training as
discussed in J17.  Methods for measuring robust classifications
even in the case where the training sample is biased (e.g. \citealp{Revsbech17}), are
also important to pursue.


%
%

\section{Cosmological Constraints with BAO and H$_0$ Priors}
\label{sec:allconstraints}

We now combine Planck \citep{Planck15}
and PS1+low-$z$ SNe with baryon acoustic
oscillation (BAO) constraints and a local prior on the value
of H$_0$ from \citet{Riess16}.
The BAO feature, the evolving size of the imprint of
acoustic waves on the distribution of cosmic matter, serves as
a standard ruler that is independent of SN\,Ia measurements. The BAO scale
is proportional to a combination of the angular diameter distance
to a given redshift and the Hubble parameter $H(z)$ at that redshift.
Following \citet{Planck15}, we use BAO constraints from the
SDSS Main Galaxy Sample (MGS; \citealp{Ross15})
and the combination of the Baryon Oscillation
Spectroscopic Survey (BOSS) and CMASS survey \citep{Anderson14}.
The BAO constraints used here give measurements of the BAO scale
to $z =$ 0.15, 0.32, and 0.57.

There is a notable internal conflict between
these priors: a 3.4$\sigma$ discrepancy between
local and CMB-inferred values of H$_0$ (\citealp{Riess16};
see also \citealp{Casertano17}, \citealp{Bonvin17}, \citealp{Jang17}).
The difference could be due to systematic uncertainties
in one or both datasets (e.g. \citealp{Addison16}),
$>$3 neutrino species, non-$\Lambda$ dark energy, or
more exotic phenomena.  We show this discrepancy
in Figure \ref{fig:h0} for a standard
$\Lambda$CDM cosmology (reionization optical depth
$\tau = 0.078$; \citealp{Planck16}).
PS1$+$low-$z$ SNe cannot explain the
disparity and therefore limit the degree to which
exotic dark energy can explain the H$_0$ tension.
Throughout this section, we remain agnostic
as to the source of the discrepancy and examine cosmological
parameters using all probes both individually and
in combination.

Following B14, we use SN data to constrain three cosmological
models: the o-$\Lambda$CDM model removes the assumption
of flatness ($\Omega_k = 0$), the $w$-CDM model allows a
fixed, non-cosmological constant value of $w$, and the $w_a$-CDM
model allows $w$ to evolve with redshift.
The constraints on these three models are presented in
Table \ref{table:allcosmo}.
All measurements of $w$ and $w_a$ are consistent with
$\Lambda$CDM (Figures \ref{fig:wcdm} and \ref{fig:wzcdm}).
With SNe+Planck+BAO+H$_0$ constraints, we find $w =$ \wbaoho\
for the $w$-CDM model and $w_a =$ \wabaoho\ for the $w_a$-CDM
model (Figure \ref{fig:wzcdm}).  With just SNe, Planck, and BAO
data, we find $w =$ \wbao\ for the $w$-CDM model and $w_a =$ \wabao\
for the $w_a$-CDM model.

Nearly all measurements
of $\Omega_k$ are consistent with a flat universe. The lone
exception is the combination of SNe, Planck and H$_0$.  This
choice of priors gives 3$\sigma$ evidence for
positive curvature, but the result is entirely due to the
local/CMB H$_0$ discrepancy and becomes insignificant
when BAO constraints are added.

As shown from the H$_0$ measurements in Table
\ref{table:allcosmo}, PS1+low-$z$ SNe and the non-$\Lambda$CDM
models considered here do not explain the local/CMB H$_0$ discrepancy.  When
H$_0$ priors are omitted, all measurements of
H$_0$ are inconsistent with \citet{Riess16} at the $\sim$2-3$\sigma$
level and would also be inconsistent with
other local measurements of H$_0$ \citep{Bonvin17,Jang17}.
When only CMB and H$_0$ priors are included, we measure
values of H$_0$ that are consistent with \citet{Riess16} only when
allowing for positive curvature or evolving $w$.  When we
combine with CMB, H$_0$ and BAO priors, all measurements of
H$_0$ are inconsistent with \citet{Riess16}
at the 2.6$\sigma$ to 2.8$\sigma$ level \textit{even
  though} H$_0$ priors are included.  Therefore,
SNe\,Ia and the models considered here do not favor
a non-$\Lambda$CDM universe and disfavor a scenario
where the H$_0$ discrepancy is due to non-cosmological constant
dark energy.


\setlength\doublerulesep{1mm}
\begin{deluxetable*}{lcccccccccccc}
\tabletypesize{\scriptsize}
\tablewidth{0pt}
\tablecaption{Cosmological Parameters from PS1, BAO, CMB, and H$_0$}
\tablehead{&\multicolumn{12}{c}{$o-\Lambda$CDM Constraints} \\*[2 pt]
  &\multicolumn{3}{c}{$\Omega_M$}&\multicolumn{3}{c}{$\Omega_{\Lambda}$}&\multicolumn{3}{c}{$\Omega_{k}$}&\multicolumn{3}{c}{H$_0$}}
\startdata
\tableline\\*[2 pt]
PS1$+$Planck$+$BAO$+$H$_0$&\multicolumn{3}{c}{0.303$\pm$0.007}&\multicolumn{3}{c}{0.694$\pm$0.008}&\multicolumn{3}{c}{0.003$\pm$0.002}&\multicolumn{3}{c}{68.682$\pm$0.694}\\
\tableline\\*[2 pt]
PS1$+$Planck&\multicolumn{3}{c}{0.330$\pm$0.045}&\multicolumn{3}{c}{0.674$\pm$0.035}&\multicolumn{3}{c}{-0.004$\pm$0.011}&\multicolumn{3}{c}{66.205$\pm$4.659}\\
PS1$+$Planck$+$BAO&\multicolumn{3}{c}{0.310$\pm$0.007}&\multicolumn{3}{c}{0.689$\pm$0.008}&\multicolumn{3}{c}{0.001$\pm$0.003}&\multicolumn{3}{c}{67.892$\pm$0.714}\\
PS1$+$Planck$+$H$_0$&\multicolumn{3}{c}{0.272$\pm$0.014}&\multicolumn{3}{c}{0.718$\pm$0.012}&\multicolumn{3}{c}{0.009$\pm$0.003}&\multicolumn{3}{c}{72.522$\pm$1.748}\\
\\*[2pt]
&\multicolumn{12}{c}{$w$-CDM Constraints}\\*[2 pt]
&\multicolumn{3}{c}{$\Omega_M$}&\multicolumn{3}{c}{$w$}&\multicolumn{3}{c}{H$_0$}&&&\\*[2 pt]
\hline \hline\\*[2 pt]
PS1$+$Planck$+$BAO$+$H$_0$&\multicolumn{3}{c}{0.299$\pm$0.008}&\multicolumn{3}{c}{-1.045$\pm$0.045}&\multicolumn{3}{c}{69.007$\pm$0.980}&&&\\
\tableline\\*[2 pt]
PS1$+$Planck&\multicolumn{3}{c}{0.317$\pm$0.017}&\multicolumn{3}{c}{-0.989$\pm$0.057}&\multicolumn{3}{c}{67.140$\pm$1.664}&&&\\
PS1$+$Planck$+$BAO&\multicolumn{3}{c}{0.312$\pm$0.010}&\multicolumn{3}{c}{-0.984$\pm$0.048}&\multicolumn{3}{c}{67.364$\pm$1.091}&&&\\
PS1$+$Planck$+$H$_0$&\multicolumn{3}{c}{0.289$\pm$0.012}&\multicolumn{3}{c}{-1.067$\pm$0.046}&\multicolumn{3}{c}{70.042$\pm$1.263}&&&\\
\\*[2pt]
&\multicolumn{12}{c}{$w_a$-CDM Constraints}\\*[2 pt]
&\multicolumn{3}{c}{$\Omega_M$}&\multicolumn{3}{c}{$w_0$}&\multicolumn{3}{c}{$w_a$}&\multicolumn{3}{c}{H$_0$}\\*[2 pt]
\hline \hline\\*[2 pt]
PS1$+$Planck$+$BAO$+$H$_0$&\multicolumn{3}{c}{0.301$\pm$0.009}&\multicolumn{3}{c}{-0.972$\pm$0.102}&\multicolumn{3}{c}{-0.372$\pm$0.452}&\multicolumn{3}{c}{69.011$\pm$0.994}\\
\tableline\\*[2 pt]
PS1$+$Planck&\multicolumn{3}{c}{0.308$\pm$0.026}&\multicolumn{3}{c}{-0.912$\pm$0.149}&\multicolumn{3}{c}{-0.513$\pm$0.826}&\multicolumn{3}{c}{68.276$\pm$2.752}\\
PS1$+$Planck$+$BAO&\multicolumn{3}{c}{0.314$\pm$0.010}&\multicolumn{3}{c}{-0.920$\pm$0.103}&\multicolumn{3}{c}{-0.313$\pm$0.418}&\multicolumn{3}{c}{67.371$\pm$1.117}\\
PS1$+$Planck$+$H$_0$&\multicolumn{3}{c}{0.277$\pm$0.012}&\multicolumn{3}{c}{-0.812$\pm$0.104}&\multicolumn{3}{c}{-1.323$\pm$0.493}&\multicolumn{3}{c}{71.611$\pm$1.365}\\
\enddata
\label{table:allcosmo}
\end{deluxetable*}

\begin{figure*}
\includegraphics[width=6.5in]{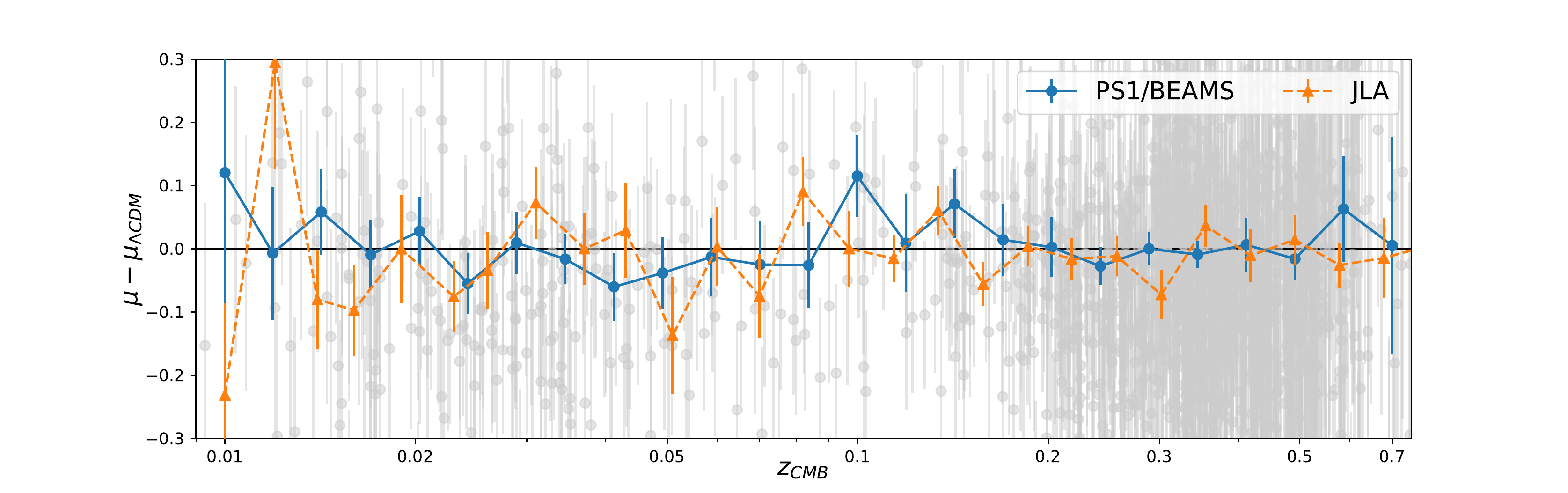}
\caption{The PS1$+$low-$z$ Hubble residual diagram, with a comparison
  to the binned SN\,Ia distances given by B14.  We see excellent agreement
  with B14 across the redshift range, with slight discrepancies at
  low-$z$ due to the addition of the CfA4 sample and a stronger
  prediction for the distance bias correction.}
\label{fig:hubbleresid}
\end{figure*}

\begin{table*}
\caption{\fontsize{9}{11}\selectfont Comparison to JLA and Pantheon Cosmological Constraints}
  \centering
%
\begin{tabular}{lrrrrrrr}
  \hline \hline\\[-1.5ex]
&This Work&&\multicolumn{2}{c}{JLA}&&\multicolumn{2}{c}{Pantheon}\\*[2pt]
&\multicolumn{1}{c}{$w$}&&\multicolumn{1}{c}{$w$}&\multicolumn{1}{c}{Diff.}&&\multicolumn{1}{c}{$w$}&\multicolumn{1}{c}{Diff.}\\
  \cline{2-2}\cline{4-5}\cline{7-8}\\[-1.5ex]
SNe$+$Planck&-0.989$\pm$0.057&&-1.017$\pm$0.056&0.028$\pm$0.080 (0.35$\sigma$)&&-1.026$\pm$0.041&0.037$\pm$0.070 (0.52$\sigma$)\\
SNe$+$Planck$+$BAO&-0.984$\pm$0.048&&-1.003$\pm$0.047&0.019$\pm$0.068 (0.28$\sigma$)&&-1.014$\pm$0.040&0.030$\pm$0.063 (0.48$\sigma$)\\
SNe$+$Planck$+$H$_0$&-1.067$\pm$0.046&&-1.064$\pm$0.051&0.010$\pm$0.068 (0.15$\sigma$)&&-1.056$\pm$0.038&-0.011$\pm$0.060 (0.19$\sigma$)\\
SNe$+$Planck$+$BAO$+$H$_0$&-1.045$\pm$0.045&&-1.038$\pm$0.047&0.012$\pm$0.065 (0.18$\sigma$)&&-1.047$\pm$0.038&0.002$\pm$0.059 (0.03$\sigma$)\\
&\multicolumn{1}{c}{$w_a$}&&\multicolumn{1}{c}{$w_a$}&\multicolumn{1}{c}{Diff.}&&\multicolumn{1}{c}{$w_a$}&\multicolumn{1}{c}{Diff.}\\
\cline{2-2}\cline{4-5}\cline{7-8}\\[-1.5ex]
SNe$+$Planck&-0.513$\pm$0.826&&-0.608$\pm$0.748&0.095$\pm$1.115 (0.09$\sigma$)&&-0.129$\pm$0.755&-0.384$\pm$1.119 (0.34$\sigma$)\\
SNe$+$Planck$+$BAO&-0.313$\pm$0.418&&-0.280$\pm$0.433&-0.033$\pm$0.602 (0.05$\sigma$)&&-0.126$\pm$0.384&-0.187$\pm$0.567 (0.33$\sigma$)\\
SNe$+$Planck$+$H$_0$&-1.323$\pm$0.493&&-1.055$\pm$0.586&-0.168$\pm$0.737 (0.23$\sigma$)&&-0.742$\pm$0.465&-0.581$\pm$0.678 (0.86$\sigma$)\\
SNe$+$Planck$+$BAO$+$H$_0$&-0.372$\pm$0.452&&-0.290$\pm$0.443&-0.073$\pm$0.648 (0.11$\sigma$)&&-0.222$\pm$0.407&-0.150$\pm$0.608 (0.25$\sigma$)\\
\hline
\end{tabular}
\label{table:b14comp}
\end{table*}

\subsection{Consistency with JLA and Pantheon Results}
\label{sec:unknowns}

The binned SN\,Ia distances from
our likelihood model are
compared to the JLA sample
in Figure \ref{fig:hubbleresid} (using
the correlated bins given by B14, Appendix F).
The agreement is close;
using weighted average Hubble residuals, PS1+low-z
distances are just 4 mmag fainter at $z > 0.2$ compared to
$z < 0.2$.

Measurements of $w$ and $\Omega_M$ in this work show excellent
agreement with B14 and the Pantheon sample (S17).
For the flat $w$-CDM model, Table \ref{table:b14comp} shows the
drift in the values of $w$ we measure with respect
to B14 and S17.  All values are consistent with B14 values to within
0.4$\sigma$.
Though these measurements are correlated, as B14 use
$\sim$75\% of the low-$z$ SNe that we do (with the exception of CfA4 and CSP2),
and we combine both SN datasets with the same CMB, BAO, and H$_0$
data, such close agreement is encouraging.

Similarly, our measurements are consistent with S17 at
$\lesssim$0.5$\sigma$.  Though these samples
are not entirely independent $-$ 34\% of the SNe\,Ia
here are included in the Pantheon sample $-$ the samples are subject
to different systematic uncertainties as well as statistical.
The 2\% discrepancy between the S17 measurement of $w$ and ours
is well within the uncertainty budget of our measurement.
In future work, we hope to combine our sample with the Pantheon data,
as this combined sample would likely provide the best current
constraints on $w$ and include just under 2,000 SNe.

Though the results presented here remain subject to
uncertainty in the population of CC\,SNe contaminating
the SN data, the agreement with other measurements is encouraging.
Our cosmological
parameter measurements also remain consistent when using
several variants as part of the BEAMS framework.
The consistency of these results with measurements
from spectroscopically confirmed PS1 SNe determined by S17
($w = -0.990\pm0.063$) gives
us additional confidence in their robustness.

In the next few years, we also expect additional CC\,SN templates
and better constraints on CC\,SN luminosity functions
will lead to even more robust simulation-based tests
for this method and other similar methods.

\section{Measuring $w$ without Spectroscopic Classifications}
\label{sec:nospec}

\begin{figure}
\includegraphics[width=3.25in]{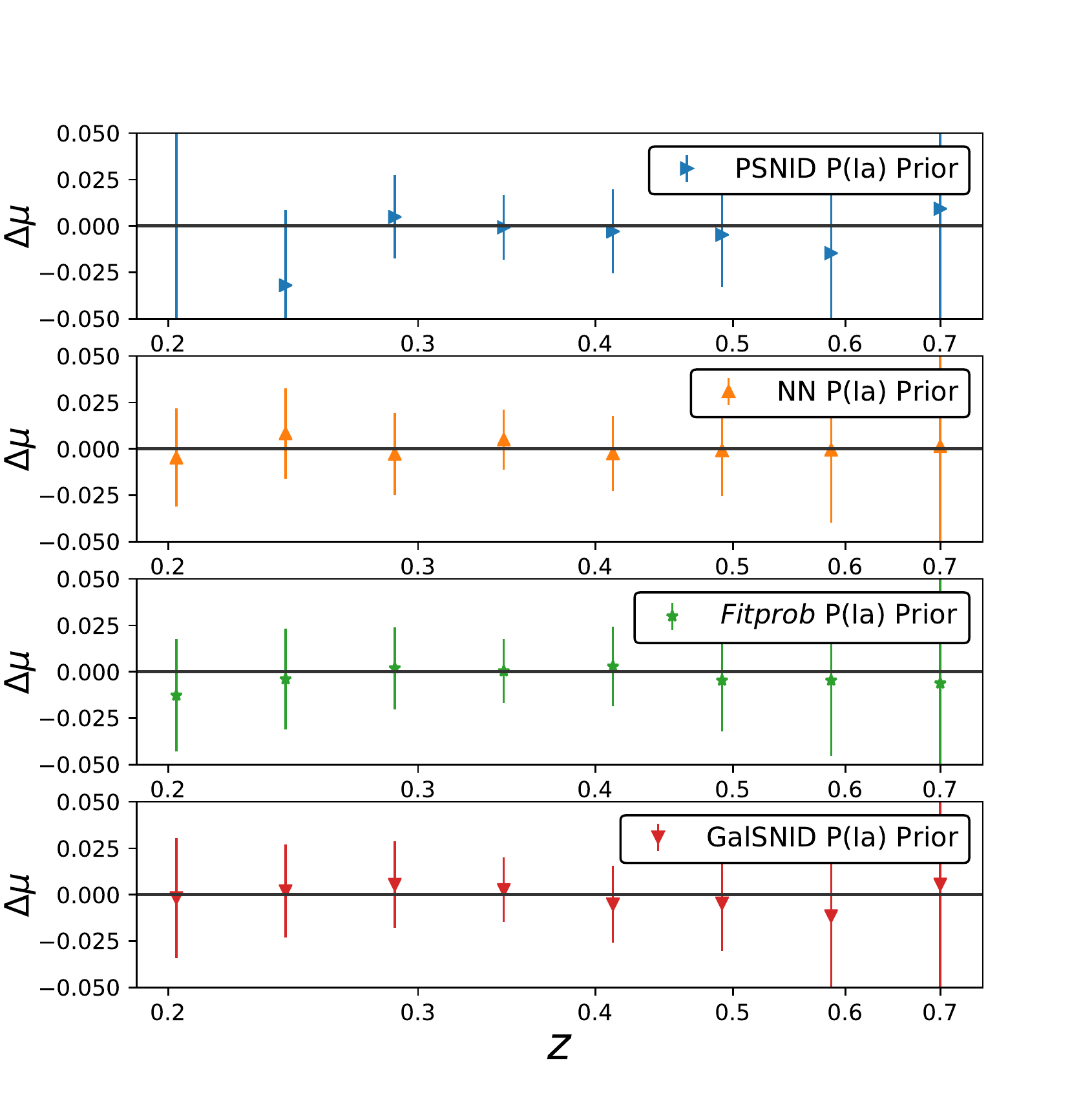}
\caption{Changes in binned distances when spectroscopic classifications
  are ignored.  $\Delta\mu$
  is the bias on distance when photometric classifications are
  used for the $\sim$13\% of the sample with spectroscopic
  classifications available.
  As we predict in J17, typical biases are $<$5 mmag for all P(Ia) priors
  at $0.2 < z < 0.5$ (the average is just 4 mmag for PSNID),
  with occasionally larger biases in bins with higher
  statistical uncertainties.  PSNID classifies few PS1 SNe at z $<$ 0.25
  as being likely SNe\,Ia, and therefore provides no meaningful
  constraints on distances at these redshifts.}
\label{fig:hostzdist}
\end{figure}

%

\begin{figure*}
\includegraphics[width=7in]{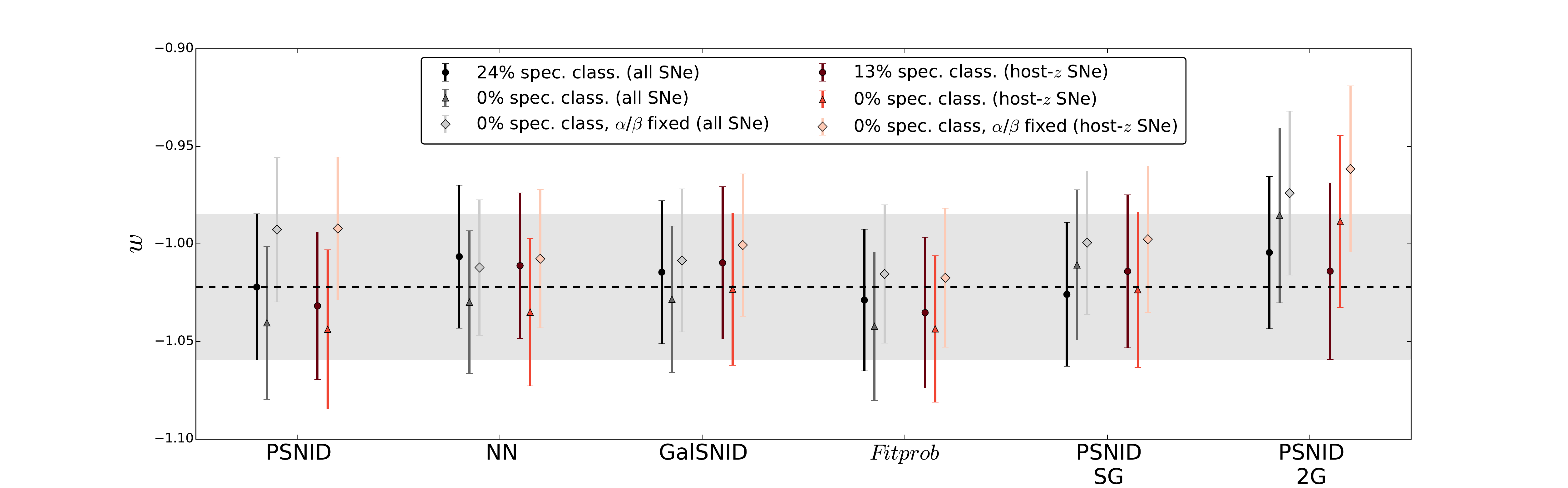}
  \caption{Changes in $w$ if photometric classifications
    are used instead of the available spectroscopic
    classifications.  The final, statistics-only measurement
    from PS1+Planck of $w = -1.022 \pm 0.037$ is shown
    with the dashed line and shaded region.  We also show the results if
    $\alpha$ and $\beta$ remain fixed to the values
    measured from spectroscopically classified
    PS1 SNe (light shading; $\alpha$ and $\beta$ are 0.155 and 2.95,
    respectively).  We find
    that biases of up to $\sim$4\% 
    can arise when spectroscopic
    classifications are not available, but are typically ameliorated by
    fixing $\alpha$ and $\beta$ to the values measured from spectroscopic samples.}
\label{fig:wclass}
\end{figure*}
  
Throughout this analysis, we have used spectroscopically
confirmed SNe\,Ia to bolster our cosmological results.
However, future samples from DES and LSST
may not have a large fraction of spectroscopic
classifications.  Here, we examine
distances, nuisance parameters, and measurements of $w$
in the case where no spectroscopic classifications
of PS1 SNe are available; we substitute photometric classifications
for the available spectroscopic classifications, apply our likelihood model, and
measure the resulting bias on $w$.
We investigate the cases of both the
full PS1 dataset and the host-$z$ sample alone
(only SNe with spectroscopic host galaxy redshifts)
to determine whether
our methodology can provide consistent results when
spectroscopic classifications are lacking.
The host-$z$ sample in particular is nearly an ideal,
magnitude-limited sample, albeit with
host galaxy selection biases.
24\% of SNe in the full PS1 dataset are spectroscopically
classified SNe\,Ia and 13\% of SNe in the host-$z$ sample are
spectroscopically classified.

When photometric classifications are used instead of
spectroscopic classifications, Figure \ref{fig:hostzdist} shows that
the binned SN\,Ia distances may occasionally change by
$>$0.05 mag where statistical uncertainties are large.
However, at $0.25 \lesssim z \lesssim 0.5$, where
$\sim$75\% of our data lie, we see median biases
less than 5 mmag for all methods.  This is in agreement with predictions
from J17, who found that in 25 samples of 1,000 high-$z$ SNe,
bias due to marginalizing over CC\,SN contamination
averaged $<$5 mmag and had sample-to-sample variations of $\sim$15 mmag in this
redshift range.  Although here we change at most 24\% of the
classifications in the sample, the results remain broadly
consistent with simulations.

In Figure \ref{fig:wclass} we
examine the change in measured $w$
if spectroscopic classifications are not used.
From every classifier, in both the full and host-$z$ samples,
we measure a $w$ consistent with the statistical
uncertainties on our
best measurement of $w$, $\sigma_w = 0.037$ (and $w$ derived from the host-$z$
sample prefers a value of $-1.032$, 1\% lower than the full sample
due to statistical fluctuations alone).
However the bias in $w$ can be nearly $\sim$4\% for the least informative classifiers
in this analysis (\fp\ and GalSNID), which constitutes a strong
argument for including a subset of spectroscopic classifications
when measuring $w$ or fixing $\alpha$ and $\beta$ to the values measured from
spectroscopic samples $-$ in this case, from PS1 spectroscopically
confirmed SNe.  Fixing $\alpha$ and $\beta$
can often improve the reliability of a methods, an effect we show in
Figure \ref{fig:wclass}.

It is reassuring that many of our results appear to
confirm what we predicted using simulations in J17.
We see likely negative biases on $w$ when using the
GalSNID, and \fp\ classifiers, and (typically) more consistent results after
fixing $\alpha$ and $\beta$.
We predicted distance biases at $0.2 < z < 0.5$
of $\lesssim$15 mmag in a given SN sample, and our results here are consistent with
that finding.
Though not statistically significant, the $\sim$few
percent differences in nuisance parameters between spectroscopically
classified SNe alone and the full sample are in the direction
we would expect.
With the advent of more robust classifiers
and better training samples, we expect the systematic uncertainties to decrease
and the reliability of simulations to improve.
Even with some modest discrepancies,
we see that the consistency level for nearly all methods is well within the
uncertainty budget on $w$, demonstrating a promising future for
SN cosmology with photometrically classified SNe.

\section{Conclusions}
\label{sec:conclusions}

The \numsnetotal\ cosmologically useful, likely
SNe\,Ia from the PS1 medium deep fields and low-$z$ surveys
constitute the largest set of SNe\,Ia assembled to date.  
Our cosmological measurement uncertainties are almost identical
to those of the JLA compilation, due
to the smaller redshift range and lower SNR of the SNe\,Ia in our sample,
but the measurements presented
here are independent of the JLA data at $z > 0.1$.
In the future, these data can be used in
conjunction with the Foundation low-$z$ SN sample \citep{Foley17}
to give independent constraints on $w$ using \textit{only} the
well-calibrated PS1 photometric system.  The SN light curves,
host galaxy spectra and host galaxy redshifts presented in this work
are available at \dataset[10.17909/T95Q4X]{https://doi.org/10.17909/T95Q4X}.

The PS1 SNe in this sample do not have spectroscopic
classifications, necessitating a Bayesian framework that
marginalizes over the CC\,SN population.  By applying this
framework, we compute binned distances from SNe\,Ia that
are an average of just 4 mmag fainter at $z > 0.2$,
compared to $z < 0.2$, than JLA distances.  From J17, we
found that this method of marginalizing over CC\,SNe in a PS1-like sample
will bias $w$ by a statistically insignificant 0.001$\pm$0.004.

From these data, we find that shape- and color-corrected SNe\,Ia in host galaxies with
$M_{\ast}/M_{\odot} > 10$ dex are \massstep\ mag (stat+sys) brighter on average than
those in $M_{\ast}/M_{\odot} < 10$ dex hosts, consistent with previous measurements.  We find no evidence for
evolution of the mass step with redshift (e.g. \citealp{Childress14})
but $\sim$1.6$\sigma$ evidence for evolution in the SALT2 $\beta$ parameter
(the correlation between SN color and luminosity).

After including CMB data, we find that PS1 SN
data are fully consistent with
a flat $\Lambda$CDM cosmology, with $w = $\w.  Combining SNe with
CMB and BAO constraints gives $w =$ \wbao\ and
adding H$_0$ constraints yields $w = $ \wbaoho.
If we allow $w$ to be parameterized by a constant component ($w_0$)
and a component that evolves with redshift ($w_a$), we find
no evidence for a $z$-dependent value of $w$.
Our constraints differ from
those of B14 by $<$0.4$\sigma$ regardless of whether
CMB, BAO, and/or H$_0$ priors are included.
They are also consistent with the constraints
from \citet{Scolnic17}.

CC\,SN contamination is the third smallest systematic
uncertainty in this analysis, and can be improved further with new
SN classification algorithms and better training
samples, as discussed in J17.  In future work, our
dominant systematics $-$ selection biases and calibration $-$ can be reduced by
combining PS1 data with Foundation and/or SNLS and SDSS data.


In carrying out this analysis, we note that we
did not blind ourselves to the cosmological results.
A blinded analysis, such as that of S17, would remove any
subconscious bias on the part of the authors to
achieve agreement (or disagreement) with $\Lambda$CDM cosmology.
We note, however that all of the photometry and most of
the bias correction simulations were undertaken before
the cosmological results were examined.  Furthermore, we
have strived for consistency with previous analyses whenever
possible, which serves to limit the number of qualitative
choices that can be tuned to yield a preferred cosmology.
Future analyses, such as DES SN\,Ia cosmology, will be fully
blinded.  As cosmology with photometrically classified SNe\,Ia
becomes a more mature subject area, the authors will feel more
comfortable undertaking blinded analyses.

In future years, SN samples from the Dark Energy Survey (DES)
and the Large Synoptic Survey Telescope (LSST) will measure
$w$ with larger, higher-SNR samples of SNe without
spectroscopic classifications.  Though CC\,SN contamination is
the second largest source of systematic uncertainty on $w$ in this
analysis, we expect that the systematic uncertainty on $w$
from CC\,SN contamination will be greatly reduced in the next few years.
Improvements will be due to larger samples of CC\,SN templates
that can be used to train SN classification algorithms
and a better understanding of the shape of the CC\,SN luminosity function.
We hope that the methods presented
here will demonstrate the robustness of measuring $w$
from photometrically classified samples as we continue to gain a better understanding of
the nature of dark energy.

\acknowledgements
We would like to thank the referee for many insightful comments,
which were helpful in improving this manuscript.
We would also like to thank Scott Fleming and the Space
Telescope Science Institute for their invaluable
assistance in making the data presented in this work
publicly available and user-friendly.  Michael Foley
also had many useful suggestions that improved this analysis.

D.O.J. is supported by a Gordon and Betty Moore Foundation
postdoctoral fellowship at the University of California, Santa Cruz.
This manuscript is based upon work supported by the National
Aeronautics and Space Administration under Contract No.\ NNG16PJ34C
issued through the {\it WFIRST} Science Investigation Teams Programme.
R.J.F.\ and D.S.\ were supported in part by NASA grant 14-WPS14-0048.
The UCSC group is supported in part by NSF grant AST-1518052 and from
fellowships from the Alfred P.\ Sloan Foundation and the David and
Lucile Packard Foundation to R.J.F.  This work was supported in part
by the Kavli Institute for Cosmological Physics at the University of
Chicago through grant NSF PHY-1125897 and an endowment from the Kavli
Foundation and its founder Fred Kavli.  D.S, gratefully acknowledges
support from NASA grant 14-WPS14-0048. D.S. is supported by NASA through
Hubble Fellowship grant HST-HF2-51383.001 awarded by the Space Telescope
Science Institute, which is operated by the Association of Universities
for Research in Astronomy, Inc., for NASA, under contract NAS 5-26555.

Many of the observations reported here were obtained at the MMT Observatory, 
a joint facility of the Smithsonian Institution and the 
University of Arizona.  This paper uses data products produced by 
the OIR Telescope Data Center, supported by the Smithsonian 
Astrophysical Observatory.  Additional data are thanks to the Anglo Australian
Telescope, operated by the Australian Astronomical Observatory, through
the National Optical Astronomy Observatory (NOAO PropID: 2014B-N0336; 
PI: D. Jones).  We also use data from observations at Kitt Peak National 
Observatory, National Optical Astronomy Observatory,
which is operated by the Association of Universities for Research in 
Astronomy (AURA) under a cooperative agreement with the National 
Science Foundation.  Also based on observations obtained with the Apache 
Point Observatory 3.5-meter telescope, which is owned and operated 
by the Astrophysical Research Consortium.

The computations in this paper used a combination 
of three computing clusters.  BEAMS analysis was performed 
using the University of Chicago Research Computing Center 
and the Odyssey cluster at Harvard University.  We are grateful 
for the support of the University of Chicago Research Computing Center for 
assistance with the calculations carried out in this work.  The 
Odyssey cluster is supported by the FAS Division of Science, Research 
Computing Group at Harvard University.  Supernova light curve 
reprocessing would not have been possible without the Data-Scope project 
at the Institute for Data Intensive Engineering and Science at 
Johns Hopkins University.

Funding for the Sloan Digital Sky Survey IV has been provided by the
Alfred P. Sloan Foundation, the U.S. Department of Energy Office of
Science, and the Participating Institutions. SDSS- IV acknowledges
support and resources from the Center for High-Performance Computing at
the University of Utah. The SDSS web site is www.sdss.org.

SDSS-IV is managed by the Astrophysical Research Consortium for the 
Participating Institutions of the SDSS Collaboration including the 
Brazilian Participation Group, the Carnegie Institution for Science, 
Carnegie Mellon University, the Chilean Participation Group, the 
French Participation Group, Harvard-Smithsonian Center for Astrophysics, 
Instituto de Astrofísica de Canarias, The Johns Hopkins University, 
Kavli Institute for the Physics and Mathematics of the Universe 
(IPMU) / University of Tokyo, Lawrence Berkeley National Laboratory, 
Leibniz Institut für Astrophysik Potsdam (AIP), Max-Planck-Institut 
für Astronomie (MPIA Heidelberg), Max-Planck-Institut für Astrophysik 
(MPA Garching), Max-Planck-Institut für Extraterrestrische Physik (MPE), 
National Astronomical Observatory of China, New Mexico State University, 
New York University, University of Notre Dame, Observatório Nacional / MCTI, 
The Ohio State University, Pennsylvania State University, Shanghai 
Astronomical Observatory, United Kingdom Participation Group, Universidad 
Nacional Autónoma de México, University of Arizona, University of Colorado 
Boulder, University of Oxford, University of Portsmouth, University of 
Utah, University of Virginia, University of Washington, University of 
Wisconsin, Vanderbilt University, and Yale University.

This research makes use of the VIPERS-MLS database, operated at CeSAM/LAM, 
Marseille, France. This work is based in part on observations obtained 
with WIRCam, a joint project of CFHT, Taiwan, Korea, Canada and France. 
The CFHT is operated by the National Research Council (NRC) of Canada, 
the Institut National des Science de l’Univers of the Centre National de 
la Recherche Scientifique (CNRS) of France, and the University of Hawaii. 
This work is based in part on observations made with the Galaxy Evolution 
Explorer (GALEX). GALEX is a NASA Small Explorer, whose mission was 
developed in cooperation with the Centre National d’Etudes Spatiales 
(CNES) of France and the Korean Ministry of Science and Technology. GALEX 
is operated for NASA by the California Institute of Technology under NASA
contract NAS5-98034. This work is based in part on data products produced at 
TERAPIX available at the Canadian Astronomy Data Centre as part of the 
Canada-France-Hawaii Telescope Legacy Survey, a collaborative project of 
NRC and CNRS. The TERAPIX team has performed the reduction of all the 
WIRCAM images and the preparation of the catalogues matched with the 
T0007 CFHTLS data release.

Funding for the DEEP2 Galaxy Redshift Survey has been
provided by NSF grants AST-95-09298, AST-0071048, AST-0507428, and 
AST-0507483 as well as NASA LTSA grant NNG04GC89G.  This research uses data 
from the VIMOS VLT Deep Survey, obtained from the VVDS database operated 
by Cesam, Laboratoire d'Astrophysique de Marseille, France.  zCosmos data 
are based on observations made with ESO Telescopes at the La Silla or 
Paranal Observatories under programme ID 175.A-0839.

\appendix
\section{A. Simulating Evolving $x_1$ and $c$ Distributions}
\label{section:x1cevol}
In this appendix, we discuss the improvement to the PS1
simulations by allowing the mean simulated $x_1$
and $c$ to evolve with redshift.
We consider the standard approach of fixed $x_1$ and $c$
populations insufficient for our
analysis, because the PS1 host-$z$ sample has
redshift-dependent host galaxy properties due to our
magnitude-limited host galaxy redshift follow-up program.
Similarly, the SN-$z$ sample consists of SNe not included
in the host-$z$ sample and therefore also has a $z$-dependent bias.
Because $x_1$ and $c$ depend on host mass, their distributions
change as a function of $z$ in a way that is not due
only to selection biases.

Using the default
simulations for the host-$z$ and SN-$z$ samples from
J17 and S17, respectively, we fit a 3rd-order polynomial
to the difference between the simulations and the
data after binning in redshift ($\Delta z = 0.05$).  We used
these polynomials as inputs to SNANA, allowing them to
define the intrinsic evolution of $x_1$
and $c$ with redshift.

\begin{figure}
\includegraphics[width=3.5in]{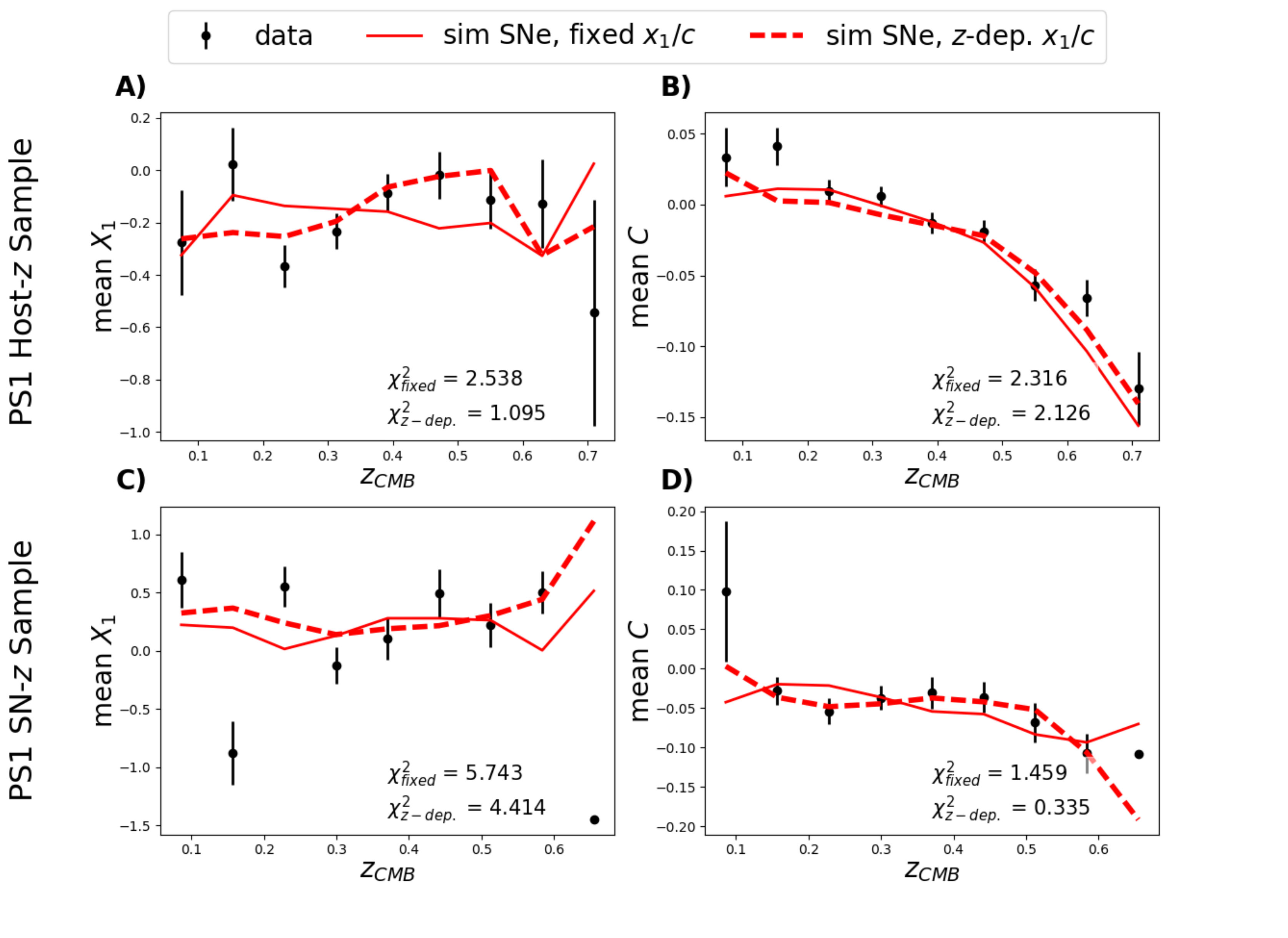}
\caption{The $z$-dependence of $x_1$ and $c$.  Allowing
  $x_1$ and $c$ to evolve with redshift allows simulations to better
  match the data.}
\label{fig:simcompvz}
\end{figure}

\begin{figure*}
\includegraphics[width=7in]{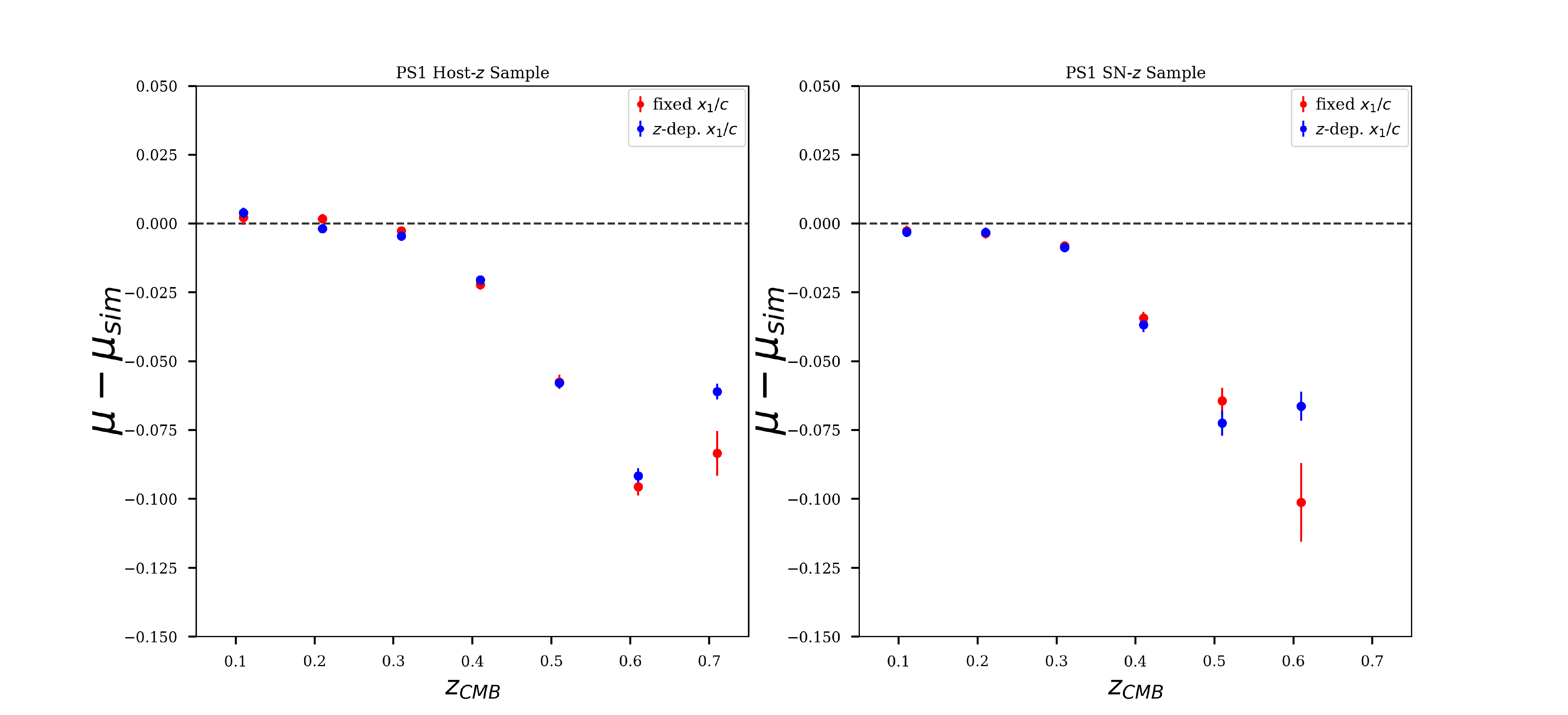}
\caption{Change in distance bias when simulating 
  $x_1$/$c$ distributions that evolve with redshift (G10 model).  The new simulations can affect the
  distance bias by $\sim$0.01-0.02 mag at high $z$.}
\label{fig:biascor}
\end{figure*}

Figure \ref{fig:simcompvz} shows the
redshift dependence of the $x_1$ and $c$ distributions
in simulations with fixed and evolving $x_1$/$c$.
Though allowing $x_1$ and $c$ to evolve with
redshift does improve the simulations, these
new simulations are only a moderately better
match to the data.

Figure \ref{fig:biascor} show the difference in
bias corrections using the G10 scatter model with
and without $z$-dependent $x_1$ and $c$ populations.
If $x_1$ and $c$ are redshift dependent, the distance
bias is slightly larger for the host-$z$ sample and
smaller by up to 0.02 mag at high $z$ for the SN-$z$ sample.

\bibliographystyle{apj}
\bibliography{ms}

\end{document}